\documentclass[journal]{IEEEtran}
\usepackage{setspace}
\usepackage{graphicx}
\usepackage{amstext}
\usepackage{algorithm}
\usepackage{algorithmic}
\usepackage{mathrsfs}
\usepackage{amssymb}
\usepackage{amsmath}
\usepackage{bm}
\allowdisplaybreaks[4]
\usepackage{epstopdf}
\usepackage{multicol}
\usepackage{stfloats}
\usepackage{url}
\usepackage{color}
\usepackage{enumerate}
\usepackage{breqn}
\usepackage{bbm}
\usepackage{cite}
\usepackage{subfig}
\usepackage{caption}
\usepackage{enumitem}
\usepackage{mathrsfs}
\usepackage{multirow}

\DeclareMathAlphabet{\mathscrbf}{OMS}{mdugm}{b}{n}

\usepackage[T1]{fontenc}

\DeclareFontFamily{OML}{zavm}{\skewchar\font=127 }
\DeclareFontShape{OML}{zavm}{m}{it}{<-> s*[.78] zavmri7m}{}
\DeclareFontShape{OML}{zavm}{b}{it}{<-> s*[.78] zavmbi7m}{}
\DeclareFontShape{OML}{zavm}{m}{sl}{<->ssub * zavm/m/it}{}
\DeclareFontShape{OML}{zavm}{bx}{it}{<->ssub * zavm/b/it}{}
\DeclareFontShape{OML}{zavm}{b}{sl}{<->ssub * zavm/b/it}{}
\DeclareFontShape{OML}{zavm}{bx}{sl}{<->ssub * zavm/b/sl}{}
\DeclareMathAlphabet{\mathsf}{OML}{zavm}{m}{it} % not `n'

\newtheorem{Theorem}{Theorem}
\newtheorem{Lemma}{Lemma}
\newtheorem{Remark}{Remark}

\newtheorem{Proposition}{Proposition}
\newtheorem{Property}{Property}
\newtheorem{Problem}{Problem}

\begin{document}
\title{  Enhancing Physical Layer Security of Random Caching in Large-Scale Multi-Antenna Heterogeneous Wireless Networks}
\author{\IEEEauthorblockN{Wanli Wen, Chenxi Liu, Yaru Fu, Tony Q. S. Quek,  Fu-Chun Zheng, and Shi Jin}
\thanks{W. Wen, Y. Fu, and T. Q. S. Quek are with the  Information Systems Technology and Design Pillar, Singapore University of Technology and Design, Singapore 487372 (email: wanli_wen@sutd.edu.sg, yaru_fu@sutd.edu.sg, tonyquek@sutd.edu.sg).}
\thanks{C. Liu is with State Key Laboratory of Networking and Switching
Technology, Beijing University of Posts and Telecommunications,
Beijing 100876, China (e-mail: chenxi.liu@bupt.edu.cn).}
\thanks{F.-C. Zheng and S. Jin are with the National Mobile Communications Research Laboratory, Southeast University, Nanjing, China (email:  fzheng@seu.edu.cn, jinshi@seu.edu.cn).}
}
\maketitle

\begin{abstract}

  In this paper, we {propose a novel secure random caching scheme for} large-scale multi-antenna heterogeneous wireless networks, where the base stations (BSs) deliver randomly cached confidential contents to the legitimate users in the presence of passive eavesdroppers as well as active jammers. In order to safeguard the content delivery, we consider that the BSs transmits the artificial noise together with the useful signals. By using tools from stochastic geometry, we first analyze the average reliable transmission probability (RTP) and the average confidential transmission probability (CTP), which take both the impact of the eavesdroppers and the impact of the jammers into consideration. We further provide tight upper and lower bounds on the average RTP. These analytical results enable us to obtain rich insights into the behaviors of the average RTP and the average CTP with respect to key system parameters. Moreover, we optimize the caching distribution of the files to maximize the average RTP of the system, while satisfying the constraints on the caching size and the average CTP. Through numerical results, we show that our proposed secure random caching scheme can effectively boost the secrecy performance of the system compared to the existing solutions.

\end{abstract}

\IEEEpeerreviewmaketitle

\setlength{\textfloatsep}{2pt} %% reduce the space after algorithm

\section{Introduction}

{With the rapid development of wireless technology, the global mobile data traffic is expected to reach 77 exabytes per month by 2022, which is a sevenfold increase over 2017 \cite{cisco2017global}.}
However, the majority of such traffic is asynchronously but repeatedly requested by many users at different times and thus a tremendous amount of mobile data traffic has actually been redundantly generated over the mobile networks \cite{Cachingattheirelessedgedesignaspectschallengesfuturedirections}. Against this backdrop, caching popular contents at the base stations (BSs) has been proposed as a promising approach for shifting the huge mobile data traffic from remote clouds to the edges in mobile networks \cite{Lisurvey, NOMAcachingWCL2019}, thereby significantly improving the users' quality of service (QoS). Motivated by this, cache-enabled wireless networks has recently been attracting great research~attention.

However, due to the equipment size and cost issues, the storage resources at BSs are usually limited. Therefore, which contents should be cached at BSs becomes a vital design problem, refereed to as the caching scheme design problem, in cache-enabled wireless networks. 
In \cite{Cache-enabledsmallcellnetworksmodelingandtradeoffs, Tamoorulhassan2015Modeling, OptimalGeographicCachingInCellularNetworks, RCKHetNetJuan2017, OptimalContentPlacementOffloadingHetNets,RCHetNetTao2018, RandomcachingCoMP2018Wen, RCCHetNetcui2017, RandomcachingDTX2018Wen,2018arXiv180102743J, D2Dcontentdeliverynetworking2014}, the authors {proposed various caching schemes to improve the content transmission reliability (i.e., the probability that the requested content can be successfully transmitted from the BSs to the users, which can reflect the users' QoS)} in large-scale cache-enabled wireless networks. 
Specifically, in \cite{Cache-enabledsmallcellnetworksmodelingandtradeoffs}, the most popular caching scheme was studied, in which each BS only stores the most popular files. {However, such a caching scheme may not yield the optimal network performance, since it cannot} provide any content diversity \cite{RCCHetNetcui2017,RandomcachingCoMP2018Wen, RandomcachingDTX2018Wen}. To {tackle this issue}, random caching scheme {was considered} in \cite{Tamoorulhassan2015Modeling,OptimalGeographicCachingInCellularNetworks,RCKHetNetJuan2017, OptimalContentPlacementOffloadingHetNets,RCHetNetTao2018, RCCHetNetcui2017, RandomcachingCoMP2018Wen,RandomcachingDTX2018Wen,2018arXiv180102743J,D2Dcontentdeliverynetworking2014}. More specifically, in \cite{Tamoorulhassan2015Modeling}, the uniform caching was studied, in which, each BS randomly stores a file according to the uniform distribution. In \cite{OptimalGeographicCachingInCellularNetworks,RCKHetNetJuan2017, OptimalContentPlacementOffloadingHetNets,RCHetNetTao2018, RCCHetNetcui2017, RandomcachingCoMP2018Wen,RandomcachingDTX2018Wen, D2Dcontentdeliverynetworking2014,2018arXiv180102743J}, the optimal random caching was examined to maximize the cache hit probability \cite{OptimalGeographicCachingInCellularNetworks,RCKHetNetJuan2017}, the successful offloading probability \cite{OptimalContentPlacementOffloadingHetNets}, and the successful transmission probability \cite{RCHetNetTao2018, RCCHetNetcui2017,RandomcachingCoMP2018Wen,RandomcachingDTX2018Wen, 2018arXiv180102743J}, or minimize the average caching failure probability~\cite{D2Dcontentdeliverynetworking2014}.

{Although significant efforts have been devoted to improve the reliability of content transmission \cite{Cache-enabledsmallcellnetworksmodelingandtradeoffs, Tamoorulhassan2015Modeling, OptimalGeographicCachingInCellularNetworks,RCKHetNetJuan2017, OptimalContentPlacementOffloadingHetNets,RCHetNetTao2018, RCCHetNetcui2017,RandomcachingCoMP2018Wen, RandomcachingDTX2018Wen,2018arXiv180102743J, D2Dcontentdeliverynetworking2014}, the content transmission secrecy is another important} design aspect in the cache-enabled wireless networks, yet receives much less attention.  {Due to the broadcast nature of wireless communications, the transmissions of the contents cached at BSs are vulnerable to potential eavesdroppers (e.g., non-paying subscribers) and jammers \cite{UAVsecureCachingZhao2018, UAVsecureCachingCheng2019, eveJamWsaadHVPoor}. By attempting to intercept the users' content transmissions, the eavesdroppers can degrade the confidentiality of content transmissions. By transmitting harmful interfering signal in the wireless networks, the jammers can reduce the reliability of content transmissions.} 
Currently, the most of mobile data services still rely on the traditional cryptographic encryption to guarantee the confidentiality of content transmission.  Nevertheless, such encryption creates an insurmountable obstacle to caching content at BSs, since the encrypted content are uniquely defined for each user's request and cannot be reused to serve other users' requests \cite{WiCachMag2016}. Against this background, physical layer security has been regarded as a promising solution to support caching and enhance the confidentiality of content transmission simultaneously in large-scale cache-enabled single-tier \cite{PLScachingSCNZheng2018, RCPLSSCNTVT2018} or multi-tier heterogeneous wireless networks (HetNets) \cite{PLScachingHetNetZheng2018}, due to the fact that it only relies on wiretap channel coding instead of source encryption. In particular, the authors in \cite{PLScachingSCNZheng2018} and \cite{PLScachingHetNetZheng2018} proposed a partition-based hybrid caching scheme, where the cache of each BS was partitioned into two halves for storing both the most popular contents and fractions of other contents, and analyzed different physical-layer security performance such as the transmission capacity and energy efficiency \cite{PLScachingSCNZheng2018} as well as the secure content delivery probability \cite{PLScachingHetNetZheng2018}.  In \cite{RCPLSSCNTVT2018}, the authors leveraged the content diversity provided by the random caching to optimize the cache hit probability subject to different content confidentiality level constraints.  

The authors in \cite{PLScachingSCNZheng2018, PLScachingHetNetZheng2018, RCPLSSCNTVT2018} only considered deploying single-antenna at each BS, and therefore, their proposed caching schemes may not achieve desirable security performance in multi-antenna wireless networks. In addition, unlike \cite{PLScachingSCNZheng2018, PLScachingHetNetZheng2018}, to deliberately confuse the eavesdroppers, an artificial-noise-aided transmission strategy was further considered in \cite{RCPLSSCNTVT2018}. The key idea of this strategy is to use part of the transmit power at the BS to inject artificial noise into the null space of the legitimate user's channel, thereby significantly reducing the link quality of the eavesdropper without interfering with the legitimate user \cite{PLSMultiantJamEveTIFS2018}. However, since the BSs only have single-antenna, in \cite{RCPLSSCNTVT2018}, each BS is only able to transmit either artificial noise or content, resulting in that some users will not be able to download their requested content in a timely manner. {Furthermore, it is worth pointing out that the authors in \cite{PLScachingSCNZheng2018, PLScachingHetNetZheng2018, RCPLSSCNTVT2018} only focused on combating the eavesdropping attack in cache-enabled wireless networks, leading to that their proposed caching schemes may not efficiently resist jamming~attack.}

In this paper, we would like to address the above issues.  
We consider a large-scale multi-antenna HetNet employing random caching and artificial-noise-aided transmission strategy, where the BSs deployed at multiple tiers transmit the randomly cached confidential content together with the artificial noise to the legitimate users in the presence of the jammers as well as eavesdroppers. 
We define and evaluate the content average reliable transmission probability (RTP) and the content average confidential transmission probability (CTP), which characterizes the reliability and the confidentiality of content transmissions, respectively.
Our key goal is to determine a secure random caching scheme that maximizes the average RTP under the  average CTP and storage resource constraints.
Our main contributions are summarized as follows:
{
\begin{itemize}
  \item We derive closed-form expressions for the average RTP and CTP in the considered HetNets. Based on which, we further derive tight upper and lower bounds on the average RTP. These analytical results allow us to reveal how the jamming and eavesdropping can have a significant impact on large-scale cache-enabled HetNets.
  \item We propose a new secure random caching scheme in which the caching distribution of the contents is judiciously determined to maximize the average RTP, while satisfying the constraints on the average CTP and the storage resource.
  \item Through numerical results, we show that the proposed secure caching scheme outperforms the existing baselines and significantly improves the average RTP, while being resist to the jamming and eavesdropping~attack.
\end{itemize}
Beyond the above contributions, this paper provides valuable insights into the design of secure large-scale cache-enable HetNets. Specifically, we show that, when the constraint on the average CTP is stringent, caching less popular contents at BSs may be helpful for maintaining the level of average RTP. In addition, we show that increasing the cache sizes at BSs does not always improve the average RTP when the secrecy constraint on the average CTP is taken into~account.}

{

The rest of the paper is organized as follows.
Section II describes the system model and formulates a secrecy performance optimization problem.
Section III analyzes the secrecy performance and derives the closed-form expressions for the average RTP and CTP.
Based on these analytical results, Section IV develops two algorithms to solve the formulated optimization problem.
Numerical results and the related discussions are provided in Section V. Finally, Section VI draws the conclusions.
Unless otherwise specified, the notations used throughout the paper are summarized in Table~\ref{tabNotations}.
}

\section{System Model}
\subsection{Network Model}
We consider {the secure downlink transmissions in a large-scale cache-enabled HetNet, in which the BSs from $K$ co-channel deployed network tiers transmit to the users in the presence of randomly distributed jammers and eavesdroppers. For ease of illustration, we denote the set of $K$ tiers by $\mathcal{K}\triangleq\{1,2,\cdots,K\}$. We assume that each BS in tier $k$ is active (as in \cite{ RCKHetNetJuan2017, RCHetNetTao2018,  RCCHetNetcui2017}) and has $M_k$ antennas, while the users, the jammers, and the eavesdroppers are equipped with a single antenna each.\footnote{ We understand that the scenario where the users, the jammers, and the eavesdroppers are all equipped with multiple antennas is more general, interesting, as well as challenging. For analytical tractability, we leave it for future work.} The eavesdroppers secretly intercept the downlink transmissions, while the jammers constantly broadcast the jamming signals into the channel to interrupt the downlink transmissions. The jammers and the eavesdroppers are not colluding, and therefore the jamming signals also interfere with the eavesdroppers. The BS locations in tier $k$ and the jammers are modeled by independent homogeneous Poisson point processes (PPPs) $\Phi_k$ and $\Phi_{\mathtt{J}}$ with densities $\lambda_k$ and $\lambda_{\mathtt{J}}$,\footnote{In practice, the density of the jammers can be estimated by using the jammer detection policies \cite{detcJammers}.} respectively. In addition, the locations of the users and eavesdroppers are modeled by some other independent stationary point processes with certain densities.   }

\begin{table}
\centering\small
\caption{ List of Notations.}\label{tabNotations}
\begin{tabular}{p{1.2cm} p{6.9cm} }%\doublespacing
  \hline\hline
  % after \\: \hline or \cline{col1-col2} \cline{col3-col4} ...
  Notation      &   Definition                                          \\ \hline
  $\mathcal{K}$ &   Set of network tiers.                               \\
  $K$           &   Number of tiers.                               \\
  $\Phi_k$      &   Point process of the BSs in tier $k$.       \\
  $\lambda_k$   &   Spatial density of the BSs in tier $k$.             \\
  $M_k$         &   Number of antennas at each BS in tier $k$.                      \\
  $P_k$         &   Transmit power of each BS in tier $k$.                      \\
  $\Phi_{\mathtt{J}}$       & Point process of the jammers.     \\
  $\lambda_{\mathtt{J}}$    & Spatial density of the jammers.     \\
  $\mathbb{C}$              & The complex number domain.           \\
  $\mathsf{u}$              & Notation representing for user.           \\
  $\mathsf{e}$              & Notation representing for eavesdropper.           \\
  %${b_k}$                           & Index of a BS in tier $k\in\mathcal{K}$.           \\
  $\mathbf{h}_{b_k,x}$    & Small-scale fading vector between BS $b_k \in \Phi_k$ and receiver $x\in\{\mathsf{u}, \mathsf{e}\}$.           \\
  $g_{j,x}$    & Small-scale fading coefficient between jammer $j \in \Phi_{\mathtt{J}}$ and receiver $x\in\{\mathsf{u}, \mathsf{e}\}$.           \\
  $\alpha$    & Path-loss exponent.           \\
  $\mathcal{N}$     & Set of files in the network.           \\
  $N$               & Number of files in the network.           \\
  $a_n$             & Popularity of file $n$. \\
  $\mathbf{a}$      & Popularity distribution of files in $\mathcal{N}$. \\
  $C_k$             & Cache size of each BS in tier $k$.           \\
  $T_{n,k}$         & Caching probability of file $n$ in tier $k$.           \\
  $\mathbf{T}$      & Caching distribution of the $N$ files.           \\
  $\Phi_{n,k}$      & Point processes of the BSs in tier $k$ which store file~$n$.           \\
  $\Phi_{-n,k}$     & Point processes of the BSs in tier $k$ which do not store file $n$.           \\
  $s_{b_k}$                         & The information-bearing signal at BS $b_k$.           \\
  $\ell_{\mathsf{u}}(b_k)$          & The served user of BS $b_k$.           \\
  $\boldsymbol{\nu}_{b_k}$          & Artificial noise vector at BS $b_k$.           \\
  $\mathbf{w}_{b_k}$                &  Beamforming vector at BS $b_k$.           \\
  $\mathbf{W}_{b_k}$                & The matrix for transmitting the artificial noise at BS~$b_k$.           \\
  $\phi_k$                          & Fraction of power allocated to the signal $s_{b_k}$.           \\
  ${\ell}_{\mathsf{u},0}$    & A typical user.   \\
  ${\ell}_{\mathsf{e},0}$    & A typical eavesdropper.           \\
  $r_{b_k,{\ell}_{\mathsf{u},0}}$    & Distance between BS $b_k$ and ${\ell}_{\mathsf{u},0}$.           \\
  $r_{b_k,{\ell}_{\mathsf{e},0}}$    & Distance between BS $b_k$ and ${\ell}_{\mathsf{e},0}$.           \\
  $d_{j,{\ell}_{\mathsf{u},0}}$      & Distance between jammer $j$ and  ${\ell}_{\mathsf{u},0}$.           \\
  $d_{j,{\ell}_{\mathsf{e},0}}$      & Distance between jammer $j$ and  ${\ell}_{\mathsf{e},0}$.           \\
  $b_{n,k_{\mathsf{u},0}}$  & Index of the serving BS of ${\ell}_{\mathsf{u},0}$ in tier $k_{\mathsf{u},0}\in\mathcal{K}$.           \\
  $b_{n,k_{\mathsf{e},0}}$                          & Index of an arbitrary BS storing file $n$ in tier $k_{\mathsf{e},0}\in\mathcal{K}$.           \\
  $R_{\mathsf{u}}$      &  Target transmission rate of the wiretap code.           \\
  $R_{\mathsf{s}}$      &  Secrecy rate of the wiretap code.           \\
  $R_{\mathsf{e}}$      &  Redundancy rate against eavesdropping.           \\
  $\epsilon$            &  A given level of confidentiality.           \\
  $\textsf{p}_{{\mathsf{u}},n}(\mathbf{T}_n)$      &  RTP of file $n$.           \\
  $\textsf{p}_{\mathsf{u}}(\mathbf{T})$            &  Average RTP.           \\
  $\textsf{p}_{{\mathsf{e}},n}(\mathbf{T}_n)$      &  CTP of file $n$.           \\
  $\textsf{p}_{\mathsf{e}}(\mathbf{T})$            &  Average RTP.           \\
  \hline\hline
\end{tabular}
\end{table}

{We assume a interference-limited communication environment, in which} the wireless channels undergo quasi-static Rayleigh fading along with a large-scale path loss. Specifically, we denote $\mathbf{h}_{b_k,x}\in\mathbb{C}^{M_k\times1}$ ($g_{j,x}\in \mathbb{C}$) as the small-scale fading vector (coefficient) between BS $b_k \in \Phi_k$ (jammer $j \in \Phi_{\mathtt{J}}$) and receiver $x\in\{\mathsf{u}, \mathsf{e}\}$, where $\mathbb{C}$ denotes the complex number domain, $\mathsf{u}$ and $\mathsf{e}$ represent the user and the eavesdropper, respectively. {As such, all the} entries of $\mathbf{h}_{b_k,x}$ and $g_{j,x}$ are identical and independent distributed (i.i.d.) circularly symmetric complex Gaussian random variables with zero mean and unit variance.  For the path loss model, we use the standard power loss propagation model, i.e., the power of {the} transmitted signal with distance $r$ is attenuated by a factor $r^{\alpha}$, where $\alpha>2$ denotes the path-loss exponent.

{We consider that there exist $N$ files, denoted by the set ${\mathcal{N}} \triangleq \{ 1,2, \cdots ,N\}$, to be cached in the considered HetNet. All the files have the same size (as in \cite{OptimalGeographicCachingInCellularNetworks, OptimalContentPlacementOffloadingHetNets,  RCCHetNetcui2017, RCKHetNetJuan2017,  RCHetNetTao2018,  RandomcachingCoMP2018Wen, RandomcachingDTX2018Wen}) and each file $n\in\mathcal{N}$ has its own popularity (i.e., the probability that file $n$ is requested by a user), denoted by $a_n\in[0,1]$, where $\sum_{n\in\mathcal{N}}a_n=1$.} Without loss of generality (w.l.o.g.), we assume that ${a_1} \ge {a_2} \ge  \cdots \ge {a_N}$. As such, the popularity distribution among $\mathcal{N}$ can be denoted by $\mathbf{a}\triangleq (a_n)_{n\in\mathcal{N}}\in[0,1]^{N\times1}$, which we assume known {\em a priori}. We note that this assumption is practical due to the fact that the file popularity evolves at a slower timescale, thus various methods can be employed to estimate the file popularity over time \cite{FemtoCachingWirelessContentDeliveryThroughDistributedCachingHelpers}. We consider a discrete-time system with time being slotted. In each time slot, each user randomly requests one file according to the file popularity $\mathbf{a}$. We study one slot in the network.

\subsection{Random Caching and {User Association}}
{We consider that each BS in tier $k$ is equipped with a cache unit of size $C_k$ (in files), $C_k\le N$. As in \cite{OptimalGeographicCachingInCellularNetworks, OptimalContentPlacementOffloadingHetNets,  RCCHetNetcui2017,  RCKHetNetJuan2017, RCHetNetTao2018,  RandomcachingCoMP2018Wen, RandomcachingDTX2018Wen},
file $n$} is stored at each BS in tier $k\in\mathcal{K}$ with a certain probability $T_{n,k}\in[0,1]$, referred to as the caching probability of file $n$ in tier $k$. We denote $\mathbf{T}\triangleq \left(\mathbf{T}_n\right)_{n\in\mathcal{N}}\in[0,1]^{NK\times 1}$, where $\mathbf{T}_n\triangleq(T_{n,k})_{k
\in\mathcal{K}}\in[0,1]^{K\times 1}$, as the caching distribution of the $N$ files in the $K$-tier HetNet.  Then, the relationship between the caching probability of file $n$ in tier $k$ and the cache sizes of the BSs in tier $k$ can be expressed as  \cite{RCCHetNetcui2017, RandomcachingCoMP2018Wen, RandomcachingDTX2018Wen}:
\begin{eqnarray}
&&0\le T_{n,k} \le 1, \quad n\in\mathcal{N}, \quad k\in\mathcal{K},\label{eqconstcachingprob}\\
&&\sum_{n\in \mathcal{N}}T_{n,k} = C_k, \quad k\in\mathcal{K}.\label{eqconst2mbcachesize}
\end{eqnarray}Let $\Phi_{n,k}\subseteq\Phi_k$ and $\Phi_{-n,k}\subseteq\Phi_k$, $n\in\mathcal{N}$ denote the point processes of the BSs in tier $k$ which store and do not store file $n$, respectively. {Then, we have} $\Phi_{n,k} \bigcup \Phi_{-n,k} = \Phi_k$. Due to the random caching and independent thinning \cite{Haenggi2012Stochastic}, we know that $\Phi_{n,k}$ and $\Phi_{-n,k}$ are two thinned and independent homogeneous PPPs with densities $\lambda_kT_{n,k}$ and $\lambda_k(1-T_{n,k})$, respectively.

We now describe the user association rule adopted in this paper. Consider a user requesting file $n$ at the beginning of a slot.  If file $n$ is not stored in the HetNet, it will not be served.\footnote{Note that in this work, we only study serving cached files at BSs to get first-order insights into the design of cache-enabled $K$-tier HetNets and characterize the benefits of caching, as in \cite{OptimalGeographicCachingInCellularNetworks, RCKHetNetJuan2017,RandomcachingDTX2018Wen}. BSs may serve those uncached files through other service mechanisms, the investigation of which is beyond the scope of this paper.}   Otherwise, it will be associated with a BS which not only stores file $n$ but also provides the maximum long-term average received information signal power (ARISP) among all BSs in the $K$-tier HetNet. This BS is referred to as the users' \emph{serving} BS and such  association mechanism is called the content-based user association.  Note that, under this association,  a user may not be associated with the BS {which provides} the maximum ARISP if it has not stored file $n$. As a result, the user usually receives a weak signal compared with the interference, and thus may not successfully receive the requested file and benefit from content diversity offered by random caching.  {To overcome this drawback, we assume that the channel state information (CSI) of each user is available at its serving BS and the serving BS  adopts the maximal ratio transmission to deliver the requested files.}\footnote{Note that, the analytical framework developed in this paper can be extended to other beamforming techniques.}

\subsection{Artificial-Noise-Aided Transmission}

In order to deliberately confuse eavesdroppers in the HetNet while guaranteeing the reliable links to the users, we consider that each BS employs the artificial-noise-aided transmission strategy. Let $s_{b_k}$ denote the information-bearing signal with $\mathbb{E}[|s_{b_k}|^2]=1$ and $\boldsymbol{\nu}_{b_k}\in \mathbb{C}^{(M_k-1) \times 1}$  denote the artificial noise vector at BS $b_k$. {As per the rules of the artificial-noise-aided transmission strategy, all the entries of {$\boldsymbol{\nu}_{b_k}$} are i.i.d. circularly symmetric complex Gaussian random variables with zero mean and
variance $\frac{1}{M_k-1}$. Then,} the transmitted signal from BS $b_k\in\Phi_k$ can be expressed as
\begin{equation}\label{eqXbk}
\mathbf{\hat{s}}_{b_k} = \sqrt{\phi_k P_k}\mathbf{w}_{b_k} s_{b_k} + \sqrt{(1-\phi_k) P_k}\mathbf{W}_{b_k} {\boldsymbol{\nu}_{b_k}},
\end{equation}where {$P_k$ denotes the transmit power of BS $b_k\in\Phi_k$,} $\phi_k\in (0,1]$  denotes the fraction of power allocated to the information-bearing signal $s_{b_k}$, $\mathbf{w}_{b_k}=\frac{\mathbf{h}_{b_k,\ell_{\mathsf{u}}(b_k)}} {\|\mathbf{h}_{b_k,\ell_{\mathsf{u}}(b_k)}\|}$ is the beamforming vector at BS $b_k$ with $\ell_{\mathsf{u}}(b_k)$ representing the served user of BS $b_k$ and $\|\mathbf{h}\|$ denoting the 2-norm of a vector $\mathbf{h}$, and $\mathbf{W}_{b_k}\in\mathbb{C}^{M_k\times(M_k-1)}$ {is the matrix for transmitting the artificial noise at BS $b_k$. We choose $\mathbf{W}_{b_k}$ as the projection matrix into the null space of ${\mathbf{h}_{b_k,\ell_{\mathsf{u}}(b_k)}}$. As such, the artificial noise will not interfere with user $\ell_{\mathsf{u}}(b_k)$.}

\subsection{Received Signal-to-Interference Ratios}
In this paper, w.l.o.g., according to Slivnyak's Theorem \cite{Haenggi2012Stochastic}, we focus on a typical receiver ${\ell}_{x,0}$, $x\in\{{\mathsf{u}},\mathsf{e}\}$ located at the origin.\footnote{A typical user (eavesdropper) is a user (an eavesdropper) that is randomly selected from all the users (eavesdroppers) in the network.}  Let $r_{b_k,{\ell}_{x,0}}$ ($d_{j,{\ell}_{x,0}}$) denote the distance between BS $b_k \in \Phi_k$ (jammer $j \in \Phi_{\mathtt{J}}$) and ${\ell}_{x,0}$. Based on (\ref{eqXbk}), the received signal at ${\ell}_{x,0}$  is given by (\ref{eqreceivedsig}), {as shown at the top of this page,}
\begin{figure*}[!t]\vspace{-8mm}
{\begingroup\makeatletter\def\f@size{9.5}\check@mathfonts
\def\maketag@@@#1{\hbox{\m@th\normalsize\normalfont#1}}\setlength{\arraycolsep}{0.0em}
\begin{eqnarray}\label{eqreceivedsig}
y_{b_{n,k_{x,0}},{\ell}_{x,0}} &=& \mathbf{h}^{\rm T}_{b_{n,k_{x,0}},{\ell}_{x,0}}\mathbf{\hat{s}}_{b_{n,k_{x,0}}} r_{{b_{n,k_{x,0}}}, {{\ell}_{x,0}}}^{-\frac{\alpha}{2}} +  \sum_{k\in\mathcal{K}}\sum_{b_k\in\Phi_{n,k}\setminus \{b_{n,k_{x,0}}\}} \mathbf{h}^{\rm T}_{{b_k}, {\ell}_{x,0}}\mathbf{\hat{s}}_{b_k} r_{{b_k}, {\ell}_{x,0}}^{-\frac{\alpha}{2}} \nonumber\\
 &&{+}\:\sum_{k\in\mathcal{K}}\sum_{{b_k}\in\Phi_{-n,k}} \mathbf{h}^{\rm T}_{{b_k}, {\ell}_{x,0}}\mathbf{\hat{s}}_{{b_k}} r_{{{b_k}}, {\ell}_{x,0}}^{-\frac{\alpha}{2}}  + \sum_{j\in\Phi_{\mathtt{J}}} \sqrt{P_{\mathtt{J}}}g_{j, {\ell}_{x,0}} d^{-\frac{\alpha}{2}}_{j,{\ell}_{x,0}}.
\end{eqnarray}\setlength{\arraycolsep}{5pt}\endgroup}\setlength{\arraycolsep}{5pt} \noindent\rule[0.25\baselineskip]{\textwidth}{0.1pt}\vspace{-6mm}\end{figure*}
where $b_{n,k_{\mathsf{u},0}}\in\Phi_{n,k_{\mathsf{u},0}}$ denotes the index of the serving BS of the typical user ${\ell}_{\mathsf{u},0}$ {in tier $k_{\mathsf{u},0}\in\mathcal{K}$,\footnote{The serving BS $b_{n,k_{\mathsf{u},0}}$ of the typical user ${\ell}_{\mathsf{u},0}$ is determined according to the content-based user association policy in Section II-B.} $b_{n,k_{\mathsf{e},0}}\in\Phi_{n,k_{\mathsf{e},0}}$ denotes the index of an arbitrary BS storing file $n$ in tier $k_{\mathsf{e},0}\in\mathcal{K}$},  and $(\cdot)^{\rm T}$ denotes the conjugate transpose operation. Then, the SIR at ${\ell}_{x,0}$ can be expressed as (\ref{eqsiruser}), {as shown at the top of this page,}
{\begin{figure*}[!t]\setlength{\arraycolsep}{0.0em}
\begingroup\makeatletter\def\f@size{9.5}\check@mathfonts
\def\maketag@@@#1{\hbox{\m@th\normalsize\normalfont#1}}
\begin{eqnarray}\label{eqsiruser}
\mathrm{SIR}_{b_{n,k_{x,0}},{\ell}_{x,0}} &=&   \frac{{{{\left| {{\bf{h}}_{b_{n,k_{x,0}},{\ell}_{x,0}}^{\rm  T}{{\bf{w}}_{b_{n,k_{x,0}}}}} \right|}^2}r_{b_{n,k_{x,0}},{\ell}_{x,0}}^{ - \alpha }}}{I_{b_{n,k_{x,0}}}^x + \sum_{k\in\mathcal{K}} (I_{n,k}^x + I_{-n,k}^x) + I_{\mathtt{J}}^x}.
\end{eqnarray}\setlength{\arraycolsep}{5pt}\endgroup\noindent\rule[0.25\baselineskip]{\textwidth}{0.1pt}\end{figure*}}where $I_{b_{n,k_{x,0}}}^x$, $I_{n,k}^x$, $I_{-n,k}^x$ and $I_{\mathtt{J}}^x$ denote the power of artificial noise from BS $b_{n,k_{x,0}}$, the interference and artificial noise from {the} BSs with storing file $n$ in tier $k$, the interference and artificial noise from {the} BSs without storing file $n$ in tier $k$, and the interference from jammers, respectively, given by (\ref{eqInt1})--(\ref{eqInt4}), {as shown at the top of the next page}.
\begin{figure*}[!t]\vspace{-6mm}
{\begingroup\makeatletter\def\f@size{9.5}\check@mathfonts
\def\maketag@@@#1{\hbox{\m@th\normalsize\normalfont#1}}\setlength{\arraycolsep}{0.0em}
\begin{align}
I_{b_{n,k_{x,0}}}^x &=   \xi_{k_{x,0}} \left\|\mathbf{h}^{\rm T}_{b_{n,k_{x,0}},{\ell}_{x,0}}\mathbf{W}_{b_{n,k_{x,0}}} \right\|^2 r_{b_{n,k_{x,0}}, {\ell}_{x,0}}^{-\alpha},\label{eqInt1}\\
I_{n,k}^x & = \sum_{{b_k}\in\Phi_{n,k}\setminus \{b_{n,k_{x,0}}\}}  \frac{\phi_kP_k } {{\phi_{k_{x,0}}}{P_{k_{x,0}}}} \left( \left| \mathbf{h}^{\rm T}_{{b_k}, {\ell}_{x,0}} {\mathbf{w}_{b_k}}\right|^2  + \xi_k \left\| \mathbf{h}^{\rm T}_{{b_k}, {\ell}_{x,0}}{\mathbf{W}_{b_k}} \right\|^2 \right) r_{{b_k}, {\ell}_{x,0}}^{-\alpha} ,\label{eqInt2}\\
I_{-n,k}^x & = \sum_{{b_k}\in\Phi_{-n,k}}  \frac{\phi_kP_k } {{\phi_{k_{x,0}}}{P_{k_{x,0}}}} \left( \left| \mathbf{h}^{\rm T}_{{b_k}, {\ell}_{x,0}} {\mathbf{w}_{b_k}}\right|^2  + \xi_k \left\| \mathbf{h}^{\rm T}_{{b_k}, {\ell}_{x,0}}{\mathbf{W}_{b_k}} \right\|^2 \right) r_{{b_k}, {\ell}_{x,0}}^{-\alpha} ,\label{eqInt3}\\
I_{\mathtt{J}}^x &= \sum_{j\in\Phi_{\mathtt{J}}}\frac{P_{\mathtt{J}}}{{\phi_{k_{x,0}}P_{k_{x,0}}}}\left|g_{j, {\ell}_{x,0}}\right|^2 d^{- \alpha}_{j,{\ell}_{x,0}}.\label{eqInt4}
\end{align}\setlength{\arraycolsep}{5pt}\endgroup}\noindent\rule[0.25\baselineskip]{\textwidth}{0.1pt}\end{figure*}Here, $\xi_k$ is defined as
\begin{equation}\label{eqxik}
\xi_k  \triangleq \begin{cases}0,&\mbox{ if } M_k=1,\\ \frac{\phi_k^{-1}-1}{M_k-1}, &\mbox{ if } M_k=2,3,\cdots.\end{cases}
\end{equation}Note that, we have $|g_{j, {\ell}_{x,0}}|^2\mathop \sim\limits^d\exp(1)$, $| \mathbf{h}^{\rm T}_{b_k, {\ell}_{x,0}} \mathbf{w}_{b_k}|^2\mathop \sim\limits^d\exp(1)$,\footnote{Note that $X\mathop \sim\limits^d Y$ means that the random variable $X$ follows the distribution $Y$.} $\| \mathbf{h}^{\rm T}_{b_k, {\ell}_{x,0}}\mathbf{W}_{b_k} \|^2\mathop \sim\limits^d\Gamma(M_k-1,1)$, {$\|\mathbf{h}^{\rm T}_{b_{n,k_{\mathsf{e},0}},\ell_{\mathsf{e},0}}\mathbf{W}_{b_{n,k_{\mathsf{e},0}}} \|^2\mathop \sim\limits^d \Gamma(M_{k_{\mathsf{e},0}}-1,1)$, and $\|\mathbf{h}^{\rm T}_{b_{n,k_{\mathsf{u},0}},{\ell}_{x,0}}\mathbf{W}_{b_{n,k_{\mathsf{u},0}}} \|^2 = 0$ due to the orthogonality between ${\mathbf{h}_{b_{n,k_{{{\mathsf{u}}},0}},\ell_{{\mathsf{u}},0}}}$ and $\mathbf{W}_{b_{n,k_{{{\mathsf{u}}},0}}}$} \cite{PLSHetNetWang2016}.

Based on (\ref{eqsiruser}), the SIR at the typical  user $\ell_{{\mathsf{u}},0}$ can be further written as
{\begingroup\makeatletter\def\f@size{9.5}\check@mathfonts
\def\maketag@@@#1{\hbox{\m@th\normalsize\normalfont#1}}\setlength{\arraycolsep}{0.0em}
\begin{eqnarray}\label{eqsirAuthorizedUser}
\mathrm{SIR}_{b_{n,k_{\mathsf{u},0}},{\ell}_{\mathsf{u},0}} &=&   \frac{{{{\left| {{\bf{h}}_{b_{n,k_{{{\mathsf{u}}},0}},\ell_{{\mathsf{u}},0}}^{\rm  T}{{\bf{w}}_{b_{n,k_{{{\mathsf{u}}},0}}}}} \right|}^2}r_{b_{n,k_{{{\mathsf{u}}},0}},\ell_{{\mathsf{u}},0}}^{ - \alpha }}}{  \sum_{k\in\mathcal{K}} ( I_{n,k}^{\mathsf{u}} + I_{-n,k}^{\mathsf{u}}) + I_{\mathtt{J}}^{\mathsf{u}}},
\end{eqnarray}\setlength{\arraycolsep}{5pt}\endgroup}where we have ${{{| {{\bf{h}}_{b_{n,k_{{{\mathsf{u}}},0}},\ell_{{\mathsf{u}},0}}^{\rm  T}{{\bf{w}}_{b_{n,k_{{{\mathsf{u}}},0}}}}} |}^2}}\mathop \sim\limits^d \Gamma(M_{k_{{{\mathsf{u}}},0}}, 1)$ and $I_{b_{n,k_{{{\mathsf{u}}},0}}}^{\mathsf{u}}=0$ due to $\|\mathbf{h}^{\rm T}_{b_{n,k_{\mathsf{u},0}},{\ell}_{x,0}}\mathbf{W}_{b_{n,k_{\mathsf{u},0}}} \|^2 = 0$. On the other hand, the SIR at the typical eavesdropper $\ell_{\mathsf{e},0}$ can be rewritten~as
{\begingroup\makeatletter\def\f@size{9.5}\check@mathfonts
\def\maketag@@@#1{\hbox{\m@th\normalsize\normalfont#1}}\setlength{\arraycolsep}{0.0em}
\begin{eqnarray}\label{eqsirusereve}
\mathrm{SIR}_{b_{n,k_{\mathsf{e},0}},{\ell}_{\mathsf{e},0}} &=&   \frac{{{{\left| {{\bf{h}}_{b_{n,k_{\mathsf{e},0}},\ell_{\mathsf{e},0}}^{\rm  T}{{\bf{w}}_{b_{n,k_{\mathsf{e},0}}}}} \right|}^2}r_{b_{n,k_{\mathsf{e},0}},\ell_{\mathsf{e},0}}^{ - \alpha }}}{I_{b_{n,k_{\mathsf{e},0}}}^{\mathsf{e}} + \sum_{k\in\mathcal{K}} (I_{n,k}^{\mathsf{e}} + I_{-n,k}^{\mathsf{e}}) + I_{\mathtt{J}}^{\mathsf{e}}},
\end{eqnarray}\setlength{\arraycolsep}{5pt}\endgroup}where we have ${{| {{\bf{h}}_{b_{n,k_{\mathsf{e},0}},\ell_{\mathsf{e},0}}^{\rm  T}{{\bf{w}}_{b_{n,k_{\mathsf{e},0}}}}} |}^2}\mathop \sim\limits^d \exp(1)$.

\subsection{{Problem Formulation}}
According to \cite{wyner,bloch,chenxi_jsac,chenxi_TWC19}, in order to perform the secure transmission in the considered HetNet, a wiretap code with the parameter pair $\left(R_{\mathsf{u}},R_{\mathsf{s}}\right)$ needs to be constructed, where $R_{\mathsf{u}}$ and $R_{\mathsf{s}}$ denote the target transmission rate and secrecy rate of the wiretap code (both in bps/Hz), respectively. Then the difference $R_{\mathsf{e}}=R_{\mathsf{u}}-R_{\mathsf{s}}$ is regarded as the redundancy rate against eavesdropping. If the channel capacity between the BS and a user is greater than or equal to $R_{\mathsf{u}}$, the user is able to decode its desired file. As such, the \emph{reliability} of file transmission is achieved. If the channel capacity between each BS to an eavesdropper is less than or equal to $R_{\mathsf{e}}$, the eavesdropper cannot decode any file. In this case, the \emph{confidentiality} of file transmission is achieved.

Based on the above discussions, we define the reliable transmission probability (RTP) of file $n$ as the probability that the transmission of file $n$ is reliable. Mathematically, it is given~by
{\begingroup\makeatletter\def\f@size{9.5}\check@mathfonts
\def\maketag@@@#1{\hbox{\m@th\normalsize\normalfont#1}}
\begin{equation}\label{eqSTPofFiln}
\textsf{p}_{{\mathsf{u}},n}(\mathbf{T}_n) \triangleq \Pr\left[ \log_2\left(1+\mathrm{SIR}_{b_{n,k_{\mathsf{u},0}},{\ell}_{\mathsf{u},0}}\right)\ge R_{\mathsf{u}}\right].
\end{equation}\endgroup}Similarly, we define the confidential transmission probability (CTP) of file $n$ as the probability that the transmission of file $n$ is confidential, which is expressed as (\ref{eqqE}), {as shown at the top of the next page}.
\begin{figure*}[!t]\vspace{-3mm}
\begingroup\makeatletter\def\f@size{9.5}\check@mathfonts
\def\maketag@@@#1{\hbox{\m@th\normalsize\normalfont#1}}
\begin{align}\label{eqqE}
\textsf{p}_{{\mathsf{e}},n}(\mathbf{T}_n)   \triangleq \Pr\left[\max_{k_{\mathsf{e},0}\in\mathcal{K}, \; b_{n,k_{\mathsf{e},0}}\in\Phi_{n,k_{\mathsf{e},0}}} \log_2\left(1+\mathrm{SIR}_{b_{n,k_{\mathsf{e},0}},{\ell}_{\mathsf{e},0}}\right) \le R_{\mathsf{e}} \right] .
\end{align}\endgroup\noindent\rule[0.25\baselineskip]{\textwidth}{0.1pt}\vspace{-6mm}\end{figure*}

\begin{Remark}[Interpretation of RTP and CTP of file $n$]
The RTP and the CTP of file $n$ measure the transmission reliability and the transmission confidentiality of file $n$, respectively. In addition, the RTP (CTP) of file $n$ can be thought of equivalently as i) the probability that a randomly chosen user (eavesdropper) succeeds in (fails to) decode file $n$, or ii) the average fraction of users (eavesdroppers) who at any time succeed in (fail to) decode file $n$.
\end{Remark}

Since each file $n\in\mathcal{N}$ is requested with probability $a_n$, according to the total probability theorem, the average RTP and CTP of a file, denoted as $\textsf{p}_{\mathsf{u}}(\mathbf{T})$ and $\textsf{p}_{\mathsf{e}}(\mathbf{T})$, respectively, are given~by
{\begingroup\makeatletter\def\f@size{9.5}\check@mathfonts
\def\maketag@@@#1{\hbox{\m@th\normalsize\normalfont#1}}\setlength{\arraycolsep}{0.0em}
\begin{eqnarray}
\textsf{p}_{\mathsf{u}}(\mathbf{T}) = \sum_{n\in\mathcal{N}}a_n \textsf{p}_{{\mathsf{u}},n}(\mathbf{T}_n), \label{eqaveragertp}\\
\textsf{p}_{\mathsf{e}}(\mathbf{T}) = \sum_{n\in\mathcal{N}}a_n \textsf{p}_{{\mathsf{e}},n}(\mathbf{T}_n), \label{eqAVERAGECTP}
\end{eqnarray}\setlength{\arraycolsep}{5pt}\endgroup}where $\mathbf{T}$ is the design parameter  related to random caching.

{From \eqref{eqaveragertp} and \eqref{eqAVERAGECTP}, we see that the caching distribution $\mathbf{T}$ significantly affects both the RTP and the CTP. The key goal of this paper is to optimize $\mathbf{T}$ to maximize the average RTP of the system, subject to the caching size constraints in \eqref{eqconstcachingprob} and \eqref{eqconst2mbcachesize} as well as the constraint on the average CTP of the system, given by
\begin{align}\label{eqconstCTP}
  \textsf{p}_{\mathsf{e}}(\mathbf{T}) \ge \epsilon,
\end{align}
where $\epsilon$ denotes a given level of confidentiality.
Mathematically, the optimization problem can be formulated as
\begin{Problem}[Secrecy Performance Optimization]\label{probOriginal}
{\setlength{\arraycolsep}{0.0em}
\begin{eqnarray}
\mathbf{T}^*\triangleq&& \arg \mathop {\max }\limits_{\mathbf{T}} \textsf{p}_{{{\mathsf{u}}}}(\mathbf{T}) \nonumber\\
\mathrm{s.t.}&& \;
(\ref{eqconstcachingprob}),\;(\ref{eqconst2mbcachesize}),\;\eqref{eqconstCTP},\nonumber
\end{eqnarray}\setlength{\arraycolsep}{5pt}}where $\mathbf{T}^*$ denotes an optimal solution.
%which is referred to as the secure random caching~scheme.
\end{Problem}
}

\section{Secrecy Performance Analysis}\label{secAnalysis}
{In order to solve Problem \ref{probOriginal}, in this section, we analyze $\textsf{p}_{\mathsf{u}}(\mathbf{T})$ and $\textsf{p}_{\mathsf{e}}(\mathbf{T})$.} Specifically, we first
derive the closed-form expression of the RTP of file $n$, i.e., $\textsf{p}_{\mathsf{u},n}(\mathbf{T})$. Then, we derive the closed-form expression of the CTP of file $n$, i.e., $\textsf{p}_{\mathsf{e},n}(\mathbf{T})$. Note that, by substituting $\textsf{p}_{\mathsf{u},n}(\mathbf{T})$ and $\textsf{p}_{\mathsf{e},n}(\mathbf{T})$ into (\ref{eqaveragertp}) and (\ref{eqAVERAGECTP}), respectively, we can directly obtain $\textsf{p}_{\mathsf{u}}(\mathbf{T})$ and $\textsf{p}_{\mathsf{e}}(\mathbf{T})$.

\subsection{Analysis of RTP}\label{labsubsecRTPana}
In this subsection, we analyze $\textsf{p}_{\mathsf{u},n}(\mathbf{T})$, using tools from stochastic geometry. To calculate $\textsf{p}_{\mathsf{u},n}(\mathbf{T})$, based on (\ref{eqsirAuthorizedUser}), we first need to analyze
the distribution of the SIR, $\mathrm{SIR}_{b_{n,k_{\mathsf{u},0}},{\ell}_{\mathsf{u},0}}$. Under random caching, we note that the interferers for the typical user $\ell_{\mathsf{u},0}$ can be classified into three {categories}, i.e., i) {the} interfering BSs storing the requested file of $\ell_{\mathsf{u},0}$ in each tier, ii) {the} interfering BSs without the requested file of $\ell_{\mathsf{u},0}$ in each tier, iii) all the jamemrs in the network. In addition, under artificial-noise-aided transmission strategy, the artificial noise is embedded into the information signal transmitted from each BS in the network. {Taking the impacts of these three categories of interferers and the artificial noise on the SIR of the typical user into account,} we can derive the distribution of $\mathrm{SIR}_{b_{n,k_{\mathsf{u},0}},{\ell}_{\mathsf{u},0}}$ and then $\textsf{p}_{\mathsf{u},n}(\mathbf{T})$, as summarized in the
following theorem.
\begin{Theorem}[RTP of file $n$]\label{TheoremSTP}
The RTP of file $n$ is given by
{\begingroup\makeatletter\def\f@size{9.5}\check@mathfonts
\def\maketag@@@#1{\hbox{\m@th\normalsize\normalfont#1}}
\begin{equation}\label{labRTPoffilen}
\textsf{p}_{\mathsf{u},n}(\mathbf{T}_n)  =  \sum\limits_{j=1}^K\lambda_jT_{n,j}\left(\phi_jP_j\right)^{\delta}\left\|\mathbf{Q}_{M_j}^{-1}(\mathbf{T}_n)\right\|_1,
\end{equation}\endgroup}where $\|\cdot\|_1$ is the $L_1$ induced matrix norm and   $\mathbf{Q}_{M}(\mathbf{T}_n)$ is a lower triangular Toeplitz matrix,~i.e.,
{\begingroup\makeatletter\def\f@size{9.5}\check@mathfonts
\def\maketag@@@#1{\hbox{\m@th\normalsize\normalfont#1}}\setlength{\arraycolsep}{0.0em}
\begin{eqnarray}\label{eqdefQMj}
\mathbf{Q}_{M}(\mathbf{T}_n) =
\begin{pmatrix}
 q_{0}(\mathbf{T}_n)&  0&  0&  0&    \\
 q_{1}(\mathbf{T}_n)&  q_{0}(\mathbf{T}_n)&  0&  0&    \\
 \vdots&  \vdots& \ddots & \vdots  &   \\
 q_{M-1}(\mathbf{T}_n)&  \cdots&  q_{1}(\mathbf{T}_n)\;\;& q_{0}(\mathbf{T}_n) &
\end{pmatrix}.
\end{eqnarray}\setlength{\arraycolsep}{5pt}\endgroup}Here, $q_{m}(\mathbf{T}_n) = f_m(\mathbf{T}_n, {\theta_{\mathsf{u}}})$, ${\theta_{\mathsf{u}}} \triangleq 2^{R_{\mathsf{u}}}-1$, $m=0,1,\cdots,M_{j}-1$, where $f_{m}(\mathbf{T}_n, \theta)$ is given by (\ref{eqfm}), {as shown at the top of the next page}, with $U_{m,M}(\xi, \theta)$, $V_{m,M}(\xi, \theta)$ and $W_m(\theta)$ given by (\ref{eqUm})--(\ref{eqWm}), respectively, {as shown at the top of the next page}.
\begin{figure*}[!t]\vspace{-3mm}
{\begingroup\makeatletter\def\f@size{9.5}\check@mathfonts
\def\maketag@@@#1{\hbox{\m@th\normalsize\normalfont#1}}
\begin{align}
f_{m}(\mathbf{T}_n, \theta) & \triangleq
\sum_{k=1}^K{ {{\lambda _k}} {{\left(  {\phi_kP_k}  \right)}^\delta }} \left({T_{n,k}} U_{m,M_k}(\xi_k, \theta) +\left(1 - {T_{n,k}}\right)V_{m,M_k}(\xi_k, \theta) \right) + W_m(\theta),\label{eqfm}\\
U_{m,M}(\xi, \theta) &\triangleq
\begin{cases}
\eta_m \left(M-1, 1, \theta\right),\mbox{ if } \xi=1,\\
\frac{1}{\left(1-\xi\right)^{M - 1}} \left( { \eta_m \left(0,1, \theta\right)  - \sum\limits_{i = 0}^{M - 2}{\left(1-\xi\right)^i} \eta_m (i, \xi, \theta) } \right), \mbox{ otherwise},
\end{cases}\label{eqUm}\\
V_{m,M}(\xi, \theta) &\triangleq \begin{cases}
\kappa_m\left(M+1,1, \theta\right)  ,\mbox{ if } \xi=1,\\
\frac{1}{\left(1-\xi\right)^{M - 1}} \left( {{\kappa_m \left(2,1, \theta\right)} - \sum\limits_{i = 0}^{M - 2} { {{{{\left(1-\xi\right)}^i}{\kappa_m(i+2,\xi, \theta)}}} } } \right), \mbox{ otherwise},
\end{cases}\label{eqVm}\\
W_m(\theta)& \triangleq  {\lambda_{{\mathtt{J}}}{P^{\delta}_{{\mathtt{J}}}}} \kappa_m\left(2, 1, \theta\right).\label{eqWm}
\end{align}\endgroup} \noindent\rule[0.25\baselineskip]{\textwidth}{0.1pt}\vspace{-6mm}\end{figure*}Here, $\eta_m (i,\xi, \theta) \triangleq \frac{   (i+1)^{(m)}  \delta }{m! (\delta-m)   }  \xi^{m+1}  {_2F_1}(m-\delta,m+i+1;m-\delta+1;  -\xi\theta)\theta^m$, $\kappa_m(i,\xi, \theta) \triangleq \frac{(-1)^m\delta_{(m)}\Gamma(i+\delta-1)}{m!\Gamma(i-1)} \Gamma(1-\delta) { \xi ^{\delta +1 }}{\theta ^{\delta}}$, ${_2F_1}\left(a,b;c;z\right)$ and $\Gamma(x,y)$ denote the Gauss hypergeometric function and Gamma function, respectively, and $x_{(m)}\triangleq x(x-1)\cdots(x-m+1)$ and $x^{(m)} \triangleq x(x+1)\cdots(x+m-1) $ denote the falling and rising factorials, respectively.\footnote{Note that when $m=0$, we define $x_{(0)}=x^{(0)} = 1$.}
\end{Theorem}

\indent\indent \textit{Proof}: See Appendix \ref{proofTheoremSTP}. $\hfill\blacksquare$

\begin{figure}[!t]
    \centering {\includegraphics[width=.5\textwidth ]{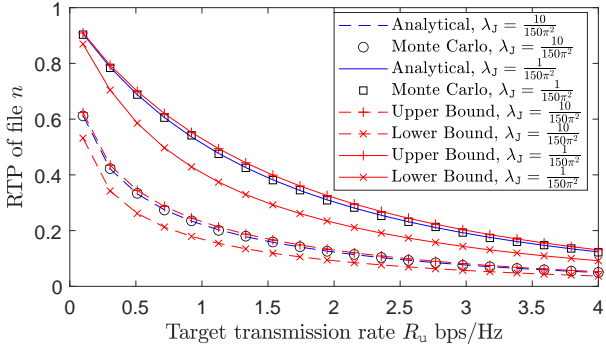}}
      \caption{{$\textsf{p}_{\mathsf{u},n}(\mathbf{T}_n)$} versus $R_{\mathsf{u}}$ at $K=2$, $M_1=4$, $M_2=2$, $\phi_1=0.9$, $\phi_2=0.5$, $P_1=20$~W, $P_2=0.13$ W, $P_{\mathtt{J}}=1$ W, $\lambda_1=\frac{1}{ 250^2\pi}$, $\lambda_2=\frac{1}{ 50^2\pi}$, $\lambda_{\mathtt{J}}=\frac{1}{150^2\pi}$, $\alpha=3.5$, $T_{n,1}=0.9$ and $T_{n,2}=0.8$.}\label{figMonteCarlo}
\end{figure}

We note that Theorem \ref{TheoremSTP} provides a closed-form expression for $\textsf{p}_{{{\mathsf{u}}},n}(\mathbf{T}_n)$. In Fig. \ref{figMonteCarlo}, we plot $\textsf{p}_{{{\mathsf{u}}},n}(\mathbf{T}_n)$ versus $R_{\mathsf{u}}$ for different values of $\lambda_{\mathtt{J}}$. We see that the ``Analytical'' curves, generated from Theorem~\ref{TheoremSTP}, accurately match the points obtained from the Monte Carlo simulations,  thus verifying the accuracy of the derived expression of $\textsf{p}_{{{\mathsf{u}}},n}(\mathbf{T}_n)$ in  Theorem \ref{TheoremSTP}.

\begin{figure}[!t]
    \centering
        \subfloat[$\lambda_{\mathtt{J}}=\frac{1}{ 150^2\pi}$.]{\includegraphics[width=.25\textwidth]{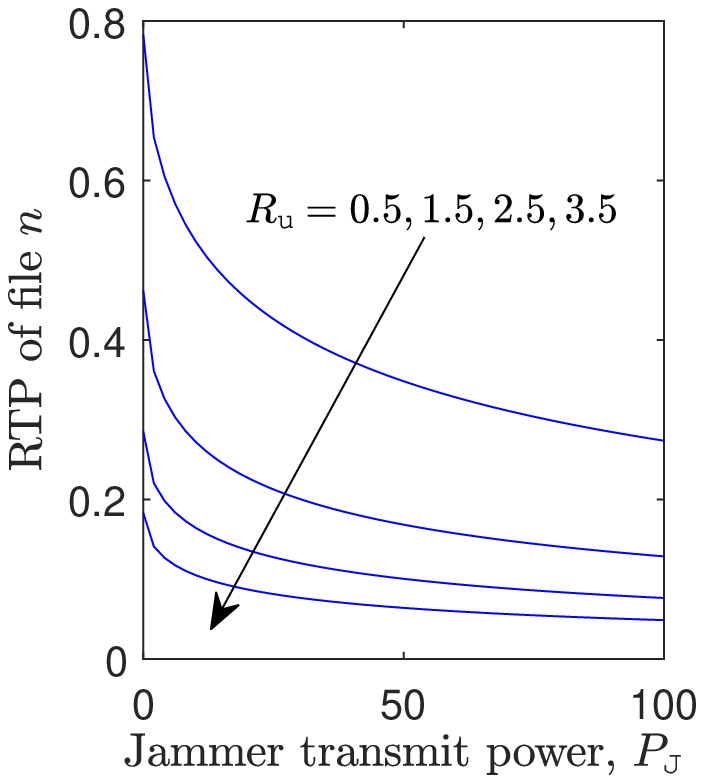}}
        \subfloat[$P_{\mathtt{J}}=1$ W.]{\includegraphics[width=.25\textwidth]{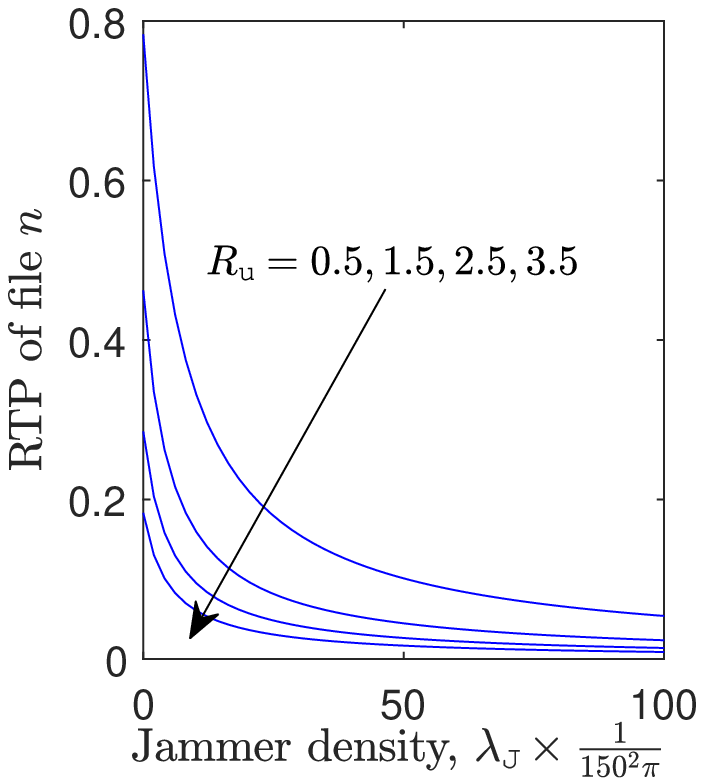}}
      \caption{{ (a) $\textsf{p}_{\mathsf{u},n}(\mathbf{T}_n)$ versus $P_{\mathtt{J}}$. (b) $\textsf{p}_{\mathsf{u},n}(\mathbf{T}_n)$ versus $\lambda_{\mathtt{J}}$. Here, $K=2$, $M_1=4$, $M_2=2$, $\phi_1=0.9$, $\phi_2=0.5$, $P_1=20$~W, $P_2=0.13$ W, $\lambda_1=\frac{1}{ 250^2\pi}$, $\lambda_2=\frac{1}{ 50^2\pi}$, $\alpha=3.5$, $T_{n,1}=0.9$ and $T_{n,2}=0.8$.} }\label{figRTPfilenVSPJLamJ}
\end{figure}

Moreover, it is important to understand how $\textsf{p}_{{\mathsf{u}},n}(\mathbf{T}_n)$ is
affected by some important system parameters, such as $P_{\mathtt{J}}$, $\lambda_{\mathtt{J}}$, $P_k$, $\lambda_k$  and $\mathbf{T}_n$. In the following, we provide some insights into the behavior of $\textsf{p}_{{{\mathsf{u}}},n}(\mathbf{T}_n)$  with respect to (w.r.t.) the above parameters.
{However,  due to} the $L_1$ induced matrix norm in the expression of $\textsf{p}_{{{\mathsf{u}}},n}(\mathbf{T}_n)$, it is extremely difficult to analyze the effects of $P_{\mathtt{J}}$, $\lambda_{\mathtt{J}}$, $P_k$, $\lambda_k$ and $\mathbf{T}_n$. To make progress,
we {first} derive the upper and lower bounds on $\textsf{p}_{{\mathsf{u}},n}(\mathbf{T}_n)$ {in the following proposition.}
\begin{Proposition}[Upper and Lower Bounds of RTP of file $n$]\label{lemmaBounds}
The RTP of file $n$ satisfies $\textsf{p}^{\textsf{L}}_{{{\mathsf{u}}},n}(\mathbf{T}_n) \le \textsf{p}_{{{\mathsf{u}}},n}(\mathbf{T}_n) \le \textsf{p}^{\textsf{U}}_{{{\mathsf{u}}},n}(\mathbf{T}_n)$, where
{\begingroup\makeatletter\def\f@size{9.5}\check@mathfonts
\def\maketag@@@#1{\hbox{\m@th\normalsize\normalfont#1}}\setlength{\arraycolsep}{0.0em}
\begin{align}
\textsf{p}^{\textsf{U}}_{{{\mathsf{u}}},n}(\mathbf{T}_n) &= 1-    \sum_{k=1}^K \sum_{m=0}^{M_k}\binom{M_k}{m} \frac{(-1)^m\lambda_kT_{n,k}\left(\phi_kP_k\right)^{\delta}} {f_0(\mathbf{T}_n, mS_{M_k}\theta_{\mathsf{u}})} \label{eqRTPub} ,\\
\textsf{p}^{\textsf{L}}_{{{\mathsf{u}}},n}(\mathbf{T}_n) &=  \sum\limits_{k=1}^K\frac{\lambda_kT_{n,k} \left(\phi_kP_k\right)^{\delta}} {\sum_{m=0}^{M_k-1} \left(1-\frac{m}{M_k}\right) f_m(\mathbf{T}_n, {\theta_{\mathsf{u}}}) }\label{eqRTPlb}.
\end{align}\setlength{\arraycolsep}{5pt}\endgroup}Here, $S_{M_k}= (M_k!)^{-M_k^{-1}}$ and $f_m (\mathbf{T}_n,  \theta)$ is given by (\ref{eqfm}).
\end{Proposition}

\indent\indent \textit{Proof}: See Appendix \ref{prooflemmaBounds}. $\hfill\blacksquare$

The tightness of $\textsf{p}^{\textsf{L}}_{{{\mathsf{u}}},n}(\mathbf{T}_n) $ and $\textsf{p}^{\textsf{U}}_{{{\mathsf{u}}},n}(\mathbf{T}_n) $ are evaluated in Fig. \ref{figMonteCarlo}. We can see that $\textsf{p}^{\textsf{L}}_{{\mathsf{u}},n}(\mathbf{T}_n)$ and $\textsf{p}^{\textsf{U}}_{{{\mathsf{u}}},n}(\mathbf{T}_n) $ have the similar trends as $\textsf{p}_{{\mathsf{u}},n}(\mathbf{T}_n)$, which show the correctness of our analysis in Proposition~\ref{lemmaBounds}.  In addition, we see that, compared to $\textsf{p}^{\textsf{L}}_{\mathsf{u}, n}(\mathbf{T}_n)$,  $\textsf{p}^{\textsf{U}}_{{{\mathsf{u}}},n}(\mathbf{T}_n) $ tightly matches $\textsf{p}_{{\mathsf{u}},n}(\mathbf{T}_n)$, and thus can serve as a good approximation for $\textsf{p}_{{\mathsf{u}},n}(\mathbf{T}_n)$. 
Based on Proposition~\ref{lemmaBounds}, some properties of $\textsf{p}_{{{\mathsf{u}}},n}(\mathbf{T}_n)$ are established~as~follows.
\begin{Property}[Effects of Jammer Transmit Power and Density]\label{propRTP2}
  $\textsf{p}^{\textsf{L}}_{{{\mathsf{u}}},n}(\mathbf{T}_n) $ increases with $P_{\mathtt{J}}$~and~$\lambda_{\mathtt{J}}$.
\end{Property}

\indent\indent \textit{Proof}: The proof is straightforward and we omit it for brevity. $\hfill\blacksquare$

Property~\ref{propRTP2} indicates that the jammers will degrade the transmission reliability of file $n$, since
the jamming signals degrade the link qualities of the typical user's channels. {Fig.~\ref{figRTPfilenVSPJLamJ} plots $\textsf{p}_{{{\mathsf{u}}},n}(\mathbf{T}_n)$ versus $P_{\mathtt{J}}$~and~$\lambda_{\mathtt{J}}$. From this figure, we see that although Property~\ref{propRTP2} is obtained based on $\textsf{p}^{\textsf{L}}_{{{\mathsf{u}}},n}(\mathbf{T}_n) $, it holds for $\textsf{p}_{{{\mathsf{u}}},n}(\mathbf{T}_n)$ as well.} To obtain more insights, we consider a special case {where all the BSs have the same number of antennas $M$ and power allocation ratio $\phi$, i.e., $M_k=M$ and $\phi_k=\phi$, for all $k\in\mathcal{K}$. In this special case, we can further obtain the following properties.}

\begin{Property}[Effects of BS Transmit Power and Density When $M_k=M$ and $\phi_k=\phi$, $k\in\mathcal{K}$]\label{propRTP3}
When $M_k=M$ and $\phi_k=\phi$, for all $k\in\mathcal{K}$, $\textsf{p}^{\textsf{L}}_{{{\mathsf{u}}},n}(\mathbf{T}_n) $ increases with $P_k$ and~$\lambda_k$ if $T_{n,k} \ge T_{n,k}^{{\mathsf{u}},\mathrm{th}}$, and {decrease} with $P_k$ and~$\lambda_k$ otherwise, where $T_{n,k}^{{\mathsf{u}},\mathrm{th}}$ is given by (\ref{property_3}), {as shown at the top of the next page}, and
\begin{figure*}[!t]{\setlength{\arraycolsep}{0.0em}\vspace{-4mm}
\begingroup\makeatletter\def\f@size{9.5}\check@mathfonts
\def\maketag@@@#1{\hbox{\m@th\normalsize\normalfont#1}}
\begin{eqnarray}\label{property_3}
T_{n,k}^{{\mathsf{u}},\mathrm{th}} = \frac{{\sum_{m = 0}^{M - 1} {\left( {1 - \frac{m}{M}} \right)} {V_{m,M}}(\xi ,{\theta _{\mathsf{u}}})\sum_{i \ne j}^K {{\lambda_iP_i^{\delta}}{T_{n,i}}{\phi ^\delta }} }}{{\sum_{m = 0}^{M - 1} {\left( {1 - \frac{m}{M}} \right)} {V_{m,M}}(\xi ,{\theta _{\mathsf{u}}})\sum_{i \ne j}^K {{\lambda_iP_i^{\delta}}{\phi ^\delta }}  + \sum_{m = 0}^{M - 1} {\left( {1 - \frac{m}{M}} \right)} {W_m}({\theta _{\mathsf{u}}})}}.
\end{eqnarray}\setlength{\arraycolsep}{5pt}\endgroup}\noindent\rule[0.25\baselineskip]{\textwidth}{0.1pt}\vspace{-6mm}\end{figure*}
 $\xi$ is given~by
\begin{equation*}
\xi  \triangleq \begin{cases}0,&\mbox{ if } M=1,\\ \frac{\phi^{-1}-1}{M-1}, &\mbox{ if } M=2,3,\cdots.\end{cases}
\end{equation*}
\end{Property}

\indent\indent \textit{Proof}: Denote $z_k\triangleq \lambda_kP_k^{\delta}$. When $M_k=M$ and $\phi_k=\phi$, for all $k\in\mathcal{K}$, we rewrite (\ref{eqRTPlb})~as
{\begingroup\makeatletter\def\f@size{9.5}\check@mathfonts
\def\maketag@@@#1{\hbox{\m@th\normalsize\normalfont#1}}\setlength{\arraycolsep}{0.0em}
\begin{eqnarray}\label{labRTPoffilenrewr}
\textsf{p}^{\textsf{L}}_{{{\mathsf{u}}},n}(\mathbf{T}_n)   =  \frac{{\sum_{k = 1}^K {{z_k}{T_{n,k}}{\phi ^\delta }} }}{f_m(\mathbf{T}_n, {\theta_{\mathsf{u}}}) }.
\end{eqnarray}\setlength{\arraycolsep}{5pt}\endgroup}Then, based on (\ref{labRTPoffilenrewr}), we have
{\begingroup\makeatletter\def\f@size{9.5}\check@mathfonts
\def\maketag@@@#1{\hbox{\m@th\normalsize\normalfont#1}}\setlength{\arraycolsep}{0.0em}
\begin{eqnarray}
\left\{ \begin{array}{l}
\frac{{\partial \textsf{p}^{\textsf{L}}_{{{\mathsf{u}}},n}(\mathbf{T}_n)}}{{\partial {P_k}}} = \frac{{\partial \textsf{p}^{\textsf{L}}_{{{\mathsf{u}}},n}(\mathbf{T}_n)}}{{\partial {z_k}}} \frac{{\partial z_k}}{{\partial {P_k}}}\\
\frac{{\partial \textsf{p}^{\textsf{L}}_{{{\mathsf{u}}},n}(\mathbf{T}_n)}}{{\partial {\lambda_k}}} = \frac{{\partial \textsf{p}^{\textsf{L}}_{{{\mathsf{u}}},n}(\mathbf{T}_n)}}{{\partial {z_k}}} \frac{{\partial z_k}}{{\partial {\lambda_k}}}
\end{array} \right.
,\label{eqparPunPk}
\end{eqnarray}\setlength{\arraycolsep}{5pt}\endgroup}where $\frac{{\partial z_k}}{{\partial {P_k}}} = \delta\lambda_k P_k^{\delta-1}$, $\frac{{\partial z_k}}{{\partial {\lambda_k}}} = P_k^{\delta}$ and
{\begingroup\makeatletter\def\f@size{9.5}\check@mathfonts
\def\maketag@@@#1{\hbox{\m@th\normalsize\normalfont#1}}\setlength{\arraycolsep}{0.0em}
\begin{eqnarray*}
&&\frac{{\partial \textsf{p}^{\textsf{L}}_{{{\mathsf{u}}},n}(\mathbf{T}_n)}}{{\partial {z_k}}}\nonumber\\ &&= \frac{{{T_{n,k}}{\phi ^\delta }\sum_{j \ne k}^K {{z_j}{T_{n,j}}{\phi ^\delta }} }}{{{{\left( {{f_m}({{\bf{T}}_n},{\theta _{\mathsf{u}}})} \right)}^2}}}\sum_{m = 0}^{M - 1} {\left( {1 - \frac{m}{M}} \right)} {V_{m,M}}(\xi ,{\theta _{\mathsf{u}}}){\phi ^\delta }\nonumber\\
&&-\frac{{{\phi ^\delta }\sum_{j \ne k}^K {{z_j}{T_{n,j}}{\phi ^\delta }} }}{{{{\left( {{f_m}({{\bf{T}}_n},{\theta _{\mathsf{u}}})} \right)}^2}}}\sum_{m = 0}^{M - 1} {\left( {1 - \frac{m}{M}} \right)} {V_{m,M}}(\xi ,{\theta _{\mathsf{u}}}).
\end{eqnarray*}\setlength{\arraycolsep}{5pt}\endgroup}Finally, by letting $\frac{{\partial \textsf{p}^{\textsf{L}}_{{{\mathsf{u}}},n}(\mathbf{T}_n)}}{{\partial {P_k}}} \ge 0$ and $\frac{{\partial \textsf{p}^{\textsf{L}}_{{{\mathsf{u}}},n}(\mathbf{T}_n)}}{{\partial {\lambda_k}}} \ge 0$ or $\frac{{\partial \textsf{p}^{\textsf{L}}_{{{\mathsf{u}}},n}(\mathbf{T}_n)}}{{\partial {P_k}}} < 0$ and $\frac{{\partial \textsf{p}^{\textsf{L}}_{{{\mathsf{u}}},n}(\mathbf{T}_n)}}{{\partial {\lambda_k}}} < 0$, {we obtain the desired result in \eqref{property_3}, which completes} the proof.$\hfill\blacksquare$

\begin{figure}[t]
    \centering
        \subfloat[$\lambda_2=\frac{1}{ 50^2\pi}$.]{\includegraphics[width=.25\textwidth]{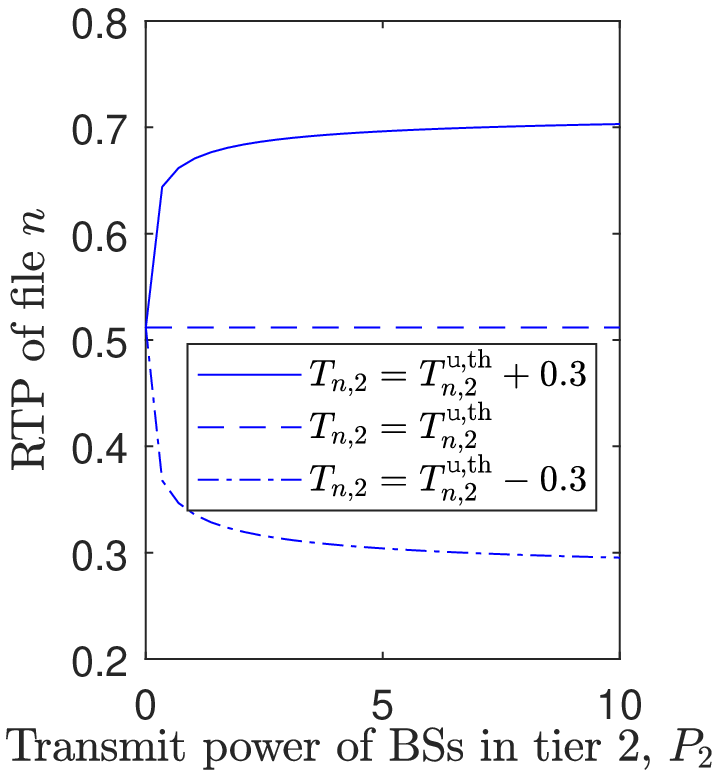}}
        \subfloat[$P_2=0.13$ W.]{\includegraphics[width=.25\textwidth]{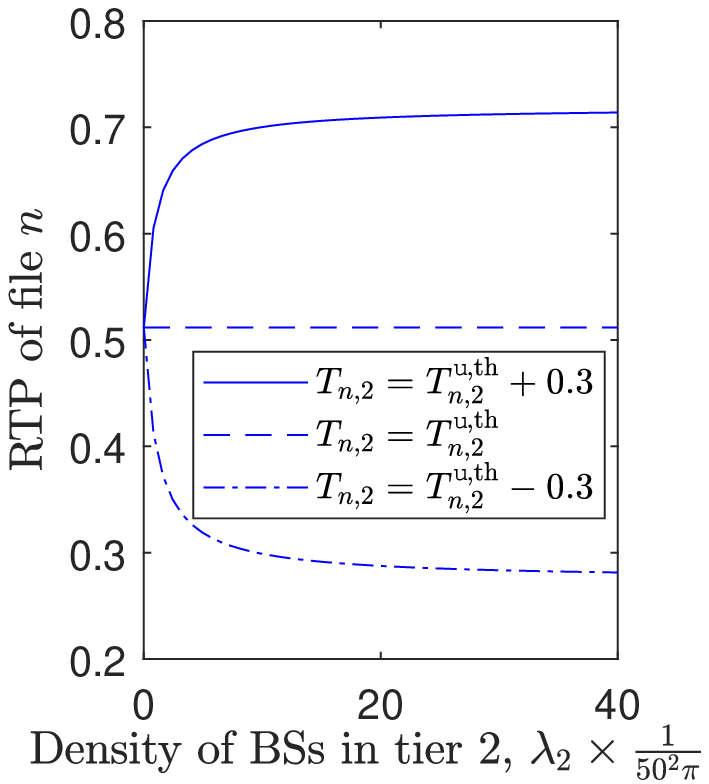}}
      \caption{{ (a) $\textsf{p}_{\mathsf{u},n}(\mathbf{T}_n)$ versus $P_2$. (b) $\textsf{p}_{\mathsf{u},n}(\mathbf{T}_n)$ versus $\lambda_2$. Here, $K=2$, $M_1=M_2=4$, $\phi_1=0.9$, $\phi_2=0.5$, $P_1=20$~W, $P_{\mathtt{J}}=1$ W, $\lambda_1=\frac{1}{ 250^2\pi}$, $\lambda_{\mathtt{J}}=\frac{1}{ 150^2\pi}$, $\alpha=3.5$, $R_{\mathsf{u}}=1.3$ bps/Hz, and $T_{n,1}=0.9$.}}\label{figRTPfilenVSP2Lam2}
\end{figure}

Property~\ref{propRTP3} indicates that if the caching probability of file $n$ in tier $k$ is relatively low, then increasing the BS transmit power or density in this tier will decrease the transmission reliability of file $n$. This can be explained as follows. If file $n$ is stored in tier $k$ with a very low probability, all the BSs in this tier will be the {interferer} of the typical user.  {As such,} the increase of the BS transmit power or density in tier $k$ will increase the interference suffered by the typical user. {Fig.~\ref{figRTPfilenVSP2Lam2} plots $\textsf{p}_{{{\mathsf{u}}},n}(\mathbf{T}_n)$ versus $P_2$~and~$\lambda_2$. From the figure, we see that Property~\ref{propRTP3} also holds for $\textsf{p}_{{{\mathsf{u}}},n}(\mathbf{T}_n)$.}

\begin{figure}[!t]
  \centering
  \includegraphics[width=0.5\textwidth]{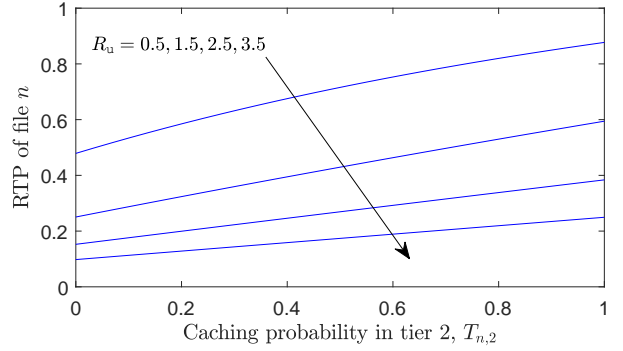}
  \caption{ $\textsf{p}_{\mathsf{u},n}(\mathbf{T}_n)$ versus $T_{n,2}$, where $K=2$, $M_1=M_2=4$, $\phi_1=0.9$, $\phi_2=0.5$, $P_1=20$~W, $P_2=0.13$ W, $P_{\mathtt{J}}=1$ W, $\lambda_1=\frac{1}{ 250^2\pi}$, $\lambda_2=\frac{1}{ 50^2\pi}$, $\lambda_{\mathtt{J}}=\frac{1}{ 150^2\pi}$, $\alpha=3.5$, and $T_{n,1}=0.9$.}\label{figRTPvsTn2}
\end{figure}

\begin{Property}[Effect of Caching Probability When $M_k=M$ and $\phi_k=\phi$,  $k\in\mathcal{K}$]\label{propRTP1}
When $M_k=M$ and $\phi_k=\phi$, for all $k\in\mathcal{K}$, $\textsf{p}^{\textsf{L}}_{{{\mathsf{u}}},n}(\mathbf{T}_n) $ {increases with}~$T_{n,k}$.
\end{Property}

\indent\indent \textit{Proof}: {See Appendix \ref{proofpropRTP1}.} $\hfill\blacksquare$

Property~\ref{propRTP1} indicates that caching a file at more BSs will always increase the transmission reliability of this file. This is because the average distance between a user requesting file $n$ and its serving BS decreases with the caching probability~$T_{n,k}$. {Fig.~\ref{figRTPvsTn2} plots $\textsf{p}_{{{\mathsf{u}}},n}(\mathbf{T}_n)$ versus $T_{n,k}$, from which we observe that Property~\ref{propRTP1} also holds for $\textsf{p}_{{{\mathsf{u}}},n}(\mathbf{T}_n)$.}

{Finally,} we study the concavity and convexity of $\textsf{p}^{\textsf{L}}_{{{\mathsf{u}}},n}(\mathbf{T}_n) $ and $\textsf{p}^{\textsf{U}}_{{{\mathsf{u}}},n}(\mathbf{T}_n) $ w.r.t. $\mathbf{T}_n$, as follows.
\begin{Property}[Concavity and Convexity of $\textsf{p}^{\textsf{L}}_{{{\mathsf{u}}},n}(\mathbf{T}_n) $ and $\textsf{p}^{\textsf{U}}_{{{\mathsf{u}}},n}(\mathbf{T}_n) $ w.r.t. Caching Probability When $M_k=M$ and $\phi_k=\phi$,  $k\in\mathcal{K}$]\label{propRTPconcavity}
When $M_k=M$ and $\phi_k=\phi$, for all $k\in\mathcal{K}$, $\textsf{p}^{\textsf{L}}_{{{\mathsf{u}}},n}(\mathbf{T}_n) $ is a concave function of~$T_{n,k}$, and $\textsf{p}^{\textsf{U}}_{{{\mathsf{u}}},n}(\mathbf{T}_n) = \textsf{p}^{\textsf{U}, 1}_{{{\mathsf{u}}},n}(\mathbf{T}_n) - \textsf{p}^{\textsf{U}, 2}_{{{\mathsf{u}}},n}(\mathbf{T}_n) +1$ is a difference-of-concave (DC) function of $T_{n,k}$, where
{\begingroup\makeatletter\def\f@size{9.5}\check@mathfonts
\def\maketag@@@#1{\hbox{\m@th\normalsize\normalfont#1}}\setlength{\arraycolsep}{0.0em}
\begin{equation}\label{eqqistatic}
\textsf{p}^{\textsf{U}, i}_{{{\mathsf{u}}}, n}(\mathbf{T}_n)  =   \sum\limits_{m\in \mathcal{M}^i}\binom{M}{m}\frac{\sum\limits_{k \in \mathcal{K}}\lambda_kT_{n,k}\left(\phi P_k\right)^{\delta}} {f_0(\mathbf{T}_n, mS_{M}\theta_{\mathsf{u}})},\; i=1,2.
\end{equation}\setlength{\arraycolsep}{5pt}\endgroup}Here, $\textsf{p}^{\textsf{U}, i}_{{{\mathsf{u}}}, n}(\mathbf{T}_n)$ is a concave function of $T_{n,k}$, $\mathcal{M}^1$  and $\mathcal{M}^2$ denote the sets of the odd and even numbers in set $\{0,1,2,\cdots,M\}$, respectively,
\end{Property}

\indent\indent \textit{Proof}: {See Appendix \ref{proofpropRTPconcavity}.} $\hfill\blacksquare$

\subsection{Analysis of CTP}
In this subsection, we analyze $\textsf{p}_{\mathsf{e},n}(\mathbf{T}_n)$, using tools from
stochastic geometry. {To derive $\textsf{p}_{{\mathsf{e}},n}(\mathbf{T}_n)$, we must derive the joint probability distribution of $ \mathrm{SIR}_{b_{n,k_{\mathsf{e},0}},{\ell}_{\mathsf{e},0}}$ for all ${k_{\mathsf{e},0}\in\mathcal{K}, \; b_{n,k_{\mathsf{e},0}}\in\Phi_{n,k_{\mathsf{e},0}}}$, which, however, is extremely challenging due to the fact that the variables $\mathrm{SIR}_{b_{n,k_{\mathsf{e},0}},{\ell}_{\mathsf{e},0}}$, for all ${k_{\mathsf{e},0}\in\mathcal{K}, \; b_{n,k_{\mathsf{e},0}}\in\Phi_{n,k_{\mathsf{e},0}}}$ are correlated with each other. In order to make progress, motivated by \cite{JGAndrewHetNetsMaxSINRass}, we focus on the high redundancy rate scenario, i.e., $R_{\mathsf{e}}>1$ bps/Hz. Note that, it shows in  \cite[Lemma 1]{JGAndrewHetNetsMaxSINRass} that in this scenario, at most one BS in the entire network can provide channel capacity greater than $R_{\mathsf{e}}$, i.e., the typical eavesdropper can successfully decode message from at most one BS.}
Then, in this scenario, by carefully characterizing the impact of random caching and {the artificial-noise-aided transmission strategy} on the distribution of $\mathrm{SIR}_{b_{n,k_{\mathsf{e},0}},{\ell}_{\mathsf{e},0}} $, for each ${k_{\mathsf{e},0}\in\mathcal{K}, \; b_{n,k_{\mathsf{e},0}}\in\Phi_{n,k_{\mathsf{e},0}}}$,  we can derive the distribution
of $\mathrm{SIR}_{b_{n,k_{\mathsf{e},0}},{\ell}_{\mathsf{e},0}} $, for each ${k_{\mathsf{e},0}\in\mathcal{K}, \; b_{n,k_{\mathsf{e},0}}\in\Phi_{n,k_{\mathsf{e},0}}}$ and then $\textsf{p}_{\mathsf{e},n}(\mathbf{T}_n)$, as summarized in the following theorem.

\begin{Theorem}[CTP of file $n$]\label{TheoremSP}
When $R_{\mathsf{e}}>1$, the CTP of file $n$ is given by
{\begingroup\makeatletter\def\f@size{9.5}\check@mathfonts
\def\maketag@@@#1{\hbox{\m@th\normalsize\normalfont#1}}\setlength{\arraycolsep}{0.0em}
\begin{eqnarray}\label{labCTPoffilen}
&&\textsf{p}_{{\mathsf{e}},n}(\mathbf{T}_n)   \nonumber \\
&& =  1-
\frac{{\sum_{k = 1}^K {{\lambda _k}{T_{n,k}}} {{\left( {{\phi _k}{P_k}} \right)}^\delta }{{(1 + {\xi _k}{\theta_{\mathsf{e}}})}^{1 - {M_k}}}}}{{{W_0}({\theta_{\mathsf{e}}}) + \sum_{k = 1}^K {{\lambda _k}{{\left( {{\phi _k}{P_k}} \right)}^\delta }{V_{0,{M_k}}}\left( {{\xi _k},{\theta_{\mathsf{e}}}} \right)} }},
\end{eqnarray}\setlength{\arraycolsep}{5pt}\endgroup}where ${\theta_{\mathsf{e}}} \triangleq 2^{R_{\mathsf{e}}}-1$, ${V_{0,M}}\left( {{\xi },{\theta}} \right)$ and ${W_0}({\theta})$ are given by (\ref{eqVm}) and (\ref{eqWm}), respectively.
\end{Theorem}

\indent\indent \textit{Proof}: See Appendix \ref{proofTheoremSP}. $\hfill\blacksquare$

\begin{figure}[!t]\vspace{-2mm}
    \centering {\includegraphics[width=.5\textwidth ]{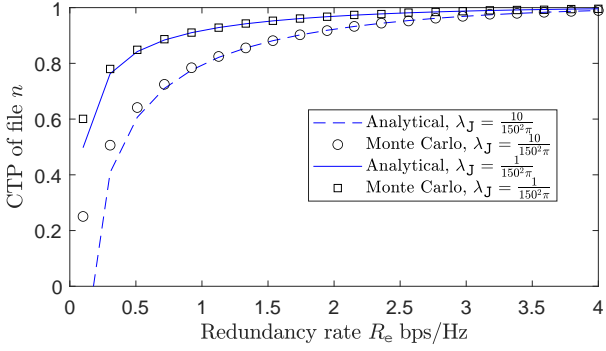}}

      \caption{{$\textsf{p}_{\mathsf{e},n}(\mathbf{T}_n)$} versus $R_{\mathsf{e}}$ at $K=2$, $M_1=4$, $M_2=2$, $\phi_1=0.9$, $\phi_2=0.5$, $P_1=20$~W, $P_2=0.13$ W, $P_{\mathtt{J}}=1$ W, $\lambda_1=\frac{1}{ 250^2\pi}$, $\lambda_2=\frac{1}{ 50^2\pi}$, $\alpha=3.5$, $T_{n,1}=0.8$ and $T_{n,2}=0.7$.}\label{figMonteCarloSP}
\end{figure}

{We note that} Theorem \ref{TheoremSP} provides a closed-form expression for $\textsf{p}_{\mathsf{e},n}(\mathbf{T}_n)$ in the high redundancy rate scenario. {In Fig.~\ref{figMonteCarloSP}, we plot $\textsf{p}_{\mathsf{e},n}(\mathbf{T}_n)$ versus $R_{\mathsf{e}}$ for different values of $\lambda_{\mathtt{J}}$.} We see that the ``Analytical'' curves, obtained from Theorem~\ref{TheoremSP}, closely match the ``Monte Carlo'' simulation points, indicating that Theorem \ref{TheoremSP} can also serve as a good approximation for $\textsf{p}_{\mathsf{e},n}(\mathbf{T}_n)$ in the low redundancy rate scenario.

{Based on Theorem \ref{TheoremSP}, we can obtain important insights into the behavior of $\textsf{p}_{{\mathsf{e}},n}(\mathbf{T}_n)$ w.r.t the system parameters (e.g., $P_{\mathtt{J}}$, $\lambda_{\mathtt{J}}$, $P_k$, $\lambda_k$ and $\mathbf{T}_n$), as follows.}

\begin{Property}[Effects of Jammer Transmit Power and Density]\label{propCTP2}
  $\textsf{p}_{{\mathsf{e}},n}(\mathbf{T}_n)$ increases with $P_{\mathtt{J}}$~and~$\lambda_{\mathtt{J}}$.
\end{Property}

\indent\indent \textit{Proof}: The proof is straightforward and we omit it for brevity. $\hfill\blacksquare$

\begin{figure}[!t]\vspace{-6mm}
    \centering
        \subfloat[$\lambda_{\mathtt{J}}=\frac{1}{ 150^2\pi}$.]{\includegraphics[width=.25\textwidth]{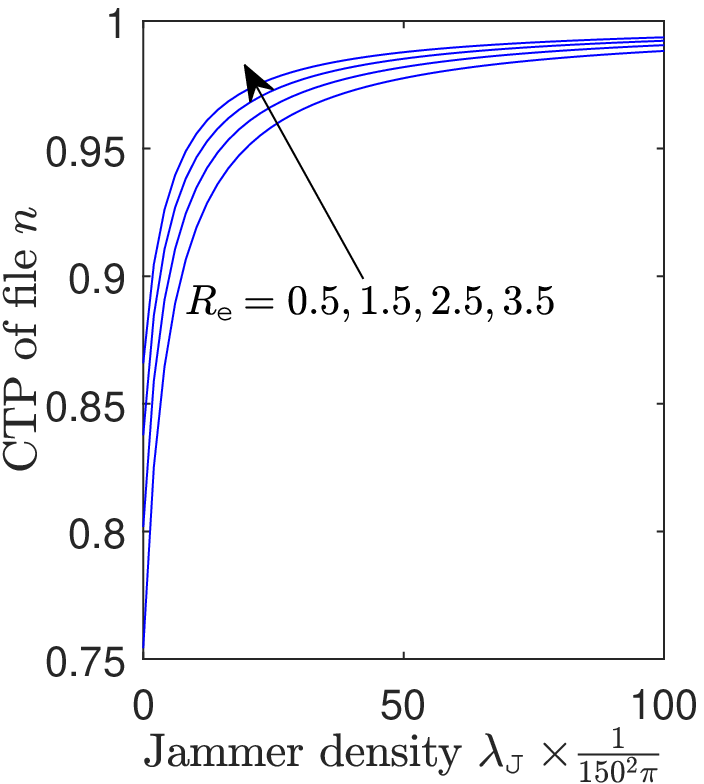}}
        \subfloat[$P_{\mathtt{J}}=1$ W.]{\includegraphics[width=.25\textwidth]{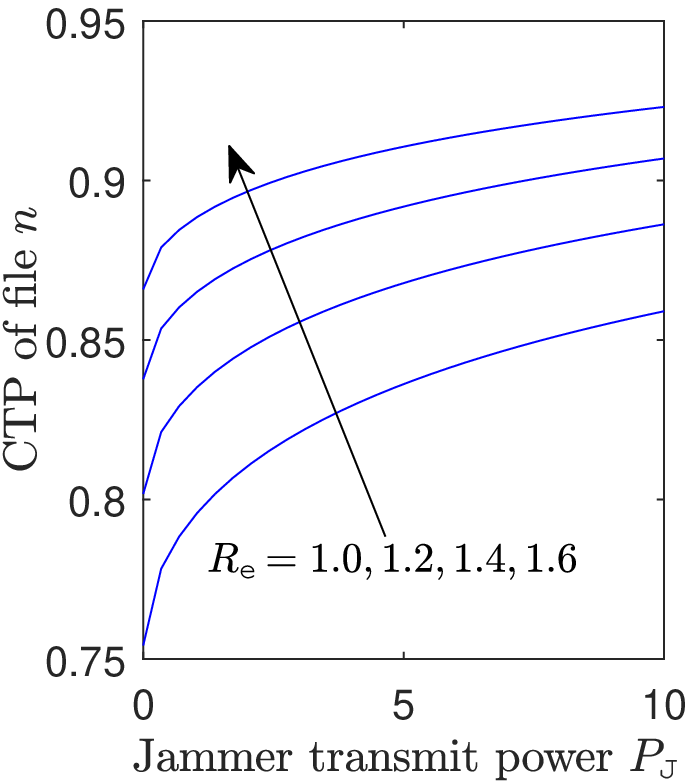}}
      \caption{{ (a) $\textsf{p}_{\mathsf{e},n}(\mathbf{T}_n)$ versus $P_{\mathtt{J}}$. (b) $\textsf{p}_{\mathsf{e},n}(\mathbf{T}_n)$ versus $\lambda_{\mathtt{J}}$. Here, $K=2$, $M_1=4$, $M_2=2$, $\phi_1=0.9$, $\phi_2=0.5$, $P_1=20$~W, $P_2=0.13$ W, $\lambda_1=\frac{1}{ 250^2\pi}$, $\lambda_2=\frac{1}{ 50^2\pi}$, $\alpha=3.5$, $T_{n,1}=0.9$ and $T_{n,2}=0.8$.}}\label{figCTPfilenVSPJLamJ}
\end{figure}

Property~\ref{propCTP2} indicates that, {although the jammers in the HetNets can} degrade the transmission reliability of file $n$, {they} can improve the transmission confidentiality of this file{, as shown in Fig.~\ref{figCTPfilenVSPJLamJ}.} This is because the jamming signals degrade the link qualities of the eavesdropper's channels.

\begin{figure}[!t]\vspace{-3mm}
    \centering
        \subfloat[$\lambda_2=\frac{1}{ 50^2\pi}$.]{\includegraphics[width=.25\textwidth]{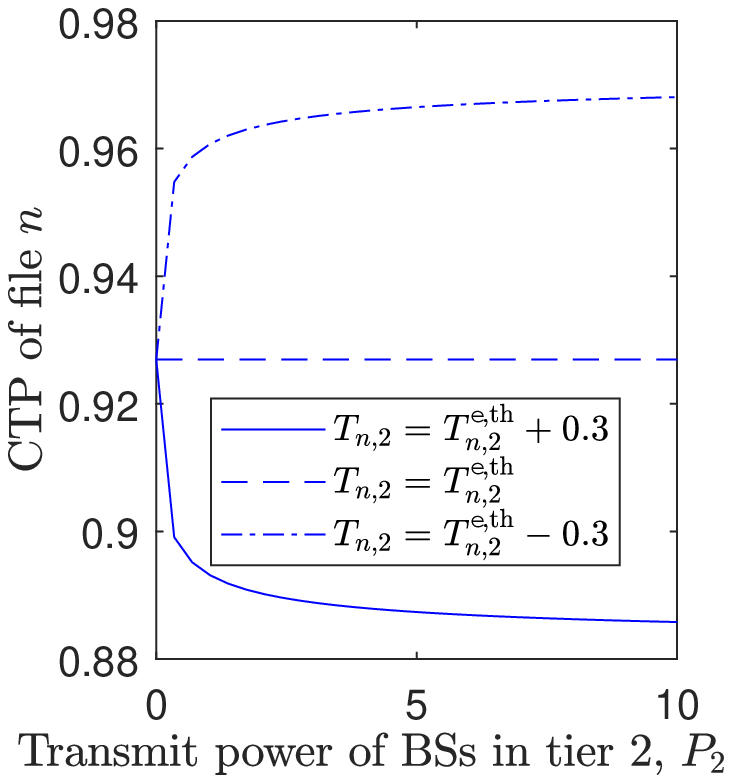}}
        \subfloat[$P_2=0.13$ W.]{\includegraphics[width=.25\textwidth]{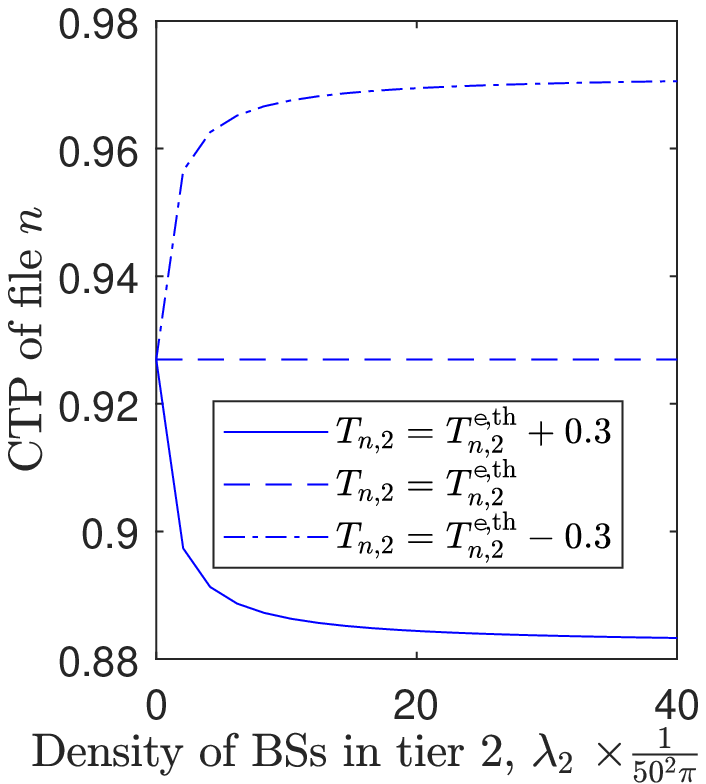}}
      \caption{{ (a) $\textsf{p}_{\mathsf{e},n}(\mathbf{T}_n)$ versus $P_2$. (b) $\textsf{p}_{\mathsf{e},n}(\mathbf{T}_n)$ versus $\lambda_2$. Here, $K=2$, $M_1=M_2=4$, $\phi_1=\phi_2=0.5$, $P_1=20$~W, $P_{\mathtt{J}}=1$ W, $\lambda_1=\frac{1}{ 250^2\pi}$, $\lambda_{\mathtt{J}}=\frac{1}{ 150^2\pi}$, $\alpha=3.5$, $R_{\mathsf{e}}=1.1$ bps/Hz, and $T_{n,1}=0.9$.}}\label{figCTPfilenVSP2Lam2}
\end{figure}

\begin{Property}[Effects of BS Transmit Power and Density]\label{propCTP3}
  $\textsf{p}_{{\mathsf{e}},n}(\mathbf{T}_n)$ decreases with $P_k$ and~$\lambda_k$ if $T_{n,k}\ge T_{n,k}^{{\mathsf{e}},\mathrm{th}}$, and increases with $P_k$ and~$\lambda_k$ otherwise, where
{\begingroup\makeatletter\def\f@size{9.5}\check@mathfonts
\def\maketag@@@#1{\hbox{\m@th\normalsize\normalfont#1}}\setlength{\arraycolsep}{0.0em}
\begin{eqnarray*}
&&T_{n,k}^{{\mathsf{e}},\mathrm{th}} \nonumber \\
&& = \frac{{{V_{0,{M_k}}}\left( {{\xi _k},{\theta _{\mathsf{e}}}} \right)\sum_{j = 1,j \ne k}^K {{\lambda_jP_j^{\delta}}{T_{n,j}}\phi _j^\delta {{\left( {1 + {\xi _j}{\theta _{\mathsf{e}}}} \right)}^{1 - {M_j}}}} }}{{{{\left( {1 + {\xi _k}{\theta _{\mathsf{e}}}} \right)}^{1 - {M_k}}}\left( {{W_0}({\theta _{\mathsf{e}}}) + \sum\limits_{j = 1,j \ne k}^K {\lambda_jP_j^{\delta}\phi _j^\delta {V_{0,{M_j}}}\left( {{\xi _j},{\theta _{\mathsf{e}}}} \right)} } \right)}}.
\end{eqnarray*}\setlength{\arraycolsep}{5pt}\endgroup}
\end{Property}
%decreases with $P_{\mathtt{J}}$ and $\lambda_{\mathtt{J}}$.
\indent\indent \textit{Proof}: The proof is similar to that in Property \ref{propRTP3} and we omit it for brevity. $\hfill\blacksquare$

Property~\ref{propCTP3} indicates that if the caching probability of file $n$ in tier $k$ is relatively low, then increasing the transmit power or density of the BSs from this tier will increase the transmission confidentiality of file $n$. This is because increasing the transmit power or density of the BSs will result in an increase in the interference received by the eavesdroppers.
{Fig.~\ref{figCTPfilenVSP2Lam2} verifies Property~\ref{propCTP3}.}

\begin{Property}[Effect of Caching Probability]\label{propCTP1}
$\textsf{p}_{{\mathsf{e}},n}(\mathbf{T}_n)$ {decreases with}~$T_{n,k}$. %Furthermore, the decreasing speed increases with $\lambda_k$ and $P_k$ but decreases with $\lambda_{\mathtt{J}}$ and~$P_{\mathtt{J}}$.
\end{Property}

\indent\indent \textit{Proof}: From (\ref{labCTPoffilen}), we have
{\begingroup\makeatletter\def\f@size{9.5}\check@mathfonts
\def\maketag@@@#1{\hbox{\m@th\normalsize\normalfont#1}}\setlength{\arraycolsep}{0.0em}
\begin{eqnarray*}
\frac{{\partial {{\textsf{p}}_{e,n}}({{\bf{T}}_n})}}{{\partial {T_{n,k}}}} =  - \frac{{{\lambda _k}{{\left( {{\phi _k}{P_k}} \right)}^\delta }{{\left( {1 + {\xi _k}{\theta _{\mathsf{e}}}} \right)}^{1 - {M_k}}}}}{{{W_0}({\theta _{\mathsf{e}}}) + \sum_{k = 1}^K {{\lambda _k}{{\left( {{\phi _k}{P_k}} \right)}^\delta }{V_{0,{M_k}}}\left( {{\xi _k},{\theta _{\mathsf{e}}}} \right)} }}.
\end{eqnarray*}\setlength{\arraycolsep}{5pt}\endgroup}Obviously, we see that $\frac{{\partial {{\textsf{p}}_{e,n}}({{\bf{T}}_n})}}{{\partial {T_{n,k}}}} < 0$ is a constant w.r.t. $T_{n,k}$, which completes the proof of this property .$\hfill\blacksquare$ 

\begin{figure}[!t]\vspace{-4mm}
  \centering
  \includegraphics[width=0.5\textwidth]{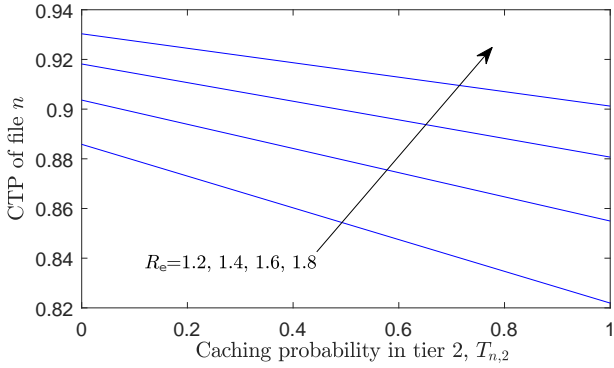}
  \caption{ $\textsf{p}_{\mathsf{e},n}(\mathbf{T}_n)$ versus $T_{n,2}$, where $K=2$, $M_1=4$, $M_2=2$, $\phi_1=0.9$, $\phi_2=0.5$, $P_1=20$~W, $P_2=0.13$ W, $P_{\mathtt{J}}=1$ W, $\lambda_1=\frac{1}{ 250^2\pi}$, $\lambda_2=\frac{1}{ 50^2\pi}$, $\lambda_{\mathtt{J}}=\frac{1}{ 150^2\pi}$, $\alpha=3.5$, $R_{\mathsf{e}}=1.2$ bps/Hz, and $T_{n,1}=0.9$.}\label{figCTPvsTn2}
\end{figure}

Property~\ref{propCTP1} indicates that caching a file at more BSs will compromise the transmission confidentiality of this file, since caching the file at more BSs incurs a higher risk of being eavesdropped. {Fig.~\ref{figCTPvsTn2} verifies Property~\ref{propCTP1}.}

\begin{Property}[Linearity of $\textsf{p}_{{\mathsf{e}},n}(\mathbf{T}_n)$ w.r.t.  Caching Probability]\label{propCTPlinearity}
$\textsf{p}_{{\mathsf{e}},n}(\mathbf{T}_n)$ is a linear function~of~$T_{n,k}$.
\end{Property}

\indent\indent \textit{Proof}: The proof is straightforward and we omit it for brevity. $\hfill\blacksquare$

In section \ref{sectionbetaST}, we shall see that this linearity of $\textsf{p}_{{\mathsf{e}},n}(\mathbf{T}_n)$ will greatly facilitate the performance optimization.

\section{Secrecy Performance Optimization}\label{sectionbetaST}
{In this section, we focus on solving Problem \ref{probOriginal}.} By substituting (\ref{labRTPoffilen}) into (\ref{eqaveragertp}) and (\ref{labCTPoffilen}) into (\ref{eqAVERAGECTP}), the RTP $\textsf{p}_{{{\mathsf{u}}}}(\mathbf{T})$ and the CTP $\textsf{p}_{{{\mathsf{e}}}}(\mathbf{T})$ are, respectively, calculated as (\ref{eqRTPOpt}) and~(\ref{eqCTPOpt}), {as shown at the top of the next page}.
\begin{figure*}[!t] {\setlength{\arraycolsep}{0.0em}
\begingroup\makeatletter\def\f@size{9.5}\check@mathfonts
\def\maketag@@@#1{\hbox{\m@th\normalsize\normalfont#1}}
\begin{eqnarray}
\textsf{p}_{\mathsf{u}}(\mathbf{T}) &=& \sum_{n\in\mathcal{N}}a_n \sum\limits_{j\in\mathcal{K}} \lambda_jT_{n,j}\left(\phi_jP_j\right)^{\delta}\left\|\mathbf{Q}_{M_j}^{-1}(\mathbf{T}_n)\right\|_1, \label{eqRTPOpt}\\
\textsf{p}_{\mathsf{e}}(\mathbf{T}) &=& \sum_{n\in\mathcal{N}}a_n \left( 1-
\frac{{\sum_{k = 1}^K {{\lambda _k}{T_{n,k}}} {{\left( {{\phi _k}{P_k}} \right)}^\delta }{{(1 + {\xi _k}{\theta_{\mathsf{e}}})}^{1 - {M_k}}}}}{{{W_0}({\theta_{\mathsf{e}}}) + \sum_{k = 1}^K {{\lambda _k}{{\left( {{\phi _k}{P_k}} \right)}^\delta }{V_{0,{M_k}}}\left( {{\xi _k},{\theta_{\mathsf{e}}}} \right)} }} \right). \label{eqCTPOpt}
\end{eqnarray}\setlength{\arraycolsep}{5pt}\endgroup}\noindent\rule[0.25\baselineskip]{\textwidth}{0.1pt}\end{figure*}

Recall that the upper bound $\textsf{p}^{\textsf{U}}_{{{\mathsf{u}}},n}(\mathbf{T}_n)$ in (\ref{eqRTPub}) provides a good approximation for $\textsf{p}_{{{\mathsf{u}}},n}(\mathbf{T}_n)$ in (\ref{labRTPoffilen}), {and $\textsf{p}^{\textsf{U}}_{{{\mathsf{u}}},n}(\mathbf{T}_n)$ is much more analytical than $\textsf{p}_{{{\mathsf{u}}},n}(\mathbf{T}_n)$.}
Hence, in the following, for simplicity and tractability, instead of maximizing $\textsf{p}_{{{\mathsf{u}}}}(\mathbf{T})$ directly, we maximize its upper bound, given by
\begin{equation}
\textsf{p}^{\textsf{U}}_{{{\mathsf{u}}}}(\mathbf{T}) = \sum_{n\in\mathcal{N}}a_n \textsf{p}^{\textsf{U}}_{{{\mathsf{u}}},n}(\mathbf{T}_n),
\end{equation}where $\textsf{p}^{\textsf{U}}_{{{\mathsf{u}}},n}(\mathbf{T}_n)$ is given by (\ref{eqRTPub}). As such, we can transform Problem \ref{probOriginal} into the following~problem
\begin{Problem}[{Simplified} Secrecy Performance Optimization]\label{probTrans}
{\setlength{\arraycolsep}{0.0em}
\begin{eqnarray}
\mathbf{T}^* &&\triangleq \arg \mathop {\max }\limits_{\mathbf{T}} \textsf{p}^{\textsf{U}}_{{{\mathsf{u}}}}(\mathbf{T}) \nonumber\\
 &&\mathrm{s.t.} \;
(\ref{eqconstcachingprob}),\;(\ref{eqconst2mbcachesize}),\;(\ref{eqconstCTP})\;\nonumber
\end{eqnarray}\setlength{\arraycolsep}{5pt}}
\end{Problem}

Problem \ref{probTrans} maximizes a non-concave objective function over a linear set, denoted by $\mathcal{T}\triangleq \{\mathbf{T}\in[0,1]^{NK}|(\ref{eqconstcachingprob}), \;(\ref{eqconst2mbcachesize}), \;(\ref{eqconstCTP})\}$, and hence is a non-convex optimization problem in general. Since the objective function $\textsf{p}^{\textsf{U}}_{{{\mathsf{u}}}}(\mathbf{T})$ is continuously differentiable on $\mathcal{T}$,  we can obtain a stationary point\footnote{Please note that for a non-convex problem, in general, there is no guarantee that an optimal solution can be obtained. Instead, obtaining a stationary
point, i.e., a point that satisfies the corresponding Karush-Kuhn-Tucker conditions, is the classic goal for solving a non-convex optimization problem \cite{BertsekasNLP}.} of Problem \ref{probTrans} by using the gradient projection method (GPM) with a diminishing step size, denoted by $\zeta^{(t)}$ with $t=0,1,2,\cdots$ being the iteration index.  The details for solving Problem~\ref{probTrans} using GPM are summarized in Algorithm~\ref{algGPM}, {where the diminishing step size satisfies $\zeta^{(t)}\to 0$, as $ t\to\infty$, $\sum_{t=1}^{\infty}\zeta^{(t)}=\infty$ and $\sum_{t=1}^{\infty}(\zeta^{(t)})^2 < \infty$ \cite[pp. 227]{BertsekasNLP}.} In addition, in Step~3 of Algorithm \ref{algGPM}, $\Pi_{\mathcal{T}}(\hat{\mathbf{T}}) \triangleq \arg\min_{\mathbf{T}\in\mathcal{T}} \|\mathbf{T}-\hat{\mathbf{T}}\|$ denotes the Euclidean projection of $\hat{\mathbf{T}}$ onto ${\mathcal{T}}$,  and $\triangledown_{\mathbf{T}}\textsf{p}^{\textsf{U}}_{{{\mathsf{u}}}}(\mathbf{T}^{(t)}) \triangleq ( {  \frac{{\partial \textsf{p}^{\textsf{U}}_{\mathsf{u}}(\mathbf{T})}} {{\partial {T_{n,k}}}}  } |_{T_{n,k}=T_{n,k}^{(t)}})_{n\in\mathcal{N}, k\in\mathcal{K}} $ is the gradient of $ \textsf{p}^{\textsf{U}}_{{{\mathsf{u}}}}(\mathbf{T}) $ at $\mathbf{T} = \mathbf{T}^{(t)}$, given by (\ref{eqGMkphik}), {as shown at the top of the next page}.
\begin{figure*}[!t]
{\begingroup\makeatletter\def\f@size{9.5}\check@mathfonts\vspace{-3mm}
\def\maketag@@@#1{\hbox{\m@th\normalsize\normalfont#1}}\setlength{\arraycolsep}{0.0em}
\begin{align} \label{eqGMkphik}
\frac{{\partial \textsf{p}^{\textsf{U}}_{{{\mathsf{u}}}}(\mathbf{T})}}{{\partial {T_{n,k}}}} &=  - {a_n}\sum\limits_{m = 1}^{{M_k}} {\binom{M_k}{m}} \frac{{{{( - 1)}^m}}}{{f_0^2({{\bf{T}}_n},m{S_{{M_k}}}{\theta _{\mathsf{u}}})}}\Bigg(\sum\limits_{i\in\mathcal{K}\setminus \{k\}} {{\lambda _i}{{\left( {{\phi _i}{P_i}} \right)}^\delta }{T_{n,i}}{\Lambda _k}({M_k},{\phi _k},m{S_{{M_k}}}{\theta _{\mathsf{u}}})} \nonumber\\
&\:  + {\lambda _k}{\left( {{\phi _k}{P_k}} \right)^\delta }\bigg( {\sum_{i\in\mathcal{K}\setminus \{k\}} {{T_{n,i}}{\Lambda _i}({M_k},{\phi _k},m{S_{{M_k}}}{\theta _{\mathsf{u}}})}  + \Psi ({M_k},{\phi _k},m{S_{{M_k}}}{\theta _{\mathsf{u}}})} \bigg)  \Bigg).
\end{align}\setlength{\arraycolsep}{5pt}\endgroup}\noindent\rule[0.25\baselineskip]{\textwidth}{0.1pt}\end{figure*}Here, ${\Lambda_k}(M_k, \phi_k, \theta ) \triangleq {\lambda _k}{\left( {{\phi _k}{P_k}} \right)^\delta }\left( {{U_{0,{M_k}}}({\xi _k},\theta ) - {V_{0,{M_k}}}({\xi _k},\theta )} \right)$ and $\Psi\left( M_k, \phi_k, \theta  \right) \triangleq \sum_{k\in\mathcal{K} } {\lambda _k}{{\left( {{\phi _k}{P_k}} \right)}^\delta }   {V_{0,{M_k}}}({\xi _k},\theta )  + {W_0}(\theta )$, where $U_{m,M}(\xi, \theta)$, $V_{m,M}(\xi, \theta)$ and $W_m(\theta)$ are given by Theorem~\ref{TheoremSTP}.  According to \cite{cvxbookDPBertsekas}, we know that the sequence $\{\mathbf{T}^{(t)}\}_{t=0}^{\infty}$ generated by Algorithm~\ref{algGPM} converges to a stationary point of Problem~\ref{probTrans}.   The computation cost of Algorithm \ref{algGPM} is dominated by the  Euclidean
projection at Step 3, which has the computational complexity $\mathcal{O}\left((KN)^3\right)$ in the use of interior point algorithm  \cite{convexoptimization}.

\begin{algorithm}[!t]%\doublespacing
\caption{Stationary Point of Problem \ref{probTrans} Based on GPM}
\begin{algorithmic}[1]\small
\STATE \textbf{initialization:} choose a feasible initial point $\mathbf{T}^{(0)}$ of Problem \ref{probTrans}, choose two positive constants $\varepsilon_{\mathrm{err}}$ and $t_{\max}$, and set $t=0$.
\REPEAT
\STATE compute  $\mathbf{T}^{(t+1)}$ according to $\mathbf{T}^{(t+1)} = \Pi_{\mathcal{T}}\left(\mathbf{T}^{(t)}- \zeta^{(t)}\triangledown_{\mathbf{T}}\textsf{p}^{\textsf{U}}_{{{\mathsf{u}}}}(\mathbf{T}^{(t)}) \right)$.
\STATE set $t\leftarrow t+1$.
\UNTIL $\|\mathbf{T}^{(t)}-\mathbf{T}^{(t-1)}\| < \varepsilon_{\mathrm{err}}$ or $t>t_{\max}$.
\end{algorithmic}\label{algGPM}
\end{algorithm}

Note that, the convergence rate of Algorithm~\ref{algGPM} is sensitive to the choice of the step size. If it is chosen improperly, Algorithm~\ref{algGPM} may need a great number of iterations to converge. Recall that in Property~\ref{propRTPconcavity}, in the special case of $M_k=M$ and $\phi_k=\phi$, for all $k\in\mathcal{K}$, $\textsf{p}^{\textsf{U}}_{{{\mathsf{u}}},n}(\mathbf{T}_n)$ has a DC structure. Thus, in this case, Problem~\ref{probTrans} becomes a DC optimization problem and thus can be solved using {convex-concave procedure (CCP)} \cite{Lipp2016}. Compared to GPM, CCP does not depend any step size and thus may lead to robust convergence performance. As such, in the following, we consider this special case and develop an iterative algorithm to obtain a stationary point of Problem \ref{probTrans} based on CCP.
The core idea of CCP is to linearize the convex term of the DC objective function to obtain a concave objective function for a maximization problem, and then solve a sequence of convex optimization problems successively.  Specifically, at iteration $t+1$, based on Property~\ref{propRTPconcavity}, we have the  following approximation problem:
\begin{Problem}[Approximation of Problem~\ref{probTrans} at Iteration $t+1$ When $M_k=M$ and $\phi_k=\phi$,~$k\in\mathcal{K}$]\label{eqsubproblem1DC}
 {\begingroup\makeatletter\def\f@size{9.5}\check@mathfonts
\def\maketag@@@#1{\hbox{\m@th\normalsize\normalfont#1}}\setlength{\arraycolsep}{0.0em}
\begin{eqnarray*}
\mathbf{T}^{(t+1)} &\triangleq  \arg \max\limits_{\mathbf{T}} \;\;\sum\limits_{n\in\mathcal{N}}a_n\left(\textsf{p}^{\textsf{U}, 1}_{\mathsf{u},n}(\mathbf{T}_n) - \tilde{\textsf{p}}^{\textsf{U}, 2}_{\mathsf{u},n} (\mathbf{T}_n;\mathbf{T}_n^{(t)})\right) + 1 \\
&\mathrm{s.t.}\;\;(\ref{eqconstcachingprob}),(\ref{eqconst2mbcachesize}),(\ref{eqconstCTP}),\;\nonumber
\end{eqnarray*}\setlength{\arraycolsep}{5pt}\endgroup}where $\tilde{\textsf{p}}^{\textsf{U}, 2}_{\mathsf{u},n} (\mathbf{T}_n;\mathbf{T}_n^{(t)}) \triangleq \textsf{p}^{\textsf{U}, 2}_{\mathsf{u},n}(\mathbf{T}_n^{(t)})+ (\mathbf{T}_n-\mathbf{T}_n^{(t)})^{\mathrm{T}}\triangledown_{\mathbf{T}_n}\textsf{p}^{\textsf{U}, 2}_{\mathsf{u},n}(\mathbf{T}_n^{(t)})$, $\textsf{p}^{\textsf{U}, i}_{\mathsf{u},n}(\mathbf{T}_n)$ is given by (\ref{eqqistatic}) and $\triangledown_{\mathbf{T}_n}\textsf{p}^{\textsf{U}, 2}_{\mathsf{u},n}(\mathbf{T}_n^{(t)})\triangleq\left( {  \frac{{\partial \textsf{p}^{\textsf{U}, 2}_{\mathsf{u},n}(\mathbf{T}_n)}} {{\partial {T_{n,k}}}}  } |_{T_{n,k}=T_{n,k}^{(t)}}\right)_{k\in\mathcal{K}}$ denotes the gradient of $\textsf{p}^{\textsf{U}, 2}_{\mathsf{u},n}(\mathbf{T}_n)$ at $\mathbf{T}_n=\mathbf{T}_n^{(t)}$, which is given by (\ref{eqpuder}), {as shown at the top of the next page}.
\begin{figure*}[!t]
{\begingroup\makeatletter\def\f@size{9.5}\check@mathfonts\vspace{-6mm}
\def\maketag@@@#1{\hbox{\m@th\normalsize\normalfont#1}}\setlength{\arraycolsep}{0.0em}
\begin{eqnarray}\label{eqpuder}
\frac{{\partial \textsf{p}^{\textsf{U}, 2}_{\mathsf{u},n}(\mathbf{T}_n)}}{{\partial {T_{n,k}}}} &=& \sum_{m \in {\cal M}^2} {\binom{M}{m}\frac{{\lambda _k}{{\left( {{\phi}{P_k}} \right)}^\delta }}{{f_0^2({{\bf{T}}_n},m{S_{{M}}}{\theta_{\mathsf{u}}})}}{\left( {\sum_{i\in\mathcal{K}\setminus\{k\}} {{T_{n,i}}{\Lambda_i}(M, \phi, m{S_{{M}}}{\theta_{\mathsf{u}}})}  + \Psi(M, \phi, m{S_{{M}}}{\theta_{\mathsf{u}}})} \right)}}  \nonumber\\
&&\:- \sum_{i\in\mathcal{K}\setminus\{k\}} {\sum_{m \in {\cal M}^2} {\binom{M}{m}\frac{{\lambda _i}{{\left( {{\phi}{P_i}} \right)}^\delta }}{{f_0^2({{\bf{T}}_n},m{S_{{M}}}{\theta_{\mathsf{u}}})}}{{T_{n,i}}{\Lambda_k}(M, \phi, m{S_{{M}}}{\theta_{\mathsf{u}}})}} }.
\end{eqnarray}\setlength{\arraycolsep}{5pt}\endgroup}\noindent\rule[0.25\baselineskip]{\textwidth}{0.1pt}\vspace{-3mm}\end{figure*}
\end{Problem}

\begin{algorithm}[!t]%\doublespacing
\caption{Stationary Point of Problem \ref{probTrans} Based on CCP}
\begin{algorithmic}[1]\small
\STATE \textbf{initialization:} choose a feasible initial point $\mathbf{T}^{(0)}$ of Problem \ref{probTrans}, choose two positive constants $\varepsilon_{\mathrm{err}}$ and $t_{\max}$, and set $t=0$.
\REPEAT
\STATE obtain  $\mathbf{T}^{(t+1)}$ by solving Problem \ref{eqsubproblem1DC}.
\STATE set $t\leftarrow t+1$.
\UNTIL $\|\mathbf{T}^{(t)}-\mathbf{T}^{(t-1)}\| < \varepsilon_{\mathrm{err}}$ or $t>t_{\max}$.
\end{algorithmic}\label{algDC}
\end{algorithm}

Due to the concave objective and linear constraints, Problem~\ref{eqsubproblem1DC} is a convex problem, and thus can be solved by an interior point method efficiently. The details for solving Problem~\ref{probTrans} using CCP are summarized in Algorithm~\ref{algDC}. {To initialize Algorithm~\ref{algDC}, we can randomly choose a point and then project it onto the linear constraint set of Problem~\ref{probTrans}.
According to \cite{Lipp2016}, we declare that the sequence $\{\mathbf{T}^{(t)}\}_{t=0}^{\infty}$ generated by Algorithm~\ref{algDC} converges to a stationary point of Problem \ref{probTrans}. Similar as Algorithm \ref{algGPM},} the computation cost of Algorithm \ref{algDC} is dominated by the solution of Problem~\ref{eqsubproblem1DC} at Step 3 as well, which has the computational complexity $\mathcal{O}\left((KN)^3\right)$ in the use of interior point algorithm  \cite{convexoptimization}.

\section{Numerical Results}\label{secSimResult}
{In this section, we provide numerical results to verify the effectiveness of our proposed secure random caching scheme. Specifically, we first demonstrate} the convergence of Algorithm \ref{algGPM} and Algorithm \ref{algDC}. Then, we compare the performance of the proposed secure random caching scheme with that of some existing baseline schemes. In the simulations, we consider a two-tier HetNet, i.e., $K=2$, consisting of a macrocell network as the 1st tier overlaid with a picocell network as the 2nd tier. Unless otherwise stated, the simulation settings are as follows: $\varepsilon_{\mathrm{err}}=10^{-6}$, $t_{\max}=100$, $P_1=20$~W, $P_2=0.13$~W, $P_{\mathtt{J}}=1$~W, $\lambda_1=\frac{1}{250^2\pi}$, $\lambda_2=\frac{1}{25^2\pi}$,  $\lambda_{\mathtt{J}}=\frac{1}{250^2\pi}$, $\alpha=4$, $R_{\mathsf{u}}=1.3$ bps/Hz, $R_{\mathsf{s}}=0.2$ bps/Hz, $\epsilon=0.7$, $M_1=M_2=10$, $\phi_1=\phi_2=0.9$, $C_1=15$, $C_2=10$, $N=20$ and $a_n=\frac{n^{-\beta}}{\sum_{i\in\mathcal{N}}i^{-\beta}}$, where $\beta=0.6$ is the Zipf exponent.

\begin{figure}[!t]\vspace{-3mm}
    \centering
        {\includegraphics[width=.4\textwidth]{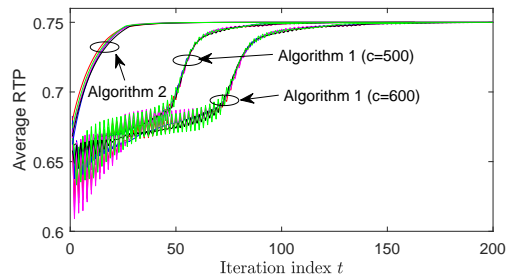}}
\caption{Convergence of Algorithm \ref{algGPM} and Algorithm \ref{algDC}. The step size for Algorithm \ref{algGPM} is $\zeta^{(t)} = \frac{c}{2+t^{0.55}}$ \cite{RandomcachingDTX2018Wen}. }\label{optAlg1convergence}
\end{figure}

\subsection{Convergence of Algorithm \ref{algGPM} and Algorithm \ref{algDC}}

In Fig. \ref{optAlg1convergence}, we present {the} convergence trajectories of Algorithm \ref{algGPM} and Algorithm \ref{algDC} as functions of iteration index for five different initial points. {We see that the convergence rate of Algorithm~\ref{algGPM} is highly sensitive to the choices of step size, while Algorithm~\ref{algDC} has more robust convergence performance, due to the fact that it does not rely on any step size. We also see that, for different initial points, both Algorithm \ref{algGPM} and Algorithm~\ref{algDC} converge to the same RTP, demonstrating their effectiveness in solving Problem~\ref{probTrans}. Moreover, compared to Algorithm~\ref{algDC}, the convergence trajectory of Algorithm~\ref{algGPM} is not monotonically decreasing, due to the Euclidean projection onto the feasible set.}

\subsection{Performance Comparisons between Proposed Scheme and Baselines}

\begin{figure}[!t]
    \centering
        \subfloat[{Average} RTP versus $\epsilon$.]{\includegraphics[width=.45\textwidth]{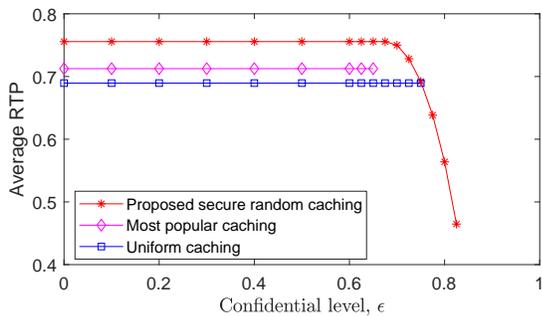}}\\
        \subfloat[Caching probability versus $n$ and $\epsilon$ of our proposed secure random caching scheme.]{\includegraphics[width=0.45\textwidth]{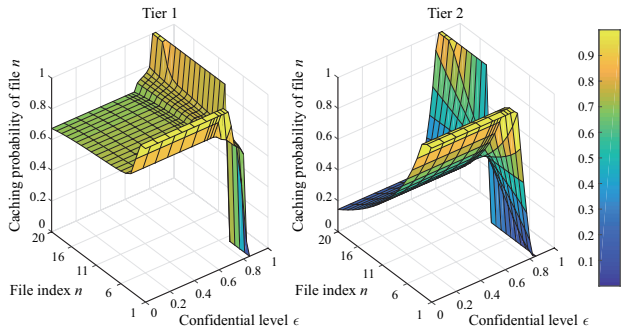}}
      \caption{Effect of confidential level.}\label{optRTPvsepsilon}
\end{figure}
{In Fig. \ref{optRTPvsepsilon} -- Fig. \ref{optRTPvscaching},  we examine the effects of system parameters and demonstrate the superiority of our proposed secure random caching schemes over two baseline schemes, in terms of the average RTP.} Specifically, Baseline 1 {adopts the most popular caching scheme, where  each BS in tier $k$ selects the $C_k$ most popular files to store \cite{Cache-enabledsmallcellnetworksmodelingandtradeoffs}, and
Baseline 2 {adopts the uniform caching scheme, where} each BS in tier $k$ randomly selects $C_k$ files to store, according to the uniform distribution \cite{Tamoorulhassan2015Modeling}.

\begin{figure}[!t]\vspace{-8mm}
    \centering
        \subfloat[{Average} RTP versus $P_{\mathtt{J}}$.]{\includegraphics[width=.25\textwidth]{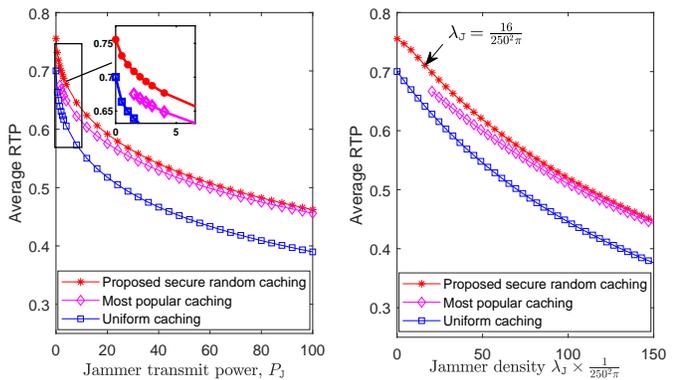}}
        \subfloat[{Average} RTP versus $\lambda_{\mathtt{J}}$ at $P_{\mathtt{J}}=0.13$ W.]{\includegraphics[width=.25\textwidth]{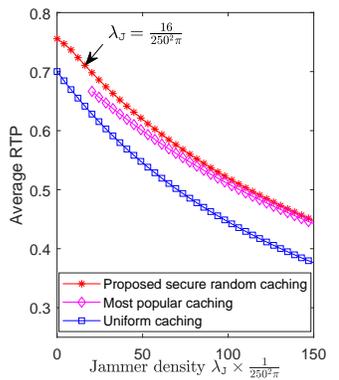}}
      \caption{(a) Effect of the jammer transmit power.  (b) Effect of the jammer density.}\label{optSTPvsbslns}
\end{figure}

Fig. \ref{optRTPvsepsilon} shows the effect of the confidential level $\epsilon$. Specifically, in Fig. \ref{optRTPvsepsilon}(a) we plot the {average} RTP versus $\epsilon$. From the figure, we can see that for a large $\epsilon$, the two baseline schemes may be infeasible (e.g., $\epsilon>0.65$ for the most popular caching and $\epsilon>0.75$ for the uniform caching), but our proposed secure random caching scheme may still be feasible (e.g., $\epsilon<0.81$), at the cost of RTP decrement (e.g., $0.65 < \epsilon<0.81$).    This can be explained as follows. The CTPs of the two baseline schemes (each with fixed caching distribution of files) are deterministic if given the system parameters. {Therefore,} the decrease of the confidential level will easily violate the confidential level constraint in (\ref{eqconstCTP}). On the contrary, our proposed scheme can wisely adjust the caching distribution of the files to increase the CTP such that the confidential level constraint is satisfied. The corresponding adjustment in caching distribution is illustrated in Fig.~\ref{optRTPvsepsilon}(b). To be specific, in the region of $\epsilon<0.65$,  a file with higher popularity has a higher caching probability, while in the region of $\epsilon \ge 0.65$, the caching probability of a file with higher popularity \textit{gradually} decreases with $\epsilon$, which in turn leads to a reduction of the average RTP.

\begin{figure}[!t]\vspace{-6mm}
    \centering
        \subfloat[{Average} RTP versus $M_1$, $M_2=\left\lceil \frac{M_1}{2} \right\rceil$ at $\epsilon=0.6$.]{\includegraphics[width=.25\textwidth]{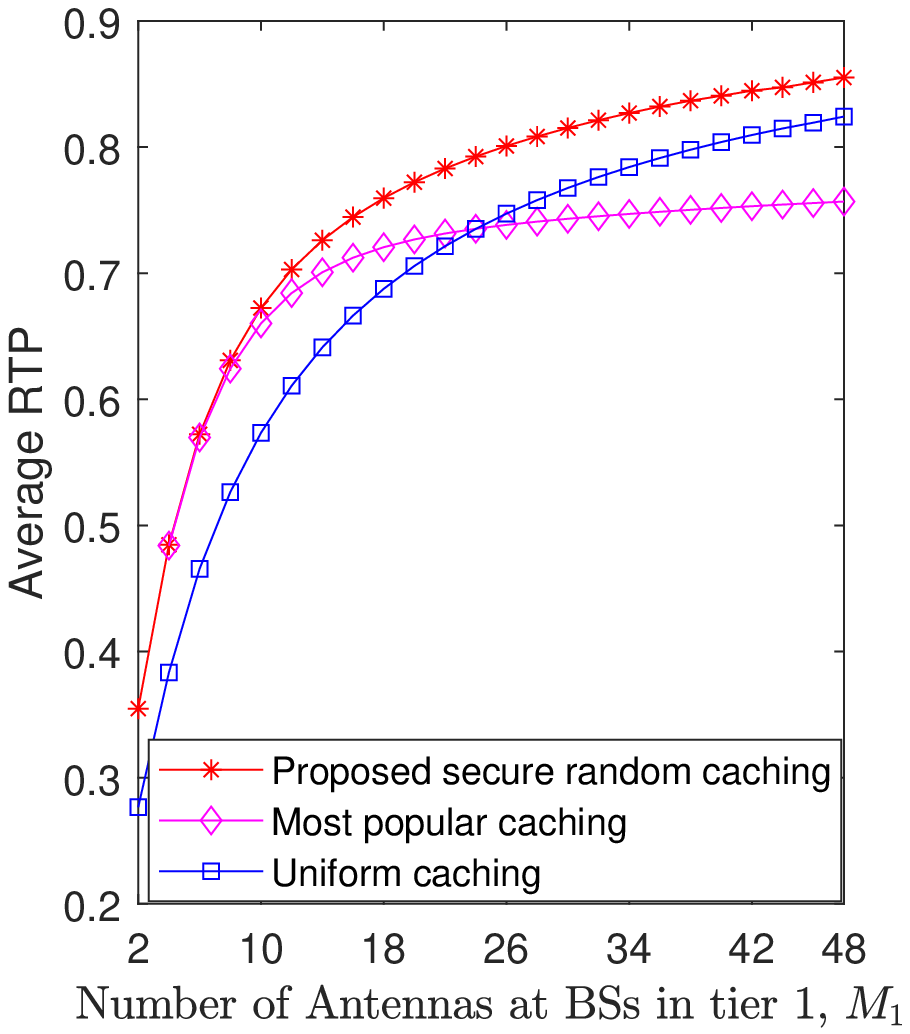}}
        \subfloat[{Average} RTP versus $\phi_1$, $\phi_2=\phi_1$.]{\includegraphics[width=.25\textwidth]{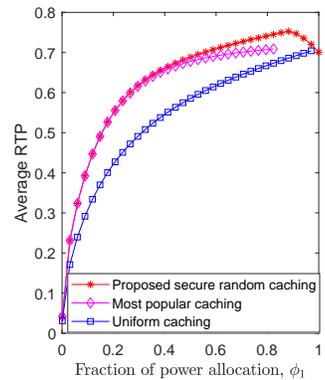}}\vspace{-3mm}\\
        \subfloat[{Average} RTP versus $R_{\mathsf{u}}$.]{\includegraphics[width=.25\textwidth]{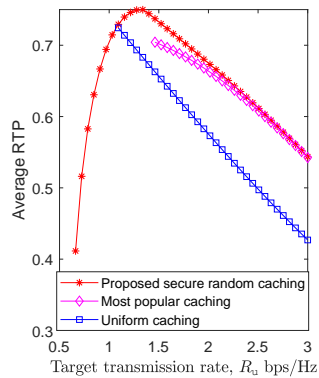}}
        \subfloat[{Average} RTP versus $R_{\mathsf{s}}$ at $R_{\mathsf{u}}=2.0$ bps/Hz.]{\includegraphics[width=.25\textwidth]{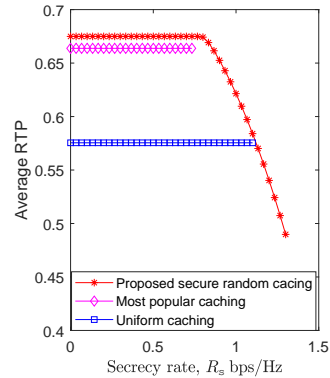}}
      \caption{(a) Effect of the number of BS antennas in tier 1. (b) Effect of the power allocation in tier 1. (c) Effect of the target transmission rate. (d) Effect of the secrecy rate.}\label{optRTPPLSpara}
\end{figure}

In Fig.~\ref{optSTPvsbslns}, we {examine the effects of the transmit power and density of the jammers, i.e., $P_{\mathtt{J}}$  and $\lambda_{\mathtt{J}}$. We first see that the average RTP of each caching scheme {decreases} as $P_{\mathtt{J}}$ or $\lambda_{\mathtt{J}}$ increases, which verifies Property~\ref{propRTP2}. We also see that, when $P_{\mathtt{J}}$ or $\lambda_{\mathtt{J}}$ is relatively high (e.g., $P_{\mathtt{J}} \ge 90$ W or  $\lambda_{\mathtt{J}} \ge \frac{180}{250^2\pi}$), the most popular caching can achieve almost the same average RTP as our secure random caching. This is because, when $P_{\mathtt{J}}$ or $\lambda_{\mathtt{J}}$ is large, the received SIR at the typical user is small. As such, our secure random caching scheme tends to cache the files with higher popularity in order to provide the largest possible average RTP. In addition, we see that, when $P_{\mathtt{J}}$ or $\lambda_{\mathtt{J}}$ is low (e.g., $P_{\mathtt{J}} \le1$ W or  $\lambda_{\mathtt{J}} \le \frac{16}{250^2\pi}$), the most popular caching may be infeasible, but our secure random caching scheme can still be feasible, demonstrating that compared to the baselines, our proposed scheme can effectively resist jamming attack.}

Next, we examine the effects of the number of BS antennas $M_k$, the power allocation $\phi_k$, the target transmission rate $R_{\mathsf{u}}$, and the secrecy rate $R_{\mathsf{s}}$ in Fig.~\ref{optRTPPLSpara}.\footnote{In Fig.~\ref{optRTPPLSpara}(a), the operation $\lceil \cdot \rceil$ denotes the ceiling function.} We see that, when $M_1$ is small, $\phi_1$ is low or $R_{\mathsf{u}}$ is large, the most popular caching scheme can achieve almost the same average RTP as our proposed secure random caching scheme, since caching the most popular files at each BS can compensate lower received SIR at the typical user. In addition, we also see that when $\phi_1$ is large, $R_{\mathsf{u}}$ is small or $R_{\mathsf{s}}$ is large, the two baseline schemes may be infeasible but our proposed schemes may still be feasible, demonstrating that our schemes can well adapt
to the changes of these parameters.

\begin{figure}[!t]
    \centering
    \includegraphics[width=.47\textwidth]{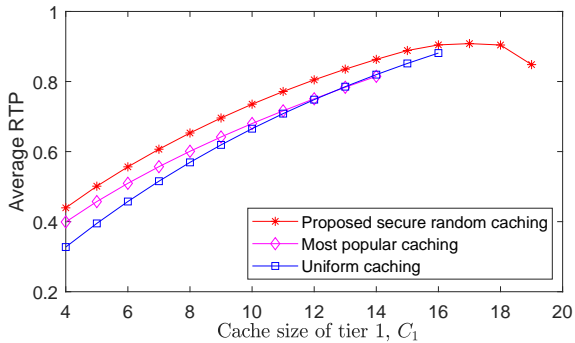}
    \caption{Effect of the cache size at $\epsilon = 0.6$. }\label{optRTPvscaching}
\end{figure}

Finally, we examine the effects of the cache size $C_k$ in Fig.~\ref{optRTPvscaching}.
{From the figure, we} can see that, when the cache size is large enough, the baselines may not be feasible, while our proposed scheme can still be feasible, but at the cost of the average RTP decrement.  This observation indicates that increasing the cache sizes at the BSs does not always increase the average RTP, which is significantly different from the observations in the existing works without considering the secrecy constraints,~e.g.,~\cite{RCHetNetTao2018,  RCCHetNetcui2017, RandomcachingCoMP2018Wen, RandomcachingDTX2018Wen}. {This can be explained as follows. If the cache size increases, for the most popular caching scheme, each BS will store more files, while for the uniform caching scheme, each file will be stored at more BSs, both of which will increase the risk of eavesdropping, i.e., the average CTP for the two baselines will decrease. If the average CTP is below the given confidential level, the baselines will be infeasible.
On the other hand, our proposed secure caching scheme can wisely adjust the caching distribution of the files (i.e., reduce the caching probability of a file with higher popularity as shown in Fig.~\ref{optRTPvsepsilon}(b)) to maintain the average CTP above the given confidential level. However, such adjustment leads to a reduction of the average RTP.} 

\section{Conclusions}

{In this paper, we examine the security issue in the cache-enabled large-scale multi-antenna HetNets, in which the multi-antenna BSs deployed at multiple independent network-tiers deliver their cached files to the requesting users in the presence of the eavesdroppers and the jammers. To confuse the eavesdroppers, the BSs transmits artificial noise as well as the useful signals, simultaneously. We first derive closed-form expressions for the average RTP and the average CTP, characterizing the impacts of the eavesdroppers and jammers on the secrecy performance of the system. We then derive closed-form upper and lower bounds for the RTP, which facilitates us to understand the effects of key system parameters. In addition, we propose a secure random caching scheme which optimizes the caching distribution of the files to maximize the average RTP of the system, while simultaneously meeting the requirements on the caching sizes at the BSs and the average CTP of the system. Numerical results show that the proposed secure random caching scheme significantly outperforms the existing baseline solutions.}

\appendices
\section{Proof of Theorem \ref{TheoremSTP}}\label{proofTheoremSTP}

Denote ${ I_n \triangleq {\sum\nolimits_{k\in\mathcal{K}} (I_{n,k}^{\mathsf{u}} + I_{-n,k}^{\mathsf{u}} ) + I_{\mathtt{J}}^{\mathsf{u}} }}$ and $\theta_{\mathsf{u}} \triangleq 2^{R_{\mathsf{u}}}-1$. Substituting (\ref{eqsirAuthorizedUser}) into (\ref{eqSTPofFiln}), we have (\ref{eqappSTP}), {as shown at the top of the next page},
\begin{figure*}[!t]{\setlength{\arraycolsep}{0.0em}\vspace{-6mm}
\begingroup\makeatletter\def\f@size{9.5}\check@mathfonts
\def\maketag@@@#1{\hbox{\m@th\normalsize\normalfont#1}}
\begin{eqnarray}\label{eqappSTP}
\textsf{p}_{{\mathsf{u}},n}(\mathbf{T}_n) &=& \mathbbm{E}_{k_{{{\mathsf{u}}},0}}\mathbbm{E}_{r_{b_{n,k_{{{\mathsf{u}}},0}},\ell_{{\mathsf{u}},0}}}\mathbbm{E}_{I_n} \left[ \Pr\left[{{{\left| {{\bf{h}}_{b_{n,k_{{{\mathsf{u}}},0}},\ell_{{\mathsf{u}},0}}^{\rm  T}{{\bf{w}}_{b_{n,k_{{{\mathsf{u}}},0}}}}} \right|}^2}} \ge \theta_{\mathsf{u}} r_{b_{n,k_{{{\mathsf{u}}},0}},\ell_{{\mathsf{u}},0}}^{\alpha} I_n \right] \right]\nonumber\\
&\mathop=\limits^{(\mathrm{a})}& \sum_{k_{{{\mathsf{u}}},0}=1}^K A_{k_{{{\mathsf{u}}},0}}\mathbbm{E}_{r_{b_{n,k_{{{\mathsf{u}}},0}},\ell_{{\mathsf{u}},0}}} \mathbbm{E}_{I_n} \left[  \sum_{m_{k_{{{\mathsf{u}}},0}}=0}^{M_{k_{{{\mathsf{u}}},0}}-1} \frac{\left(\theta_{\mathsf{u}} r_{b_{n,k_{{{\mathsf{u}}},0}},\ell_{{\mathsf{u}},0}}^{\alpha} I_n\right)^{m_{k_{{{\mathsf{u}}},0}}}}{{m_{k_{{{\mathsf{u}}},0}}}!} e^{-\theta_{\mathsf{u}} r_{b_{n,k_{{{\mathsf{u}}},0}},\ell_{{\mathsf{u}},0}}^{\alpha} I_n} \right]\nonumber\\
&\mathop=\limits^{(\mathrm{b})}&  \sum_{j=1}^K A_{j}\mathbbm{E}_{r_{b_{n,j},\ell_{{\mathsf{u}},0}}}  \left. {\left[   \sum_{m_{j}=0}^{M_{j}-1} \frac{(-s)^{{m_j}}}{{m_{j}}!} \frac{\mathrm{d}^{m_j}}{\mathrm{d}s^{m_j}}\mathcal{L}_{I_n}\left( s \right) \right]} \right|_{s= \theta_{\mathsf{u}} r_{b_{n,j},\ell_{{\mathsf{u}},0}}^{\alpha}}\nonumber\\
& \mathop=\limits^{(\mathrm{c})} &  \sum_{j=1}^K A_{j}\int_{0}^{\infty}f_{r_{b_{n,j},\ell_{{\mathsf{u}},0}}}(r_0)  \left. {\left[   \sum_{m_{j}=0}^{M_{j}-1} \frac{(-s)^{{m_j}}}{{m_{j}}!} \frac{\mathrm{d}^{m_j}}{\mathrm{d}s^{m_j}}\mathcal{L}_{I_n}\left( s \right) \right]} \right|_{s= \theta_{\mathsf{u}} r_0^{\alpha}} \mathrm{d}r_0.
\end{eqnarray}\setlength{\arraycolsep}{5pt}\endgroup}\noindent\rule[0.25\baselineskip]{\textwidth}{0.1pt}\end{figure*}where (a) is due to the fact that ${{{| {{\bf{h}}_{b_{n,k_{{{\mathsf{u}}},0}},\ell_{{\mathsf{u}},0}}^{\rm  T}{{\bf{w}}_{b_{n,k_{{{\mathsf{u}}},0}}}}} |}^2}}\mathop \sim\limits^d \Gamma(M_{k_{{{\mathsf{u}}},0}}, 1)$ and $A_{k_{{{\mathsf{u}}},0}}$ is the probability that the typical user $\ell_{{\mathsf{u}},0}$ is associated with tier $k_{{{\mathsf{u}}},0}$; (b) follows from $\mathbbm{E}_{I_n}\left[I_n^me^{-sI_n}\right]=(-1)^m\frac{\mathrm{d}^m}{\mathrm{d}s^{m}}\mathcal{L}_{I_n}(s)$ with $\frac{\mathrm{d}^m}{\mathrm{d}s^{m}}\mathcal{L}_{I_n}(s)$ denoting the $m$-order derivative of the Laplace transform of random variable $I_n$, i.e.,  $\mathcal{L}_{I_n}(s)\triangleq\mathbbm{E}_{I_n}\left[e^{-sI_n}\right]$; and (c) uses the probability density function (p.d.f.) of the distance ${r_{b_{n,j},\ell_{{\mathsf{u}},0}}}$, i.e., $f_{r_{b_{n,j},\ell_{{\mathsf{u}},0}}}(r_0)$, which is given by
{\begingroup\makeatletter\def\f@size{9.5}\check@mathfonts
\def\maketag@@@#1{\hbox{\m@th\normalsize\normalfont#1}}\setlength{\arraycolsep}{0.0em}
\begin{eqnarray}\label{eqPDF}
&&f_{r_{b_{n,{j}},{\ell}_{\mathsf{u},0}}}(r_0) \nonumber \\
&&= \frac{2\pi\lambda_{j}T_{n,j}}{A_{j}} r_0 \exp\left(-r_0^2\sum_{k=1}^{K}\pi\lambda_kT_{n,k}\left(\frac{{\phi}_k{P}_k}{{\phi}_j{P}_j}\right)^\delta\right).
\end{eqnarray}\setlength{\arraycolsep}{5pt}\endgroup}Thus, to calculate $\textsf{p}_{{\mathsf{u}},n}(\mathbf{T}_n)$, we only need to calculate $\frac{\mathrm{d}^m}{\mathrm{d}s^{m}}\mathcal{L}_{I_n}(s)$.

First, we calculate $\mathcal{L}_{I_n}(s)$. Note that we have ${{\cal L}_{I_n}}(s) = \prod_{k\in\mathcal{K}} {{{\cal L}_{{I_{n,k}^{\mathsf{u}}}}}(s)}  {{{\cal L}_{{I_{ - n,k}^{\mathsf{u}}}}}(s)}{{\cal L}_{{I_{\mathtt{J}}^{\mathsf{u}}}}}(s)$. In the following, we calculate ${{{\cal L}_{{{I}_{n,k}^{\mathsf{u}}}}}(s)}$, ${{{\cal L}_{{I_{ - n,k}^{\mathsf{u}}}}}(s)}$ and ${{\cal L}_{{I_{\mathtt{J}}^{\mathsf{u}}}}}(s)$, respectively. Let ${X_b} \triangleq   \frac{\phi_kP_k}{\phi_jP_j} ( {{| {{\bf{h}}_{b,{\ell _{{\mathsf{u}},0}}}^{\rm{T}}{{\bf{w}}_b}} |}^2}   + {\xi _k}{{\| {{\bf{h}}_{b,{\ell _{{\mathsf{u}},0}}}^{\rm{T}}{{\bf{W}}_b}} \|}^2} )$ with the p.d.f. being given by \cite[Lemma 1]{TIFS2013ZhangPLSAdHoc}. We calculate ${{{\cal L}_{{{I}_{n,k}^{\mathsf{u}}}}}(s)}$ as (\ref{eqLInku}), {as shown at the top of the next page},
\begin{figure*}[!t]{\setlength{\arraycolsep}{0.0em}\vspace{-6mm}
\begingroup\makeatletter\def\f@size{9.5}\check@mathfonts
\def\maketag@@@#1{\hbox{\m@th\normalsize\normalfont#1}}
\begin{eqnarray}
{\mathcal{L}_{I_{n,k}^{\mathsf{u}}}}(s) = {\mathbb{E}_{I_{n,k}^{\mathsf{u}}}}\left[ {{e^{ - sI_{n,k}^{\mathsf{u}}}}} \right]
%& = & \mathbb{E}_{{\Phi _{n,k}},\{X_b\}}\left[ {\exp \left( { - s\sum\limits_{b \in {\Phi _{n,k}} { \setminus }\{ {b_{n,{k_{{\mathsf{u}},0}}}}\} } X_b r_{b,{\ell _{{\mathsf{u}},0}}}^{ - \alpha }} \right)} \right]\nonumber\\
%&=& \mathbb{E}{_{{\Phi _{n,k}}}}\left[ {\prod\limits_{b \in {\Phi _{n,k}}{ \setminus }\{ {b_{n,{k_{\mathsf{u},0}}}}\} } {{\mathbb{E}_{{X_b}}}\left[ {\exp \left( { - s{X_b}r_{b,{\ell _{\mathsf{u},0}}}^{ - \alpha }} \right)} \right]} } \right]\nonumber\\
%&= & \mathbb{E}_{{\Phi _{n,k}}}\left[ {\prod\limits_{b \in {\Phi _{n,k}}{ \setminus }\{ {b_{n,{k_{\mathsf{u},0}}}}\} } {{{{\cal L}}_{{X_b}}}(sr_{b,{\ell _{\mathsf{u},0}}}^{ - \alpha })} } \right]\nonumber\\
&\mathop=\limits^{(\mathrm{d})}& \exp \left( { - 2\pi {\lambda _{n,k}}{T_{n,k}}\int_{{{\left(\frac{\phi_kP_k}{\phi_jP_j} \right)}^{\frac{\delta}{2 }}}{r_0}}^\infty  {\left( {1 - {{{\cal L}}_{{X_b}}}(s{v^{ - \alpha }})} \right)} v\mathrm{d}v} \right)\nonumber\\
&\mathop=\limits^{(\mathrm{e})}& \begin{cases}
      \exp \left( { - \pi {\lambda _k}{T_{n,k}}{{\left(\frac{\phi_kP_k}{\phi_jP_j} \right)}^\delta }r_0^2\left( {{}_2{F_1}\left( { - \delta ,{M_k};1 - \delta ; - sr_0^{ - \alpha }} \right) - 1} \right)} \right), & \mbox{if } \xi_k=1, \\
      \exp \left( { - \pi {\lambda _k}{T_{n,k}}\left( {\left(\frac{\phi_kP_k}{\phi_jP_j} \right)^\delta }A(s) - r_0^2B(s) \right)} \right), & \mbox{otherwise}.
    \end{cases}\label{eqLInku}
\end{eqnarray}\setlength{\arraycolsep}{5pt}\endgroup}\noindent\rule[0.25\baselineskip]{\textwidth}{0.1pt}\end{figure*}where (d) follows from the probability generating functional (PGFL) over PPP and ${{{\cal L}}_{{X_b}}}(s)$ is  the Laplace transform of $X_b$; (e) follows from \cite[eq. (49)]{2017TIFSWangPLS} and $A(s)$ and $B(s)$ are, respectively, given by (\ref{eqappAs}) and (\ref{eqappBs}), {as shown at the top of the next page}, with
\begin{figure*}[!t]{\setlength{\arraycolsep}{0.0em}\vspace{-6mm}
\begingroup\makeatletter\def\f@size{9.5}\check@mathfonts
\def\maketag@@@#1{\hbox{\m@th\normalsize\normalfont#1}}
\begin{eqnarray}
A(s)&\triangleq& {s^\delta }\left( {\frac{{\Gamma(1+\delta)\Gamma(1-\delta)}}{{{{\left( {1 - {\xi _k}} \right)}^{{M_k} - 1}}}} - \sum\limits_{i = 0}^{{M_k} - 2} {\frac{{\xi _k^{1 + \delta }{\Gamma(i+\delta+1)\Gamma(1-\delta)}}}{{{{\left( {1 - {\xi _k}} \right)}^{{M_k} - i - 1}}}\Gamma(i+1)}} } \right) ,\label{eqappAs}\\
B(s)&\triangleq&  1 - \frac{{\delta \varphi (1,{\frac{\phi_kP_k}{\phi_jP_j} }sr_0^{ - \alpha })}}{{\left( {1 + \delta } \right){{\left( {1 - {\xi _k}} \right)}^{{M_k} - 1}}{\xi _k}{\frac{\phi_kP_k}{\phi_jP_j} }sr_0^{ - \alpha }}} + \sum_{i = 0}^{{M_k} - 2} {\frac{{\delta {\xi _k}\varphi (i + 1,{\xi _k}{\frac{\phi_kP_k}{\phi_jP_j} }sr_0^{ - \alpha })}}{{\left( {i + \delta + 1 } \right){{\left( {1 - {\xi _k}} \right)}^{{M_k} - i - 1}}{{\left( {{\xi _k}{\frac{\phi_kP_k}{\phi_jP_j} }sr_0^{ - \alpha }} \right)}^{i + 1}}}}} .\label{eqappBs}
\end{eqnarray}\setlength{\arraycolsep}{5pt}\endgroup}\noindent\rule[0.25\baselineskip]{\textwidth}{0.1pt}\end{figure*} $\varphi(i,x)\triangleq{}_2{F_1}\left( {i,i + \delta;i + \delta + 1; - {x^{ - 1}}} \right)$. Similarly, ${{{\cal L}_{{I_{ - n,k}^{\mathsf{u}}}}}(s)}$ and ${{\cal L}_{{I_{\mathtt{J}}^{\mathsf{u}}}}}(s)$ are calculated as (\ref{eqLInnku}) and (\ref{eqLIJ}), respectively, {as shown at the top of the next page}.
\begin{figure*}[!t]{\setlength{\arraycolsep}{0.0em}\vspace{-6mm}
\begingroup\makeatletter\def\f@size{9.5}\check@mathfonts
\def\maketag@@@#1{\hbox{\m@th\normalsize\normalfont#1}}
\begin{eqnarray}
{\mathcal{L}_{I_{-n,k}^{\mathsf{u}}}}(s)
&=& \begin{cases}
      \exp \left( { - \pi {\lambda _k}\left( {1 - {T_{n,k}}} \right){{\left( {\frac{\phi_kP_k}{\phi_jP_j} } \right)}^\delta }s^{\delta}\frac{\Gamma(M_k+\delta)}{\Gamma(M_k)}\Gamma(1-\delta)} \right), & \mbox{if } \xi_k=1, \\
      \exp \left( { - \pi {\lambda _k}\left( {1 - {T_{n,k}}} \right){{\left( {\frac{\phi_kP_k}{\phi_jP_j} } \right)}^\delta A(s) }} \right), & \mbox{otherwise},
    \end{cases}\label{eqLInnku}\\
{{\cal L}_{{I_{\mathtt{J}}}}}(s) &=& \exp \left( { - \pi {\lambda _{\mathtt{J}}}{{\left( {\frac{{{P_{\mathtt{J}}}}}{{{\phi _j}{P_j}}}} \right)}^\delta }{s^\delta }\Gamma(1+\delta)\Gamma(1-\delta)} \right).\label{eqLIJ}
\end{eqnarray}\setlength{\arraycolsep}{5pt}\endgroup}\noindent\rule[0.25\baselineskip]{\textwidth}{0.1pt}\end{figure*}Based on (\ref{eqLInku}), (\ref{eqLInnku}) and (\ref{eqLIJ}), we have
{\setlength{\arraycolsep}{0.0em}
\begin{eqnarray}\label{eqappLIns}
\mathcal{L}_{I_n}(s)  = \exp \left( {\eta (s)} \right),
\end{eqnarray}\setlength{\arraycolsep}{5pt}}where $\eta (s)$ is given by (\ref{eqetas}), {as shown at the top of the next page},
\begin{figure*}[!t]{\setlength{\arraycolsep}{0.0em}\vspace{-6mm}
\begingroup\makeatletter\def\f@size{9.5}\check@mathfonts
\def\maketag@@@#1{\hbox{\m@th\normalsize\normalfont#1}}
\begin{eqnarray}\label{eqetas}
\eta (s) = \begin{cases}
              - \pi {\lambda _{j}}{T_{n,j}}\left( {\sum\limits_{k = 1}^K {\frac{{{\lambda _k}{T_{n,k}}}}{{{\lambda _{j}}{T_{n,j}}}}{{\left( {\frac{\phi_kP_k}{\phi_jP_j} } \right)}^\delta }{g_k}(s)}  + \frac{{{\lambda _{\mathtt{J}}}}}{{{\lambda _{j}}{T_{n,j}}}}{{\left( {\frac{{{P_{\mathtt{J}}}}}{{{\phi _{j}}{P_{j}}}}} \right)}^\delta }{s^\delta }\Gamma(1+\delta)\Gamma(1-\delta)} \right), & \mbox{if } \xi_k=1, \\
             - \pi {\lambda _{j}}{T_{n,j}}\left( {\sum\limits_{k = 1}^K {\frac{{{\lambda _k}{T_{n,k}}}}{{{\lambda _{j}}{T_{n,j}}}}{f_k}(s) + \frac{{{\lambda _{\mathtt{J}}}}}{{{\lambda _{j}}{T_{n,j}}}}{{\left( {\frac{{{P_{\mathtt{J}}}}}{{{\phi _{j}}{P_{j}}}}} \right)}^\delta }{s^\delta }\Gamma(1+\delta)\Gamma(1-\delta)} } \right), & \mbox{otherwise}.
           \end{cases}
\end{eqnarray}\setlength{\arraycolsep}{5pt}\endgroup}
\noindent\rule[0.25\baselineskip]{\textwidth}{0.1pt}\end{figure*}with ${f_k}(s) \triangleq \frac{1}{{{T_{n,k}}}}{\left( {\frac{\phi_kP_k}{\phi_jP_j} } \right)^\delta }A(s) - r_0^2B(s)$ and ${g_k}(s) \triangleq r_0^2\left( {{}_2{F_1}\left( { - \delta ,{M_k};1 - \delta ; - sr_0^{ - \alpha }} \right) - 1} \right) + \left( {\frac{1}{{{T_{n,k}}}} - 1} \right) {s^\delta }\frac{\Gamma(M_k+\delta)}{\Gamma(M_k)}\Gamma(1-\delta)$. %with{\setlength{\arraycolsep}{0.0em}
%\begin{eqnarray*}
%{g_k}(s) &\triangleq& r_0^2\left( {{}_2{F_1}\left( { - \delta ,{M_k};1 - \delta ; - sr_0^{ - \alpha }} \right) - 1} \right) \nonumber\\
%&&\; + \left( {\frac{1}{{{T_{n,k}}}} - 1} \right){s^\delta }\frac{\Gamma(M_k+\delta)}{\Gamma(M_k)}\Gamma(1-\delta),\nonumber\\
%{f_k}(s) &\triangleq& \frac{1}{{{T_{n,k}}}}{\left( {\frac{\phi_kP_k}{\phi_jP_j} } \right)^\delta }A(s) - r_0^2B(s).
%\end{eqnarray*}\setlength{\arraycolsep}{5pt}}% are defined as
%{\setlength{\arraycolsep}{0.0em}
%\begin{eqnarray*}
%{g_k}(s) &\triangleq& r_0^2\left( {{}_2{F_1}\left( { - \delta ,{M_k};1 - \delta ; - sr_0^{ - \alpha }} \right) - 1} \right) + \left( {\frac{1}{{{T_{n,k}}}} - 1} \right){s^\delta }\frac{\Gamma(M_k+\delta)}{\Gamma(M_k)}\Gamma(1-\delta),\nonumber\\
%{f_k}(s) &\triangleq& \frac{1}{{{T_{n,k}}}}{\left( {\frac{\phi_kP_k}{\phi_jP_j} } \right)^\delta }A(s) - r_0^2B(s).
%\end{eqnarray*}\setlength{\arraycolsep}{5pt}}

Next, we calculate $\frac{\mathrm{d}^{m_j}}{\mathrm{d}s^{m_j}}\mathcal{L}_{I_n}(s)$. Note that, directly computing the derivatives will lead to intractable expressions. To {address this issue}, according to \cite[Lemma 1]{MultiAntFrwkChangli2018}, we obtain the following recursive relations
{\begingroup\makeatletter\def\f@size{9.5}\check@mathfonts
\def\maketag@@@#1{\hbox{\m@th\normalsize\normalfont#1}}\setlength{\arraycolsep}{0.0em}
\begin{eqnarray}\label{appeqrecu}
&&\frac{(-s)^{m}}{{m}!}\frac{\mathrm{d}^{m}}{\mathrm{d}s^{m}}\mathcal{L}_{I_n}(s)\\
&&=
\begin{cases}\displaystyle
\mathcal{L}_{I_n}(s), & \mbox{if } {m}=0, \\
\sum\limits_{i=0}^{{m}-1}\frac{{m}-i}{{m}}z_{{m}-i}(s) \frac{(-s)^i}{i!}\frac{\mathrm{d}^i}{\mathrm{d}s^{i}}\mathcal{L}_{I_n}(s), & \mbox{if }{m}=1,2,\cdots,\nonumber
\end{cases}
\end{eqnarray}\setlength{\arraycolsep}{5pt}\endgroup}where
{\setlength{\arraycolsep}{0.0em}
\begin{eqnarray}\label{appeqtms}
z_m(s) = \frac{(-s)^m}{m!}\frac{\mathrm{d}^m}{\mathrm{d}s^{m}}\eta(s).
\end{eqnarray}\setlength{\arraycolsep}{5pt}}From (\ref{appeqrecu}), we can see that in order to calculate $\frac{\mathrm{d}^m}{\mathrm{d}s^{m}}\mathcal{L}_{I_n}(s)$, we only need to calculate $z_m(s)$, which are related to the derivatives of $\eta(s)$. As shown in \cite{MultiAntFrwkChangli2018}, obtaining a closed-form solution for $\frac{\mathrm{d}^m}{\mathrm{d}s^{m}}\eta(s)$ is generally much easier than for $\frac{\mathrm{d}^m}{\mathrm{d}s^{m}}\mathcal{L}_{I_n}(s)$. Specifically, we have (\ref{eqdm}), {as shown at the top of this page,}
\begin{figure*}[!t]{\setlength{\arraycolsep}{0.0em}
\begingroup\makeatletter\def\f@size{9.5}\check@mathfonts\vspace{-6mm}
\def\maketag@@@#1{\hbox{\m@th\normalsize\normalfont#1}}
\begin{eqnarray}\label{eqdm}
\frac{{{{\rm{d}}^{{m}}}}}{{{\rm{d}}{s^{{m}}}}}\eta (s) =
\begin{cases}
{ - \pi {\lambda _j}{T_{n,j}}\left( {\sum\limits_{k = 1}^K {\frac{{{\lambda _k}{T_{n,k}}}}{{{\lambda _j}{T_{n,j}}}}{{\left( {\frac{\phi_kP_k}{\phi_jP_j} } \right)}^\delta }\frac{{{{\rm{d}}^{{m}}}}}{{{\rm{d}}{s^{{m}}}}}{g_k}(s)}  + \frac{{{\lambda _{\mathtt{J}}}}}{{{\lambda _j}{T_{n,j}}}}{{\left( {\frac{{{P_{\mathtt{J}}}}}{{{\phi _j}{P_j}}}} \right)}^\delta }\Gamma \left( {1 + \delta } \right)\Gamma \left( {1 - \delta } \right)\frac{{{{\rm{d}}^{{m}}}}}{{{\rm{d}}{s^{{m}}}}}{s^\delta }} \right)}, & \mbox{if } \xi_k=1, \\
{ - \pi {\lambda _j}{T_{n,j}}\left( {\sum\limits_{k = 1}^K {\frac{{{\lambda _k}{T_{n,k}}}}{{{\lambda _j}{T_{n,j}}}}\frac{{{{\rm{d}}^{{m}}}}}{{{\rm{d}}{s^{{m}}}}}{f_k}(s) + \frac{{{\lambda _{\mathtt{J}}}}}{{{\lambda _j}{T_{n,j}}}}{{\left( {\frac{{{P_{\mathtt{J}}}}}{{{\phi _j}{P_j}}}} \right)}^\delta }\Gamma \left( {1 + \delta } \right)\Gamma \left( {1 - \delta } \right)\frac{{{{\rm{d}}^{{m}}}}}{{{\rm{d}}{s^{{m}}}}}{s^\delta }} } \right)}, & \mbox{otherwise}.
\end{cases}
\end{eqnarray}\setlength{\arraycolsep}{5pt}\endgroup}\noindent\rule[0.25\baselineskip]{\textwidth}{0.1pt}\vspace{-6mm}\end{figure*}where $\frac{{{{\rm{d}}^{{m}}}}}{{{\rm{d}}{s^{{m}}}}}s^{\delta} = {{\delta _{\left( {{m_j}} \right)}}{s^{\delta  - {m_j}}}}$, $\frac{{{{\rm{d}}^{{m}}}}}{{{\rm{d}}{s^{{m}}}}}{g_k}(s)$ and $\frac{{{{\rm{d}}^{{m}}}}}{{{\rm{d}}{s^{{m}}}}}{f_k}(s)$ are given by (\ref{eqgder}) and (\ref{eqfder}), respectively, {as shown at the top of the next page}.
\begin{figure*}[!t]{\setlength{\arraycolsep}{0.0em}\vspace{-6mm}
\begingroup\makeatletter\def\f@size{9.5}\check@mathfonts
\def\maketag@@@#1{\hbox{\m@th\normalsize\normalfont#1}}
\begin{eqnarray}
\frac{{{{\rm{d}}^{{m}}}}}{{{\rm{d}}{s^{{m}}}}}{g_k}(s)&=&r_0^{2 - \alpha {m}}\frac{{{{\delta}^{({m})}}{M_k^{({m})}}}}{{{{(1 - \delta )}^{({m})}}}}{}_2{F_1}\left( { - \delta  + {m},{M_k} + {m};1 - \delta  + {m}; - sr_0^{ - \alpha }} \right) + \left( {\frac{1}{{{T_{n,k}}}} - 1} \right)\frac{{\Gamma \left( {{M_k} + \delta } \right)\Gamma \left( {1 - \delta } \right)}}{{\Gamma \left( {{M_k}} \right){s^{m-\delta}}}}{\delta_{\left( {{m}} \right)}}, \label{eqgder}\\
\frac{{{{\rm{d}}^{{m}}}}}{{{\rm{d}}{s^{{m}}}}}{f_k}(s) &=& \frac{1}{{{T_{n,k}}}}{\left( {\frac{\phi_kP_k}{\phi_jP_j} } \right)^\delta }\left( {\frac{{\Gamma \left( {1 + \delta } \right)\Gamma \left( {1 - \delta } \right)}}{{{{\left( {1 - {\xi _k}} \right)}^{{M_k} - 1}}}} - \sum\limits_{i = 0}^{{M_k} - 2} {\frac{{\xi _k^{1 + \delta }\Gamma \left( {i + 1 + \delta } \right)\Gamma \left( {1 + \delta } \right)}}{{{{\left( {1 - {\xi _k}} \right)}^{{M_k} - i - 1}}\Gamma \left( {i + 1} \right)}}} } \right){\delta _{\left( {{m}} \right)}}{s^{\delta  - {m}}} \nonumber \\
 &&\:+ r_0^2\left( {\frac{\delta }{{\left( {1 + \delta } \right){{\left( {1 - {\xi _k}} \right)}^{{M_k} - 1}}{\xi _k}{\frac{\phi_kP_k}{\phi_jP_j} }r_0^{ - \alpha }}}\frac{{{{\rm{d}}^{{m}}}}}{{{\rm{d}}{s^{{m}}}}}\frac{1}{s}\varphi \left(1,{\frac{\phi_kP_k}{\phi_jP_j} } sr_0^{ - \alpha }\right)} \right) \nonumber \\
 &&\:- r_0^2\left( {\frac{{\delta {\xi _k}}\sum_{i = 0}^{{M_k} - 2} {\frac{{{{\rm{d}}^{{m}}}}}{{{\rm{d}}{s^{{m}}}}}\frac{1}{{{s^{i + 1}}}}\varphi (i + 1,{\xi _k} {\frac{\phi_kP_k}{\phi_jP_j} }sr_0^{ - \alpha })}}{{\left( {i + 1 + \delta } \right){{\left( {1 - {\xi _k}} \right)}^{{M_k} - i - 1}}{{\left( {{\xi _k} {\frac{\phi_kP_k}{\phi_jP_j} }r_0^{ - \alpha }} \right)}^{i + 1}}}} } \right).\label{eqfder}
\end{eqnarray}\setlength{\arraycolsep}{5pt}\endgroup}\noindent\rule[0.25\baselineskip]{\textwidth}{0.1pt}\end{figure*}Here, $\frac{{{\mathrm{d}^{{m}}}}}{{\mathrm{d}{s^{{m}}}}}\frac{1}{s}\varphi (1,{\frac{\phi_kP_k}{\phi_jP_j} }sr_0^{ - \alpha })$ and $\frac{{{\mathrm{d}^{{m}}}}}{{\mathrm{d}{s^{{m}}}}}\frac{1}{{{s^{i + 1}}}}\varphi (i + 1,{\xi _k} {\frac{\phi_kP_k}{\phi_jP_j} }sr_0^{ - \alpha })$ can be calculated as (\ref{eqd1s1}) and (\ref{eqd1s2}), respectively, {as shown at the top of the next page}.
\begin{figure*}[!t]{\setlength{\arraycolsep}{0.0em}\vspace{-6mm}
\begingroup\makeatletter\def\f@size{9.5}\check@mathfonts
\def\maketag@@@#1{\hbox{\m@th\normalsize\normalfont#1}}
\begin{eqnarray}
\frac{{{\mathrm{d}^{{m}}}}}{{\mathrm{d}{s^{{m}}}}}\frac{1}{s}\varphi \left(1,{\frac{\phi_kP_k}{\phi_jP_j} }sr_0^{ - \alpha }\right) &=& \Gamma \left( {2 + \delta } \right)\Gamma \left( { - \delta } \right) {\frac{\phi_kP_k}{\phi_jP_j} }r_0^{ - \alpha  - 2}{\delta_{({m})}}{s^{\delta  - {m}}}+ \frac{{\Gamma \left( {2 + \delta } \right)\Gamma \left( \delta  \right)}}{{\Gamma \left( {1 + \delta } \right)\Gamma \left( {1 + \delta } \right)}}\frac{{{m}!{\delta ^{({m})}}}}{{{{\left( {1 - \delta } \right)}^{({m})}}}}{\left( {{\frac{\phi_kP_k}{\phi_jP_j} } r_0^{ - \alpha }} \right)^{{m} + 1}}\nonumber \\
 &&\:\times{}_2{F_1}\left( {1 + {m},{m} - \delta ;{m} + 1 - \delta ; -  {\frac{\phi_kP_k}{\phi_jP_j} }sr_0^{ - \alpha }} \right),\label{eqd1s1} \\
\frac{{{\mathrm{d}^{{m}}}}}{{\mathrm{d}{s^{{m}}}}}\frac{1}{{{s^{i + 1}}}}\varphi \left(i + 1,{\xi _k} {\frac{\phi_kP_k}{\phi_jP_j} }sr_0^{ - \alpha }\right) &=& \frac{{\Gamma \left( {i + 2 + \delta } \right)\Gamma \left( { - \delta } \right)}}{{\Gamma \left( {i + 1} \right)}}{\left( {{\xi _k} {\frac{\phi_kP_k}{\phi_jP_j} }r_0^{ - \alpha }} \right)^{\delta  + i + 1}}{\delta _{({m})}}{s^{\delta  - {m}}} \nonumber \\
 &&\:+ \frac{{\Gamma \left( {i + 2 + \delta } \right)\Gamma \left( \delta  \right)}}{{\Gamma \left( {i + 1 + \delta } \right)\Gamma \left( {1 + \delta } \right)}}{\left( {{\xi _k} {\frac{\phi_kP_k}{\phi_jP_j} }r_0^{ - \alpha }} \right)^{{m} + i + 1}}\frac{{{{(i + 1)}^{({m})}}{\delta ^{({m})}}}}{{{{\left( {1 - \delta } \right)}^{({m})}}}}\nonumber\\
 &&\:\times{}_2{F_1}\left( {i + 1 + k,k - \delta ;k + 1 - \delta ; - {\xi _k} {\frac{\phi_kP_k}{\phi_jP_j} }sr_0^{ - \alpha }} \right).\label{eqd1s2}
\end{eqnarray}\setlength{\arraycolsep}{5pt}\endgroup}\noindent\rule[0.25\baselineskip]{\textwidth}{0.1pt}\vspace{-6mm}\end{figure*}Based on \cite[Theorem 2]{MultiAntFrwkChangli2018}, we have
{\begingroup\makeatletter\def\f@size{9.5}\check@mathfonts
\def\maketag@@@#1{\hbox{\m@th\normalsize\normalfont#1}}\setlength{\arraycolsep}{0.0em}
\begin{eqnarray}\label{eqappeTMj}
\left. {\left[   \sum_{m_{j}=0}^{M_{j}-1} \frac{(-s)^{{m_j}}}{{m_{j}}!} \frac{\mathrm{d}^{m_j}}{\mathrm{d}s^{m_j}}\mathcal{L}_{I_n}\left( s \right) \right]} \right|_{s= \theta_{\mathsf{u}} r_0^{\alpha}} = \left\|e^{\mathbf{Z}_{M_j}}\right\|_1,
\end{eqnarray}\setlength{\arraycolsep}{5pt}\endgroup}where $\mathbf{Z}_M$ is the following $M\times M$ lower triangular Toeplitz matrix, i.e.,
{\begingroup\makeatletter\def\f@size{9.5}\check@mathfonts
\def\maketag@@@#1{\hbox{\m@th\normalsize\normalfont#1}}\setlength{\arraycolsep}{0.0em}
\begin{eqnarray}
\mathbf{Z}_{M} = {\begin{pmatrix}
 z_{0}(\theta_{\mathsf{u}} r_0^{\alpha})&  0&  0&  0&    \\
 z_{1}(\theta_{\mathsf{u}} r_0^{\alpha})&  z_{0}(\theta_{\mathsf{u}} r_0^{\alpha})&  0&  0&    \\
 \vdots&  \vdots& \ddots & \vdots  &   \\
 z_{M-1}(\theta_{\mathsf{u}} r_0^{\alpha})&  \cdots&  z_{1}(\theta_{\mathsf{u}} r_0^{\alpha})\;\;& z_{0}(\theta_{\mathsf{u}} r_0^{\alpha}) &
\end{pmatrix}} ,
\end{eqnarray}\setlength{\arraycolsep}{5pt}\endgroup}and its non-zero entries are determined by (\ref{appeqtms}) with ${s= \theta_{\mathsf{u}} r_0^{\alpha}}$,~i.e.,
{\begingroup\makeatletter\def\f@size{9.5}\check@mathfonts
\def\maketag@@@#1{\hbox{\m@th\normalsize\normalfont#1}}\setlength{\arraycolsep}{0.0em}
\begin{eqnarray*}
z_m(\theta_{\mathsf{u}} r_0^{\alpha}) &=& \frac{{{{\left( { - {\theta_{\mathsf{u}}}r_0^\alpha } \right)}^{{m}}}}}{{{m}!}} {\left. {\frac{{{{\rm{d}}^{{m}}}}}{{{\rm{d}}{s^{{m}}}}}\eta (s)} \right|_{s = {\theta_{\mathsf{u}}}r_0^\alpha }} \nonumber \\
& = &  - \pi X_j r_0^2\left( {{c_{{m}}} - {\mathbbm{1}}[{m} = 0]} \right).
\end{eqnarray*}\setlength{\arraycolsep}{5pt}\endgroup}Here, $X_j\triangleq {\sum_{k = 1}^K {{\lambda _k}{T_{n,k}}{{\left( {\frac{{{\phi _k}{P_k}}}{{{\phi _j}{P_j}}}} \right)}^\delta }} }$ and ${c_m} \triangleq \frac{f_m(\mathbf{T}_n,\theta_{\mathsf{u}})}{{{{\left( {{\phi _j}{P_j}} \right)}^\delta }{X_j}}}$, where $f_m(\mathbf{T}_n,\theta_{\mathsf{u}})$ is given by Theorem~\ref{TheoremSTP}.

Finally, substituting (\ref{eqappeTMj}) into (\ref{eqappSTP}), we have
{\begingroup\makeatletter\def\f@size{9.5}\check@mathfonts
\def\maketag@@@#1{\hbox{\m@th\normalsize\normalfont#1}}\setlength{\arraycolsep}{0.0em}
\begin{eqnarray*}
\textsf{p}_{{\mathsf{u}},n}(\mathbf{T}_n)
 &=& \sum\limits_{j = 1}^K {\frac{{{\lambda _j}{T_{n,j}}}}{{{X_j}}}\int_0^\infty  {2\pi {X_j}{r_0}\exp \left( { - \pi {X_j}r_0^2} \right){{\left\| {{e^{{{\bf{Z}}_{{M_j}}}}}} \right\|}_1}} {\rm{d}}{r_0}}\nonumber\\
 &\mathop=\limits^{(\mathrm{f})}&  \sum\limits_{j = 1}^K {{\lambda _j}{T_{n,j}}{{\left( {{\phi _j}{P_j}} \right)}^\delta }{{\left\| {{\bf{Q}}_{{M_j}}^{ - 1}} \right\|}_1}} .
\end{eqnarray*}\setlength{\arraycolsep}{5pt}\endgroup}where (f) follows from \cite[Proposition 1]{MultiAntFrwkChangli2018} and uses the definition of ${{\bf{Q}}_{{M}}}$ in (\ref{eqdefQMj}). The proof of Theorem~\ref{TheoremSTP} is completed.~$\hfill\blacksquare$

\section{Proof of Proposition \ref{lemmaBounds}}\label{prooflemmaBounds}

The lower bound on $\textsf{p}_{{\mathsf{u}},n}(\mathbf{T}_n)$ can be directly obtained by using the Lemma 1 in \cite{ThoutEESCNMultiAntChang2014}, and thus we omit the details due to page limitation. In the following, we focus on the calculation of the upper bound on $\textsf{p}_{{\mathsf{u}},n}(\mathbf{T}_n)$. Specifically, based on the derivation in (\ref{eqappSTP}), we have (\ref{eqappSTPUB}), {as shown at the top of the next page},
\begin{figure*}[!t]{\setlength{\arraycolsep}{0.0em} \vspace{-15mm}
\begingroup\makeatletter\def\f@size{9.5}\check@mathfonts
\def\maketag@@@#1{\hbox{\m@th\normalsize\normalfont#1}}
\begin{eqnarray}\label{eqappSTPUB}
\textsf{p}_{{\mathsf{u}},n}(\mathbf{T}_n) &=& 1- \mathbbm{E}_{k_{{{\mathsf{u}}},0}}\mathbbm{E}_{r_{b_{n,k_{{{\mathsf{u}}},0}},\ell_{{\mathsf{u}},0}}}\mathbbm{E}_{I_n} \left[ \Pr\left[{{{\left| {{\bf{h}}_{b_{n,k_{{{\mathsf{u}}},0}},\ell_{{\mathsf{u}},0}}^{\rm  T}{{\bf{w}}_{b_{n,k_{{{\mathsf{u}}},0}}}}} \right|}^2}} \le \theta_{\mathsf{u}} r_{b_{n,k_{{{\mathsf{u}}},0}},\ell_{{\mathsf{u}},0}}^{\alpha} I_n \right] \right]\nonumber\\
&\mathop  = \limits^{({\rm{a}})}& 1 - \sum\limits_{{k_{\mathsf{u}, 0}} = 1}^K {{A_{{k_{\mathsf{u}, 0}}}}} {{\mathbbm{E}}_{{r_{{b_{n,{k_{\mathsf{u}, 0}}}},{\ell _{\mathsf{u}, 0}}}}}}{\mathbbm{E}_{{I_n}}}\left[ {\frac{{\gamma ({M_{{k_{\mathsf{u}, 0}}}},{\theta_{\mathsf{u}}}r_{{b_{n,{k_{\mathsf{u}, 0}}}},{\ell _{\mathsf{u}, 0}}}^\alpha {I_n})}}{{\Gamma ({M_{{k_{\mathsf{u}, 0}}}})}}} \right] \nonumber \\
&\mathop  < \limits^{({\rm{b}})}& 1 - \sum\limits_{k = 1}^K {{A_k}} {\mathbbm{E}_{{r_{{b_{n,k}},{\ell _{\mathsf{u}, 0}}}}}}{\mathbbm{E}_{{I_n}}}\left[ {{{\left( {1 - {e^{ - {S_{{M_k}}}{\theta_{\mathsf{u}}}r_{{b_{n,k}},{\ell _{\mathsf{u}, 0}}}^\alpha {I_n}}}} \right)}^{{M_k}}}} \right] \nonumber\\
&\mathop  = \limits^{({\rm{c}})}& 1 - \sum\limits_{k = 1}^K {\sum\limits_{m = 0}^{{M_k}} {\binom{M_k}{m}{{\left( { - 1} \right)}^m}} {A_k}{\mathbbm{E}_{{r_{{b_{n,k}},{\ell _{\mathsf{u}, 0}}}}}}\left[ {{\mathcal{L}_{{I_n}}}\left( {m{S_{{M_k}}}{\theta_{\mathsf{u}}}r_{{b_{n,k}},{\ell _{\mathsf{u}, 0}}}^\alpha } \right)} \right]} \nonumber\\
&=& 1 - \sum\limits_{k = 1}^K {\sum\limits_{m = 0}^{{M_k}} {\binom{M_k}{m}{{\left( { - 1} \right)}^m}} {A_k}} \int_0^\infty  {{f_{{r_{{b_{n,k}},{\ell _{\mathsf{u}, 0}}}}}}({r_0}){\mathcal{L}_{{I_n}}}\left( {m{S_{{M_k}}}{\theta_{\mathsf{u}}}r_0^\alpha } \right){\rm{d}}{r_0}}.
\end{eqnarray}\setlength{\arraycolsep}{5pt}\endgroup}\noindent\rule[0.25\baselineskip]{\textwidth}{0.1pt}\vspace{-4mm}\end{figure*}where (a) is due to the fact that ${{{| {{\bf{h}}_{b_{n,k_{{{\mathsf{u}}},0}},\ell_{{\mathsf{u}},0}}^{\rm  T}{{\bf{w}}_{b_{n,k_{{{\mathsf{u}}},0}}}}} |}^2}}\mathop \sim\limits^d \Gamma(M_{k_{{{\mathsf{u}}},0}}, 1)$ and $A_{k_{{{\mathsf{u}}},0}}$ is the probability that the typical user $\ell_{{\mathsf{u}},0}$ is associated with tier $k_{{{\mathsf{u}}},0}$; (b) follows from a lower bound on the incomplete gamma function, i.e., $\frac{\gamma(a,b)}{\Gamma(a)}>\left(1-e^{-S_ab}\right)^a$ for $a>1$; and (c)  follows from the binomial expansion. Substituting (\ref{eqPDF}) and (\ref{eqappLIns}) into (\ref{eqappSTPUB}) and using some algebraic manipulations, we obtain the upper bound on $\textsf{p}_{{\mathsf{u}},n}(\mathbf{T}_n)$, as shown in (\ref{eqRTPub}).  Thus, we complete the proof.$\hfill\blacksquare$

\section{Proof of Property \ref{propRTP1}}\label{proofpropRTP1}

When $M_k=M$ and $\phi_k=\phi$, for all $k\in\mathcal{K}$, based on (\ref{eqRTPlb}), we have (\ref{eqAppplun}), {as shown at the top of the next page,}
\begin{figure*}[!t]\vspace{-3mm}
\begin{align}\label{eqAppplun}
\frac{\partial \textsf{p}^{\textsf{L}}_{{{\mathsf{u}}},n}(\mathbf{T}_n)}{{\partial {T_{n,k}}}} =  \frac{{{\lambda _k}{{\left( {\phi {P_k}} \right)}^\delta }}}{{{{\left( {{f_m}({{\bf{T}}_n},{\theta _{\mathsf{u}}})} \right)}^2}}}  \sum_{m=0}^{M-1}\left(1-\frac{m}{M}\right)\left(V_{m,M}(\xi ,{\theta _{\mathsf{u}}})\sum_{j = 1}^K {{\lambda _j}{{\left( {\phi {P_j}} \right)}^\delta }}  + {W_m}({\theta _{\mathsf{u}}})\right).
\end{align}
\noindent\rule[0.25\baselineskip]{\textwidth}{0.1pt}
\end{figure*}where $V_{m,M}(\xi ,{\theta _{\mathsf{u}}})$ and ${W_m}({\theta _{\mathsf{u}}})$ are given by (\ref{eqVm}) and (\ref{eqWm}), respectively.
First, it is easy to see that $\sum_{m=0}^{M-1}\left(1-\frac{m}{M}\right){W_m}({\theta _{\mathsf{u}}})\ge 0$ due to ${W_m}({\theta _{\mathsf{u}}})\ge 0$. Denote $Y\triangleq \sum_{m=0}^{M-1}\left(1-\frac{m}{M}\right)V_{m,M}(\xi ,{\theta _{\mathsf{u}}})$. Then, based on the following Lemma~\ref{lemmaSignY}, we have $Y \ge 0$. Hence, we have $\frac{\partial \textsf{p}^{\textsf{L}}_{{{\mathsf{u}}},n}(\mathbf{T}_n)}{{\partial {T_{n,k}}}} > 0$, which completes the~proof.~$\hfill\blacksquare${
\begin{Lemma}[Sign of $Y$]\label{lemmaSignY}
We have $Y\ge0$.
\end{Lemma}

\indent\indent \emph{Proof:} According to (\ref{eqVm}), we consider the cases of $\xi=1$ and $\xi\neq 1$, respectively. If $\xi=1$, we have $Y = {\sum_{m = 0}^{M - 1} {\left( {1 - \frac{m}{M}} \right)} \frac{{\Gamma (M + \delta )}}{{\Gamma (M)}}\Gamma (1 - \delta ){\theta ^\delta }}$ and thus $Y\ge 0$ holds clearly. If $\xi \neq 1$, for ease of illustration, we rewrite $Y$ as $Y = Y_1 Y_2 $, where $Y_1$ is given by (\ref{eqyxi}), {as shown at the top of the next page}, and $Y_2\triangleq \sum_{m=0}^{M-1}\left(1-\frac{m}{M}\right) \frac{(-1)^m\delta_{(m)}}{m!}$.
We first show $Y_1 \ge 0$. To be specific, we further rewrite $Y_1$ as $Y_1= \Gamma (1 - \delta ){\theta ^\delta }\frac{1}{{{{\left( {1 - \xi } \right)}^{M - 1}}}} y(\xi)$, where $y(\xi)\triangleq {\Gamma (1 + \delta ) - \sum_{i = 0}^{M - 2} {{{\left( {1 - \xi } \right)}^i}\frac{{\Gamma (i + 1 + \delta )}}{{\Gamma (i + 1)}}{\xi ^{\delta  + 1}}} } $ and $\frac{\rm d}{{\rm d}\xi}y(\xi) = - {\xi^\delta }{(1 - \xi)^{M - 2}}\frac{{\Gamma (M  + \delta )}}{{\Gamma (M - 1)}}$. Clearly, if $\xi < 1$, we have $Y_1 \ge 0$ due to $\Gamma (1 - \delta ){\theta ^\delta }\frac{1}{{{{\left( {1 - \xi } \right)}^{M - 1}}}} \ge 0$, $\frac{\rm d}{{\rm d}\xi}y(\xi) \le 0$, $y(0)=\Gamma(1+\delta)>0$ and $y(1)=0$; if $\xi > 1$ and $M$ is odd, i.e., $M=1,3,\cdots$, we have $Y_1 \ge 0$ due to $\Gamma (1 - \delta ){\theta ^\delta }\frac{1}{{{{\left( {1 - \xi } \right)}^{M - 1}}}} \ge 0$, $\frac{\rm d}{{\rm d}\xi}y(\xi) \ge 0$, and $y(1)=0$; if $\xi > 1$ and $M$ is even, i.e., $M=2,4,\cdots$, we have $Y_1 \ge 0$ due to $\Gamma (1 - \delta ){\theta ^\delta }\frac{1}{{{{\left( {1 - \xi } \right)}^{M - 1}}}} \le 0$, $\frac{\rm d}{{\rm d}\xi}y(\xi) \le 0$, and $y(1)=0$.
Next, we show $Y_2 > 0$. Specifically, we have $Y_2 \ge \sum_{m=0}^{M-1} \frac{(-1)^m\delta_{(m)}}{m!} = (-1)^{M-1}\binom{\delta-1}{M-1} > 0$. Thus, if $\xi\neq 1$, we still have $Y\ge 0$. By considering both cases of $\xi=1$ and $\xi\neq1$, we complete the proof.~$\hfill\blacksquare$

\begin{figure*}[!t]\vspace{-3mm}
\begin{equation}\label{eqyxi}
Y_1 \triangleq  {\Gamma (1 - \delta ){\theta ^\delta }\frac{1}{{{{\left( {1 - \xi } \right)}^{M - 1}}}}\left( {\Gamma (1 + \delta ) - \sum\limits_{i = 0}^{M - 2} {{{\left( {1 - \xi } \right)}^i}\frac{{\Gamma (i + 1 + \delta )}}{{\Gamma (i + 1)}}{\xi ^{\delta  + 1}}} } \right)}.
\end{equation}\noindent\rule[0.25\baselineskip]{\textwidth}{0.1pt}
\end{figure*}

}

\section{Proof of Property~\ref{propRTPconcavity}}\label{proofpropRTPconcavity}

{Due to the space limitations, we only prove the property that $\textsf{p}^{\textsf{U}}_{{{\mathsf{u}}},n}(\mathbf{T}_n) = \textsf{p}^{\textsf{U}, 1}_{{{\mathsf{u}}},n}(\mathbf{T}_n) - \textsf{p}^{\textsf{U}, 2}_{{{\mathsf{u}}},n}(\mathbf{T}_n) +1$ is a DC function of $T_{n,k}$. Note that the property that $\textsf{p}^{\textsf{L}}_{{{\mathsf{u}}},n}(\mathbf{T}_n) $ is a concave function of~$T_{n,k}$ can be proved by following similar steps. Denote $T_n \triangleq \sum_{k\in\mathcal{K}}\lambda_k(\phi P_k)^{\delta}T_{n,k}$. By using the definition of $f_0(\mathbf{T}_n, \theta)$ in (\ref{eqfm}), we rewrite $\textsf{p}^{\textsf{U}, i}_{{{\mathsf{u}}}, n}(\mathbf{T}_n)$ as (\ref{eqPUiunTnrew}), {as shown at the top of this page},
\begin{figure*}[!t]\vspace{-6mm}
\begin{align}\label{eqPUiunTnrew}
\textsf{p}^{\textsf{U}, i}_{{{\mathsf{u}}}, n}(\mathbf{T}_n)  &= \sum\limits_{m \in {{\cal M}^i}} \binom{M}{m} \frac{{{T_n}}}{{\left( {{{\tilde U}_m} - {{\tilde V}_m}} \right){T_n} + \sum_{k \in {\cal K}} {{\lambda _k}{{(\phi {P_k})}^\delta }{{\tilde V}_m}}  + {W_m}(m{S_M}{\theta_{\mathsf{u}}})}}.
\end{align}
\noindent\rule[0.25\baselineskip]{\textwidth}{0.1pt}\end{figure*}where $\tilde{U}_m\triangleq U_{0,M}(\xi, mS_{M}\theta_{\mathsf{u}})$ and $\tilde{V}_m\triangleq V_{0,M}(\xi, mS_{M}\theta_{\mathsf{u}})$. Then, we have (\ref{eqGradeqPUiunTnrew}), {as shown at the top of this page}.
\begin{figure*}[!t]\vspace{-6mm}
\begin{align}\label{eqGradeqPUiunTnrew}
\frac{{{\partial ^2}\textsf{p}_{u,n}^{\textsf{U},i}({{\bf{T}}_n})}}{{\partial T_n^2}} = \frac{{ - 2\left( {\sum_{k \in \cal K} {{\lambda _k}{{(\phi {P_k})}^\delta }{{\tilde V}_m}}  + {W_m}(m{S_M}{\theta_{\mathsf{u}}})} \right)\left( {{{\tilde U}_m} - {{\tilde V}_m}} \right)}}{{{{({f_0}({{\bf{T}}_n},m{S_M}{\theta_{\mathsf{u}}}))}^2}}}.
\end{align}
\noindent\rule[0.25\baselineskip]{\textwidth}{0.1pt}\vspace{-5mm}\end{figure*}
By following similar proof steps for Lemma~1 in Appendix~C, one can show that ${{{\tilde U}_m} - {{\tilde V}_m}} \ge 0$, ${{\tilde V}_m}\ge 0$ and ${W_m}(m{S_M}{\theta_{\mathsf{u}}}) \ge 0$, and thus, we have $\frac{\partial\textsf{p}^{\textsf{U}, i}_{{{\mathsf{u}}}, n}(\mathbf{T}_n)}{{\partial T_n}} \le 0$, implying that $\textsf{p}^{\textsf{U}, i}_{{{\mathsf{u}}}, n}(\mathbf{T}_n)$ is concave w.r.t. $T_n$. Finally, since $T_n$ is an affine mapping of $T_{n,k}$, according to \cite{convexoptimization}, we can conclude that $\textsf{p}^{\textsf{U}, i}_{{{\mathsf{u}}}, n}(\mathbf{T}_n)$ is also a concave function of $T_{n,k}$, which completes the proof. $\hfill\blacksquare$}

\section{Proof of Theorem \ref{TheoremSP}}\label{proofTheoremSP}

Denote ${ I_n^{\mathsf{e}} \triangleq I_{b_{n,k_{\mathsf{e},0}}}^{\mathsf{e}} +  {\sum_{k\in\mathcal{K}} (I_{n,k}^{\mathsf{e}} + I_{-n,k}^{\mathsf{e}} ) + I_{\mathtt{J}}^{\mathsf{e}} }}$, {$\mathbf{SIR}_n\triangleq\left(\mathbf{SIR}_{n,k_{\mathsf{e},0}}\right)_{k_{\mathsf{e},0}\in\mathcal{K}}$ with $\mathbf{SIR}_{n,k_{\mathsf{e},0}} \triangleq (\mathrm{SIR}_{b_{n,k_{\mathsf{e},0}},{\ell}_{\mathsf{e},0}})_ {b_{n,k_{\mathsf{e},0}}\in\Phi_{n,k_{\mathsf{e},0}}}$} and $\theta_{{{\mathsf{e}}}} \triangleq 2^{R_{{{\mathsf{e}}}}}-1$. Then, substituting (\ref{eqsirusereve}) into (\ref{eqqE}), we~have
{\begingroup\makeatletter\def\f@size{9.5}\check@mathfonts
\def\maketag@@@#1{\hbox{\m@th\normalsize\normalfont#1}}\setlength{\arraycolsep}{0.0em}
\begin{align}
&\textsf{p}_{{\mathsf{e}},n}(\mathbf{T}_n) \nonumber\\
 &= \Pr\left[\max_{k_{\mathsf{e},0}\in\mathcal{K}, \; b_{n,k_{\mathsf{e},0}}\in\Phi_{n,k_{\mathsf{e},0}}} \log_2\left(1+\mathrm{SIR}_{b_{n,k_{\mathsf{e},0}},{\ell}_{\mathsf{e},0}}\right) \le R_{\mathsf{e}} \right]\nonumber\\
& = \Pr\left[\bigcap_{k_{\mathsf{e},0}\in\mathcal{K}, \; b_{n,k_{\mathsf{e},0}}\in\Phi_{n,k_{\mathsf{e},0}}}  \left(\mathrm{SIR}_{b_{n,k_{\mathsf{e},0}},{\ell}_{\mathsf{e},0}} \le \theta_{\mathsf{e}} \right)\right]\nonumber\\
&\mathop=\limits^{(\mathrm{a})} 1 - \Pr\left[\bigcup_{k_{\mathsf{e},0}\in\mathcal{K}, \; b_{n,k_{\mathsf{e},0}}\in\Phi_{n,k_{\mathsf{e},0}}} \left(\mathrm{SIR}_{b_{n,k_{\mathsf{e},0}},{\ell}_{\mathsf{e},0}} > \theta_{\mathsf{e}} \right) \right]\nonumber\\
&= 1 - \mathbb{E}_{\mathbf{SIR}_n}\left[\mathbbm{1}\left[\bigcup_{k_{\mathsf{e},0}\in\mathcal{K}, \; b_{n,k_{\mathsf{e},0}}\in\Phi_{n,k_{\mathsf{e},0}}} \left(\mathrm{SIR}_{b_{n,k_{\mathsf{e},0}},{\ell}_{\mathsf{e},0}} > \theta_{\mathsf{e}} \right) \right]\right]\nonumber\\
&\mathop=\limits^{(\mathrm{b})} 1 - \sum_{k_{\mathsf{e},0}\in\mathcal{K}}\mathbb{E}_{\mathbf{SIR}_{n,k_{\mathsf{e},0}}}\left[\sum_{{ b_{n,k_{\mathsf{e},0}}\in\Phi_{n,k_{\mathsf{e},0}}}}\mathbbm{1}\left[ \mathrm{SIR}_{b_{n,k_{\mathsf{e},0}},{\ell}_{\mathsf{e},0}} > \theta_{\mathsf{e}} \right]\right] \nonumber \\
&\mathop=\limits^{(\mathrm{c})}1 - \sum\limits_{j \in {\cal K}} {2\pi {\lambda _{n,j}}{T_{n,j}}\int_0^\infty  {\Pr \left[ {{{\left| {{\bf{h}}_{{b_{n,j}},{\ell _{\mathsf{e},0}}}^{\rm{T}}{{\bf{w}}_{{b_{n,j}}}}} \right|}^2} > {\theta_{\mathsf{e}}}{r^\alpha }I_n^{\mathsf{e}}} \right]}  } r{\rm{d}}r \nonumber\\
&\mathop=\limits^{(\mathrm{d})}1 - \sum\limits_{j \in {\cal K}} {2\pi {\lambda _{n,j}}{T_{n,j}}\int_0^\infty  {{{\cal L}_{I_n^{\mathsf{e}}}}\left( {{\theta_{\mathsf{e}}}{r^\alpha }} \right)} } r{\rm{d}}r,\label{eqappctp}
\end{align}\setlength{\arraycolsep}{5pt}\endgroup}where {(a) follows from the De Morgan's laws in the generalized form, and $\mathbbm{1}[X]$ denotes the indicator function of the event $X$, which takes value 1 when the event $X$ happens and value 0 when the event $X$ does not happen; (b) is in general an upper bound (by union bound) but holds with equality if at most one of the BSs can provide channel capacity greater than $R_{\mathsf{e}}$, which is precisely the case when we consider $R_{\mathsf{e}}>1$;} (c) is obtained by using the Campbell Mecke Theorem \cite{Haenggi2012Stochastic}; and (d) follows from ${{\left| {{\bf{h}}_{{b_{n,j}},{\ell _{\mathsf{e},0}}}^{\rm{T}}{{\bf{w}}_{{b_{n,j}}}}} \right|}^2} \mathop \sim\limits^d \exp(1)$. Thus, to calculate $\textsf{p}_{{{{\mathsf{e}}}},n}(\mathbf{T}_n, R_{{{\mathsf{e}}}})$, we only need to calculate ${{\cal L}_{I_n^{\mathsf{e}}}}(s)$. Note that, we have ${{\cal L}_{I_n^{\mathsf{e}}}}(s) = \mathcal{L}_{I_{b_{n,k_{\mathsf{e},0}}}^{\mathsf{e}}}(s) {{\cal L}_{{I_{\mathtt{J}}^{\mathsf{e}}}}}(s)  \prod_{k\in\mathcal{K}} {{{\cal L}_{{I_{n,k}^{\mathsf{e}}}}}(s)} {{{\cal L}_{{I_{ - n,k}^{\mathsf{e}}}}}(s)}$. By following the similar steps as in the derivation of $\mathcal{L}_{I_n}^{\mathsf{u}}(s)$ in Appendix~A, we have $\mathcal{L}_{I_{b_{n,k_{\mathsf{e},0}}}^{\mathsf{e}}}(s) = {\left( {1 + {\xi _j}s{r^{ - \alpha }}} \right)^{1 - {M_j}}}$, ${{\cal L}_{{I_{\mathtt{J}}^{\mathsf{e}}}}}(s)={{\cal L}_{{I_{\mathtt{J}}^{\mathsf{u}}}}}(s)$ and
{\begingroup\makeatletter\def\f@size{9.5}\check@mathfonts
\def\maketag@@@#1{\hbox{\m@th\normalsize\normalfont#1}}\setlength{\arraycolsep}{0.0em}
\begin{eqnarray}
&&\prod\limits_{k\in\mathcal{K}} {{{\cal L}_{{I_{n,k}^{\mathsf{e}}}}}(s)}    {{{\cal L}_{{I_{ - n,k}^{\mathsf{e}}}}}(s)} \nonumber \\
&&=
\begin{cases}
 \exp \left( { - \pi \sum\limits_{k = 1}^K {{\lambda _k}{{\left( {\frac{{{\phi _k}{P_k}}}{{{\phi _j}{P_j}}}} \right)}^\delta }\frac{{\Gamma ({M_k} + \delta )}}{{\Gamma ({M_k})}}\Gamma (1 - \delta ){s^\delta }} } \right), & \mbox{if } \xi_k=1, \\
 \exp \left( { - \pi \sum\limits_{k = 1}^K {{\lambda _k}{{\left( {\frac{{{\phi _k}{P_k}}}{{{\phi _j}{P_j}}}} \right)}^\delta }A(s)} } \right), & \mbox{otherwise},
\end{cases}\nonumber
\end{eqnarray}\setlength{\arraycolsep}{5pt}\endgroup}where $A(s)$ is given by (\ref{eqappAs}). Then, we have
{\begingroup\makeatletter\def\f@size{9.5}\check@mathfonts
\def\maketag@@@#1{\hbox{\m@th\normalsize\normalfont#1}}\setlength{\arraycolsep}{0.0em}
\begin{align}\label{eqappLIne}
{{\cal L}_{I_n^{\mathsf{e}}}}\left( {{\theta_{\mathsf{e}}}{r^\alpha }} \right) =
\frac{{{{\left( {{\phi _j}{P_j}} \right)}^\delta }{{\left( {1 + {\xi _j}{\theta _e}} \right)}^{1 - {M_j}}}}}{{{W_0}\left( {\theta _e } \right) + \sum_{k = 1}^K {{\lambda _k}{{\left( {{\phi _k}{P_k}} \right)}^\delta }{V_{0,{M_k}}}\left( {1,{\theta _e}} \right)} }},
\end{align}\setlength{\arraycolsep}{5pt}\endgroup}where ${W_0}({\theta})$ and ${V_{0,M}}\left( {{\xi },{\theta}} \right)$ are given by Theorem~\ref{TheoremSTP}. Substituting (\ref{eqappLIne}) into (\ref{eqappctp}), we complete the proof of Theorem~\ref{TheoremSP}.~$\hfill\blacksquare$

%\section{Proof of the Lemma \ref{lemmacontiProblem}}\label{prooflemmacontiProblem}

%\bibliographystyle{IEEEtran}
%\bibliography{IEEEabrv,my_refs}

% Generated by IEEEtran.bst, version: 1.14 (2015/08/26)

\end{document}